\let\ifarxiv=\iftrue

\documentclass[12pt,a4paper]{article}
\ifarxiv\pdfoutput=1\fi

\usepackage{amsmath,amssymb}

\ifnum\pdfoutput=1
\usepackage[bookmarks=true,hyperfigures=true]{hyperref}
\usepackage[final]{graphicx}
\else
\usepackage[bookmarks=true,hyperfigures=true,hypertex]{hyperref}
\usepackage[draft]{graphicx}
\fi

\usepackage[nosort]{cite}
\usepackage[hyperref,bulletsep]{collect}

\ifx\hypersetup\sadfkjashdfkxja\else
\hypersetup{pdftitle={Scattering in Mass-Deformed N>=4 Chern-Simons Models}}
\hypersetup{pdfauthor={Abhishek Agarwal, Niklas Beisert, Tristan McLoughlin}}
\hypersetup{pdfsubject={}}
\hypersetup{pdfkeywords={Mass-Deformed Super-Poincare}}
\hypersetup{plainpages=false}
\hypersetup{bookmarksnumbered=true}
\hypersetup{pdfstartview=FitH}
\hypersetup{pdfpagemode=UseNone}
\hypersetup{colorlinks=false}
\hypersetup{citebordercolor={.5 1 .5}}
\hypersetup{urlbordercolor={.5 1 1}}
\hypersetup{linkbordercolor={1 .7 .7}}
\fi

\numberwithin{equation}{section}

\usepackage[a4paper,left=2.8cm,right=2.8cm,top=2.8cm,bottom=2.8cm]{geometry}

\allowdisplaybreaks[3]

\usepackage[font={small},labelfont={bf},width=0.85\textwidth]{caption}

\makeatletter
\let\old@startsection=\@startsection
\renewcommand{\@startsection}[6]{\old@startsection{#1}{#2}{#3}{#4}{#5}{#6\mathversion{bold}}}
\makeatother

\def\[{\begin{equation}}
\def\]{\end{equation}}
\def\<{\begin{eqnarray}}
\def\>{\end{eqnarray}}

\newcommand{\nn}{\nonumber}
\newcommand{\nln}{\nonumber\\}
\newcommand{\nl}[1][0pt]{\nonumber\\[#1]&\hspace{-4\arraycolsep}&\mathord{}}

\newcommand{\earel}[1]{\mathrel{}&\hspace{-2\arraycolsep}#1\hspace{-2\arraycolsep}&\mathrel{}}
\newcommand{\eq}{\earel{=}}

\ifx\genfrac\sdflkaj

\else

\fi
\newcommand{\sfrac}[2]{{\textstyle\frac{#1}{#2}}}
\newcommand{\half}{\sfrac{1}{2}}
\newcommand{\ihalf}{\sfrac{i}{2}}

\newcommand{\order}[1]{\mathcal{O}(#1)}

\newcommand{\Lagr}{\mathcal{L}}
\newcommand{\superN}{\mathcal{N}}

\newcommand{\Tr}{\mathop{\mathrm{Tr}}}
\newcommand{\diag}{\mathop{\mathrm{diag}}}

\newcommand{\sign}{\mathop{\mathrm{sign}}}
\renewcommand{\Im}{\mathop{\mathrm{Im}}}
\newcommand{\smat}{\mathcal{S}}
\newcommand{\scat}{\mathcal{T}}
\newcommand{\singlet}{\mathbf{1}}

\newcommand{\twist}{\tilde}
\newcommand{\eps}{\epsilon}

\newcommand{\cstree}{\mathnormal{\Upsilon}^{(0)}}
\newcommand{\csbox}{\mathnormal{\Upsilon}^{(1)\,\square}}
\newcommand{\csbubu}{\mathnormal{\Upsilon}^{(1)\,\circ}}
\newcommand{\csbubt}{\mathnormal{\Upsilon}^{(1)\,\twist\circ}}

\newcommand{\intbub}{I_m}

\newcommand{\Integers}{\mathbb{Z}}
\newcommand{\Complex}{\mathbb{C}}
\newcommand{\Projspc}{\mathbb{P}}
\newcommand{\Reals}{\mathbb{R}}
\newcommand{\Quats}{\mathbb{H}}

\let\oldPhi=\Phi
\let\oldPsi=\Psi
\let\oldGamma=\Gamma
\let\oldDelta=\Delta
\let\oldSigma=\Sigma
\let\oldLambda=\Lambda
\let\oldTheta=\Theta
\let\oldPi=\Pi
\renewcommand{\Phi}{\mathnormal{\oldPhi}}
\renewcommand{\Psi}{\mathnormal{\oldPsi}}
\renewcommand{\Gamma}{\mathnormal{\oldGamma}}
\renewcommand{\Sigma}{\mathnormal{\oldSigma}}
\renewcommand{\Delta}{\mathnormal{\oldDelta}}
\renewcommand{\Theta}{\mathnormal{\oldTheta}}
\renewcommand{\Lambda}{\mathnormal{\oldLambda}}
\renewcommand{\Pi}{\mathnormal{\oldPi}}

\newcommand{\indup}[1]{_{\mathrm{#1}}}

\newcommand{\supup}[1]{^{\mathrm{#1}}}

\newcommand{\matr}[2]{\left(\begin{array}{#1}#2\end{array}\right)}
\newcommand{\alg}[1]{\mathfrak{#1}}
\newcommand{\grp}[1]{\mathrm{#1}}
\newcommand{\gen}[1]{\mathfrak{#1}}
\newcommand{\eom}[1]{\mathrm{E}[#1]}

\newcommand{\lrbrk}[1]{\left(#1\right)}
\newcommand{\bigbrk}[1]{\bigl(#1\bigr)}
\newcommand{\Bigbrk}[1]{\Bigl(#1\Bigr)}
\newcommand{\brk}[1]{(#1)}

\newcommand{\bigvev}[1]{\bigl\langle#1\bigr\rangle}
\newcommand{\bigcomm}[2]{\big[#1,#2\big]}
\newcommand{\comm}[2]{[#1,#2]}
\newcommand{\tprod}[2]{\langle#1,#2\rangle}
\newcommand{\tprods}[2]{\langle#1#2\rangle}

\newcommand{\acomm}[2]{\{#1,#2\}}

\newcommand{\bigabs}[1]{\bigl|#1\bigr|}

\newcommand{\set}[1]{\{#1\}}
\newcommand{\state}[1]{\mathopen{|}#1\mathclose{\rangle}}
\newcommand{\bra}[1]{\langle #1|}
\newcommand{\ket}[1]{|#1\rangle}
\newcommand{\braket}[2]{\langle #1|#2\rangle}

\ifx\href\asklfhas\newcommand{\href}[2]{#2}\fi
\newcommand{\arxivlink}[1]{\href{http://arxiv.org/abs/#1}{arxiv:#1}}

\makeatletter
\def\mr@ignsp#1 {\ifx\:#1\@empty\else #1\expandafter\mr@ignsp\fi}%
\newcommand{\multiref}[1]{\begingroup
\xdef\mr@no@sparg{\expandafter\mr@ignsp#1 \: }%
\def\mr@comma{}%
\@for\mr@refs:=\mr@no@sparg\do{\mr@comma\def\mr@comma{,}\ref{\mr@refs}}%
\endgroup}
\makeatother

\newcommand{\hypref}[2]{\ifx\href\asklfhas #2\else\href{#1}{#2}\fi}
\newcommand{\Secref}[1]{Section~\multiref{#1}}
\newcommand{\secref}[1]{Sec.~\multiref{#1}}
\newcommand{\Appref}[1]{Appendix~\multiref{#1}}
\newcommand{\appref}[1]{App.~\multiref{#1}}

\newcommand{\Figref}[1]{Figure~\multiref{#1}}
\newcommand{\figref}[1]{Fig.~\multiref{#1}}
\renewcommand{\eqref}[1]{(\multiref{#1})}

\makeatletter
\newlength{\apb@width}
\newcommand{\autoparbox}[2][c]{\settowidth{\apb@width}{#2}\parbox[#1]{\apb@width}{#2}}
\newcommand{\includegraphicsbox}[2][]{\autoparbox{\includegraphics[#1]{#2}}}
\makeatother

\begin{document}


\thispagestyle{empty}
\begin{flushright}\footnotesize
\texttt{\arxivlink{0812.3367}}\\
\texttt{AEI-2008-090}%
\end{flushright}
\vspace{0.5cm}

\begin{center}
{\Large\textbf{\mathversion{bold}%
Scattering in Mass-Deformed\\$\mathcal{N}\geq 4$ Chern--Simons Models}\par}
\vspace{1cm}

\textsc{A.~Agarwal, N.~Beisert and T.~McLoughlin}
\vspace{5mm}

\textit{Max-Planck-Institut f\"ur Gravitationsphysik\\
Albert-Einstein-Institut\\
Am M\"uhlenberg 1, D-14476 Potsdam, Germany}\vspace{3mm}

\verb+{abhishek,nbeisert,tmclough}@aei.mpg.de+
\par\vspace{1cm}

\ifarxiv\smash{\raisebox{-7cm}{\makebox[0cm][c]{\hspace*{26cm}\includegraphics[height=6cm]{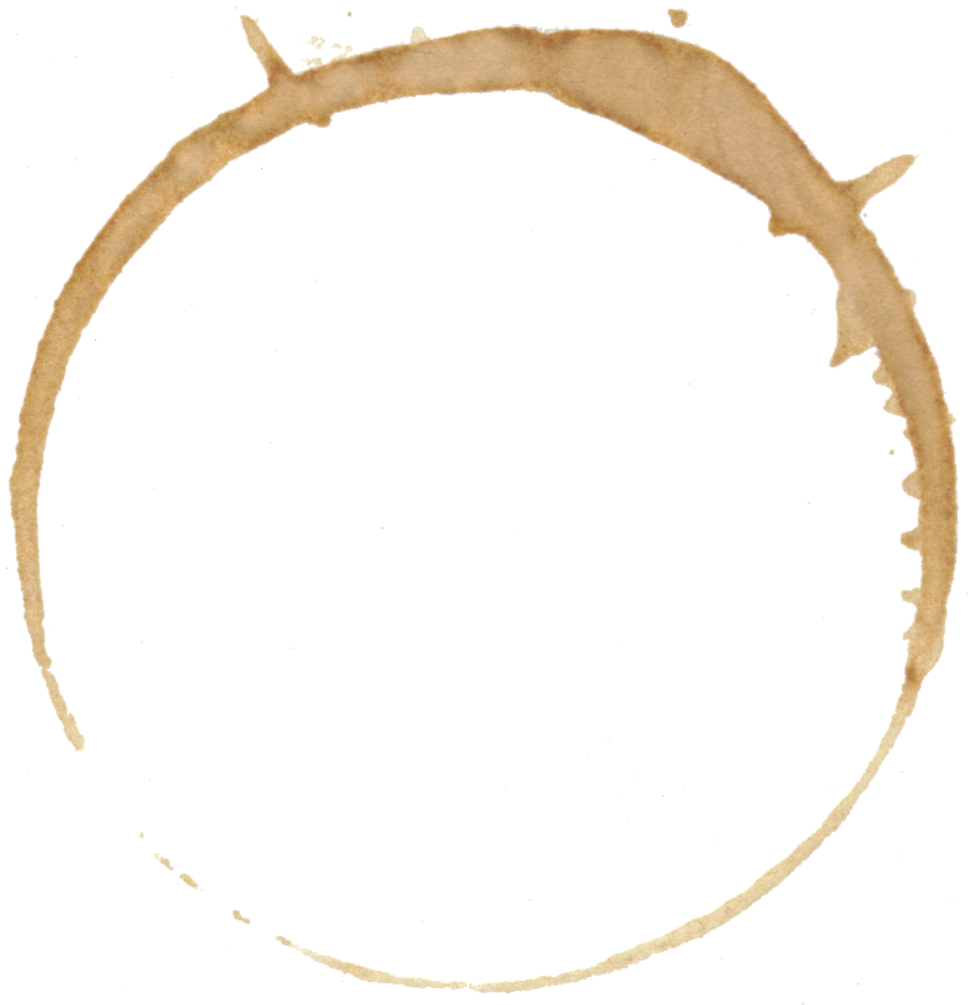}}}}\fi

\vfill

\textbf{Abstract}\vspace{5mm}

\begin{minipage}{12.7cm}
We investigate the scattering matrix in
mass-deformed $\mathcal{N}\geq 4$ Chern--Simons models
including as special cases the BLG and ABJM theories of multiple M2 branes.
Curiously the structure of this scattering matrix in three spacetime dimensions is equivalent to
(a) the two-dimensional worldsheet matrix found in the context of AdS/CFT integrability
and (b) the R-matrix of the one-dimensional Hubbard model.
The underlying reason is that all three models are based on
an extension of the $\mathfrak{psu}(2|2)$ superalgebra
which constrains the matrix completely.
We also compute scattering amplitudes in one-loop field theory
and find perfect agreement with scattering unitarity.
\end{minipage}

\end{center}
\vfill

\newpage

\section{Introduction}

The calculation of scattering matrices in
supersymmetric gauge theories has seen significant recent progress.
Exciting new insights, such as connections between scattering amplitudes
and Wilson loops \cite{Alday:2007hr,Drummond:2007aua,Brandhuber:2007yx,Alday:2007mf, Alday:2007he},
and the discovery of novel hidden symmetries,
namely dual superconformal invariance \cite{Drummond:2006rz, Bern:2006ew,Bern:2007ct,Drummond:2007cf,Drummond:2007au},
have led to a deeper understanding of the gauge theory.
Many of these results have been possible because of the development of powerful new tools,
such as twistor methods \cite{Witten:2003nn},
recursion relations \cite{Cachazo:2004kj,Britto:2004ap, Britto:2005fq}
and generalized unitarity \cite{Bern:1994zx,Bern:1994cg,Bern:1997sc,Britto:2004nc,Forde:2007mi},
for calculating on-shell quantities in Yang--Mills theories.
Further, in the specific case of
maximally supersymmetric four-dimensional $\superN=4$ Yang--Mills theory (YM)
the existence of a string dual has provided a tractable strong-coupling
description and has resulted in several impressive results and conjectures,
for a recent review of the  subject see \cite{Alday:2008yw}.

It is interesting to see if one can extend these results
to a broader class of theories particularly those with less supersymmetry.
One such class is the $\superN\geq 4$ supersymmetric Chern--Simons (SCS)
matter theories constructed by Hosomichi et al.\ (HLLLP)
in \cite{Hosomichi:2008jd,Hosomichi:2008jb}
which builds upon the construction of Gaiotto and Witten \cite{Gaiotto:2008sd}.
In the construction of Gaiotto and Witten,
the gauge group was chosen to be a particular subgroup of the symplectic group $\grp{Sp}(2n)$,
with no particular restrictions imposed on the representations
of the matter fields and where there is an $\alg{su}(2)\times \alg{su}(2)=\alg{so}(4)$ R-symmetry
which is required for the $\mathcal{N}= 4$ supersymmetry.
In \cite{Hosomichi:2008jd,Hosomichi:2008jb},
the matter content was augmented by twisted hypermultiplets
where the action of the  $\alg{su}(2)$'s
on the bosonic and fermionic degrees of freedom is interchanged relative
to the untwisted case.
In the absence of further constraints on the representations
of the two matter multiplets, this construction also results
in $\mathcal{N}= 4$ supersymmetry.
It was further shown that if both matter multiplets
are in the same representation
the supersymmetry extends to $\mathcal{N}=5$.
In the special case, where the representations of the matter multiplets
can be decomposed into a complex representation and its conjugate,
such as bifundamental representations of $\grp{SU}(N)\times \grp{SU}(M)$,
the supersymmetry was shown to be enhanced to  $\mathcal{N}= 6$.
When the representations are furthermore real corresponds to $\mathcal{N}= 8$.

The theories with $\superN=6$ and $8$ had been previously
found in the context of the low energy effective action
of multiple membranes
by Bagger, Lambert, Gustavsson (BLG)
and Aharony, Bergman, Jafferis, Maldacena (ABJM)
\cite{Bagger:2006sk,Bagger:2007jr, Gustavsson:2007vu, Bagger:2007vi, Bagger:2008se,Gustavsson:2008bf, Aharony:2008ug}.
The $\superN=6$ theory, of which there is an infinite $\grp{SU}(N)\times\grp{SU}(N)$ family \cite{Aharony:2008ug},
has been conjectured for finite values of $N$ and $k$,
the Chern--Simons level number, to describe the low energy dynamics of $N$ M2 branes
on $\Reals^{1,2}\times \Complex^4/\Integers_k$.
The gauge theory also possesses a well defined large $N$ limit,
which is obtained by taking both $N$ and $k$ to be large while $\lambda = N/k$ is held fixed.
In this limit, the theory is expected to be dual to string theory on $AdS_4\times \Complex\Projspc^3$,
which shares many similarities with the well studied case of string theory on $AdS_5\times S^5$.
Indeed, the conformal $\superN=6$ SCS gauge theory also shares some vital qualitative features
with its four dimensional $\superN=4$ YM counterpart,
including the fact that its spectrum at weak coupling is described by
an integrable quantum spin chain \cite{Minahan:2008hf}.
Furthermore, for the $\superN=4$ Yang--Mills theory the planar integrability,
which may be thought of as feature of the `world sheet physics' of the gauge theory,
is intimately tied to dual superconformal symmetry:
a property of the spacetime scattering matrix
of the gauge theory \cite{Beisert:2008iq,Berkovits:2008ic}.
It is interesting to ask if similar relationships and structures appear
in the $\superN=6$ Chern--Simons scattering amplitudes.

With these motivations in mind, we explore the scattering amplitudes
for a class of SCS theories with $\mathcal{N}\geq 4$ supersymmetry
which includes the $\superN=6$ and $8$ theories as special cases.
In fact we will study the
supersymmetry preserving massive deformations of the $4\leq \mathcal{N}\leq 8$ superconformal theories.
For this there are several independent and equally important reasons.
The mass-deformed theories are natural generalizations that subsume the superconformal SCS theories as special cases.  Since this is the largest class of SCS theories that can be analyzed with the methods that we use in the paper, we shall choose to study the massive models rather than their conformal limits.
Other than reasons of generality, the massive models also enable us to have well defined asymptotic
scattering states and the infra-red divergences associated with the on-shell external particles are
regularized (the infra-red divergences due to the Chern-Simons gauge field remain and
are treated, where necessary, separately). Finally, introducing the mass-deformation gives rise
to the supergroup $\grp{PSU}(2|2)$ as part of the space-time symmetry. This provides
an interesting  parallel with the spin-chain/world-sheet S-matrix that appeared in the recent
computation of the planar spectrum of the single trace operators in $\mathcal{N}=4$ SYM and, relatedly, in
the one-dimensional Hubbard model.

As in the study of most conformal field theories,
the three dimensional SCS models are plagued with problems of infra-red divergences,
which typically manifest themselves in the collinear and low momentum behavior of massless propagators.
One might hope that such problems can be remedied by making the theory massive.
In extended supersymmetric gauge theories,
making the dynamical matter fields massive is impossible
to achieve without violating some or all the supersymmetries of the massless cases.
However, in the special case of superconformal Chern--Simons models,
it is indeed possible to add masses to the matter fields while preserving
all the supersymmetries of the massless models, at the cost of losing (super)conformal invariance.
This  was established for $\mathcal{N}\geq 4$ SCS theories
in \cite{Hosomichi:2008jd,Hosomichi:2008jb} and in previous work
for the mass-deformation of the $\mathcal{N}=8$ M2 brane theory, \cite{Gomis:2008cv}
(more recent analyses of the mass deformations of SCS and M2 brane theories theory
can be found in \cite{Song:2008bi, Honma:2008un,Gomis:2008vc,Bergshoeff:2008ta}).
Unfortunately, the only degree of freedom contributed by the gauge filed,
namely its zero mode, does lead to residual mild infra-red divergences:
These divergences are apparent in the amplitudes when the
momenta of two external states become coincident 
and the propagator of an internal gluon becomes singular.
The main difference to massless models is the absence
of collinear divergences which have a larger phase space.
The infra-red divergences in our massive model are thus of a mild nature.
To the order at which we carry out the computations in this paper,
these additional potential divergences are largely irrelevant, see \secref{sec:UniUntwist}, 
however we expect them to play in important role at higher orders in perturbation theory,
and elaborate upon this issue later in the paper.

Rendering the SCS theories massive amounts to adding non-central extensions
of the  supersymmetry algebras, which generically take on the form
$\acomm{\gen{Q}}{\gen{Q}}\sim \gen{P}+m\gen{R}$:
$\gen{R}$ denoting the internal symmetry generators.
The massive theories that we study, typically
have the  mass deformed Poincar\'e algebras
\cite{Bergshoeff:2008ta}
\[\label{eq:TheGroup}
\grp{SL}(2,\Reals)\ltimes\grp{PSU}(2|2)\ltimes\Reals^3
\]
or
\[\label{eq:TheGroup8}
\grp{SL}(2,\Reals)\ltimes\grp{PSU}(2|2)^2\ltimes\Reals^3
\]
as their underlying symmetries which are among the exceptional
super-Poincar\'e algebras discussed in \cite{Nahm:1977tg}.
The appearance of the mass $m$ in the supersymmetry algebra
adds a new parameter to the theory.
Being part of the fundamental anti-commutation relations of the supercharges,
prevents the mass from `running', in the sense of renormalization group flows.
One may be tempted to view the mass-deformed theories as a one parameter family
of models extending each of the conformal $\mathcal{N} \geq 4$ SCS theories,
to which they reduce in the massless limit.
However, note that $m$ is the only mass scale
in the theory and thus all models which differ only in $m$ only
are expected to be related by an overall rescaling of dimensionful quantities.
In this sense the massless limit is singular and it
involves enhancement to superconformal symmetry.
For physical quantities the limit may nevertheless be smooth
as we shall observe in this paper.
Still one has to keep in mind that the IR singular behavior of some
quantities may be different in the limit and one has to replace the mass
by an alternative IR regulator.

It is worth noting that these massive supersymmetry algebras
have played an important role in a number of recent studies of supersymmetric gauge theories.
For instance, in the case of $\mathcal{N} = 4$ supersymmetric Yang--Mills theory in four dimensions,
$\alg{su}(2|2)$ played a crucial role as the symmetry of the scattering matrix
\cite{Staudacher:2004tk}
of the spin chain describing the planar limit of the gauge theory \cite{Beisert:2005tm}.
It was shown that the symmetry algebra uniquely fixes the spin chain scattering matrix
up to an overall prefactor. This S-matrix is, by the AdS/CFT correspondence,
the worldsheet S-matrix for the dual string theory \cite{Hofman:2006xt, Klose:2006zd, Arutyunov:2006yd}
but it was also shown that it is equivalent
to Shastry's R-matrix for the one-dimensional Hubbard model \cite{Beisert:2006qh}.
In the case of the  spin chain, the $\alg{sl}(2)$ automorphism of its symmetry algebra,
does not translate into a real symmetry of the system, as the quantum spin chain
is not relativistically invariant. As was pointed out in \cite{Lin:2005nh,Hofman:2006xt},
the $\alg{su}(2|2)$ in conjunction with its $\alg{sl}(2)$ automorphism,
is nothing but a mass-deformed supersymmetric extension of the three dimensional Lorentz algebra.
From the point of view of the three dimensional algebra,
the non-Lorentz invariance implies that the physical spectrum of the spin chain corresponds
to a preferred reference frame, see \cite{Hofman:2006xt}.

Various other interesting supersymmetric three dimensional Yang--Mills theories
with mass deformed super-Poincar\'e algebras as their symmetries have also recently been studied.
In particular, the D2 brane worldvolume theory, namely, $\mathcal{N} = 8$ super Yang--Mills
on $\Reals\times S^2$ and its spectrum was studied in \cite{Lin:2005nh}.%
\footnote{This theory can be viewed as a dimensional reduction of  $\mathcal{N} = 8$
super Yang--Mills on $\Reals\times S^3$ to  $\Reals\times S^2$}
In the same paper,  supersymmetric Yang--Mills Chern--Simons theories,
with various degrees of supersymmetry, were also formulated.
A salient feature of these gauge theories is that they have massive spectra,
as well as propagating gluonic degrees of freedom.
The gluons are rendered massive by either requiring the spacial geometry to be $S^2$,
or by introducing Chern--Simons terms in their actions.
In this regard, the theories studied in the present paper
depart substantially from the examples of the super Yang--Mills theories mentioned above.
In our case, there are no propagating gluons, the `gauge' part of the theories
being described by pure Chern--Simons terms.
Furthermore, the spacial part of the geometries underlying
the gauge theories will be taken to be $\Reals^2$.

In this paper, we study the $2\leftrightarrow 2$ scattering processes
in all the massive SCS theories mentioned above.
One of the main observations is that, as in the case of scattering processes
in the spin chain corresponding to  $\mathcal{N} =4$ Yang--Mills theory in four dimensions,
the matrix structure of the two particle (spacetime) scattering matrix
is completely fixed by supersymmetry. For pure $\mathcal{N} =4$ SCS  theories,
without twisted hypermultiplets,
this means that relevant scattering matrix is completely determined by supersymmetry
up to a single undetermined function.
Indeed the scattering matrix for the Chern--Simons theory is formally
identical to the spin chain S-matrix.
For the more general case of mixed $\mathcal{N}= 4$ supersymmetry,
i.e.\ SCS theories with twisted hypermultiplets,
supersymmetry leaves one with three undetermined functions.
As shown later in the paper, supersymmetry enhancement to $5\leq \mathcal{N} \leq 8$
can be obtained by imposing suitable constraints on the mixed $\mathcal{N}=4$ theories.
The number of undetermined functions reduces from three to two or one upon supersymmetry enhancement.
Importantly, being a direct consequence of the supersymmetry algebra,
the structure of the scattering matrix derived in this fashion is expected
to hold to \emph{all} orders in perturbation theory.%
\footnote{We should point out that divergences in
scattering amplitudes may potentially deform the
supersymmetry transformation laws
in analogy to what happens for conformal theories.
For example, the supersymmetry algebra
requires the dimension of spacetime to be exactly three,
while in dimensional regularization it is $3-2\epsilon$.
Subleading terms in the $\epsilon$ expansion may not have the same structure.}
This result has a parallel in the spacetime scattering matrix
of $\mathcal{N}=4$ super Yang--Mills theory in four dimensions,
where all the four particle scattering matrix elements
can be determined in terms of a single function \cite{Parke:1985pn,Mangano:1990by}.

Apart from establishing the matrix structure of the scattering matrix,
we compute the undetermined functions for the mass-deformed SCS theories
at the tree and one-loop level.
The perturbative calculations also lead to independent checks
that the relations between the various elements of the scattering matrix
predicted by supersymmetry are indeed satisfied.
We compute the one-loop correction to the scattering matrix in two different ways,
i.e.\ by the use of standard Feynman rules as well as by using unitarity.
As is well known, perturbative corrections to scattering amplitudes
can be computed efficiently by `gluing' lower order amplitudes together
using relations derived from unitarity.
However, in principle this only determines the piece of the amplitude
with branch cuts and so suffers from the `polynomial ambiguity'
whereby there are undetermined rational functions of the kinematical variables.
We demonstrate explicitly, by calculating specific elements using standard off-shell methods,
that the amplitudes can indeed be completely evaluated
using the discontinuities across the cuts of the integrands,
and that there are no rational functions unrelated to terms with branch cuts.
We then use this simplifying feature to compute all the two particle scattering matrix elements,
at the one loop order, using unitarity.
Interestingly, while our calculations yield non-trivial one loop corrections
to the two particle scattering matrix of $\mathcal{N} = 4$ SCS theories,
with or without additional hypermultiplets,
we find that all such corrections vanish identically for the cases of $\mathcal{N} \geq 5$ supersymmetry.

The paper is organized as follows.
In \Secref{sec:SusyPart} we elaborate upon the realization of extended supersymmetry algebras
and their mass-deformations in supersymmetric Chern--Simons theories.
In particular, we discuss the realization on the supersymmetry algebra
on the asymptotic/scattering states of the gauge theories in question.
We also introduce a particular basis for spinors in three dimensions
that closely  resembles the  often employed  spinor-helicity basis
in the case of four dimensional theories.
Following this discussion, \Secref{sec:SusyScat},
we set-up the four particle scattering picture in terms of the asymptotic states
and derive the constraints imposed upon the scattering matrix elements by supersymmetry.
We show that the constraints can be solved,
leading to a complete determination of the matrix structure of the $2\leftrightarrow 2$ scattering matrix
of all the massive $\mathcal{N} \geq 4$ SCS theories.
Further, the explicit correspondence between two-dimensional worldsheet/spin chain scattering matrix
and the Chern--Simons $S$-matrix is described.
\Secref{sec:color_structures} is devoted to the analysis of color structure
of the scattering amplitudes, which is left unspecified in the sections outlined above.
Specifically, we focus on the interpretations of color ordering and planarity,
while leaving the choice of the gauge group to be as general as possible.
In the final two sections, \Secref{sec:fieldtheory} and \Secref{sec:unitarity},
perturbative calculations that verify the predictions of the supersymmetry algebra
as well as compute the unspecified functions at the tree and one-loop order, are presented.
As mentioned before, the perturbative computations are carried out
using both Feynman rules as well as unitarity methods whose one-loop validity is thus established.
We end the paper with an elaborate appendix,
where most of the details relevant to the Lagrangian formulation
of the massive SCS theories as well as useful details
regarding the supersymmetry algebra are contained.

\section{Supersymmetry and Asymptotic States}
\label{sec:SusyPart}

\subsection{Extended Supersymmetry in Chern--Simons Theories}
\label{sec:susy}

Let us start with a brief review of $\superN\geq 4$ supersymmetry
in three-dimensional quantum field theory coupled to a Chern--Simons gauge field.

\begin{figure}
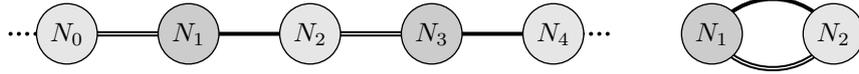
\centering
\includegraphicsbox{FigQuiver.mps}\qquad%
\includegraphicsbox{FigQuiverExt.mps}
\caption{The quiver structure of a generic
$\superN=4$ Chern--Simons theory with unitary gauge groups
(left) and of a $\superN=5,6,8$ Chern--Simons theory (right).
The solid and double lines represent
untwisted and twisted matter, respectively.
The circles represent gauge groups $\grp{U}(N_k)$ and gauge fields.}
\label{fig:structquiver}
\end{figure}

The study of the conformal case with $\grp{OSp}(\superN|4,\Reals)$ symmetry
was initiated for $\superN=4$ supersymmetry in \cite{Gaiotto:2008sd}; it was
extended to include additional twisted matter
and $\superN=5,6$ supersymmetry by \cite{Hosomichi:2008jd, Hosomichi:2008jb}.
This is in addition to the very many earlier and parallel developments in the
$\superN=6,8$ case briefly described earlier
\cite{Aharony:2008ug,Bagger:2006sk, Bagger:2007jr, Gustavsson:2007vu, Bagger:2007vi, Bagger:2008se, Gustavsson:2008bf, Schnabl:2008wj,Schwarz:2004yj}.
It was shown that there is a correspondence between the
permissible field content of such a model and the classification
of Lie superalgebras:
In general, the even part of the superalgebra specifies the gauge symmetry
for the Chern--Simons fields while the odd part specifies the matter content.
For $\superN\geq 5$ supersymmetry there is only one type of matter multiplet
and a simple Lie superalgebra fixes the model completely.
For $\superN=4$ supersymmetry, however,
there are two types of matter multiplets,
so-called untwisted and twisted hypermultiplets.
The field content in each of the two matter sectors is specified
by a semi-simple Lie superalgebra.
The even part of the two superalgebras must coincide
in order for the Chern--Simons sector to be defined consistently.
In particular, this leads to certain quivers
of simple Lie superalgebras, see \cite{Hosomichi:2008jd,Hosomichi:2008jb}.
The general structure of these quivers is illustrated in
\figref{fig:structquiver} for the example of unitary gauge groups:
Considering only the untwisted matter fields one finds
a direct sum of superalgebras $\alg{su}(N_{2k-1}|N_{2k})$.
Likewise the twisted matter fields define
a direct sum of superalgebras $\alg{su}(N_{2k}|N_{2k+1})$.
For orthosymplectic superalgebras the nodes must
alternate between orthogonal and symplectic algebras.%
\footnote{There may also be some more exotic quivers using
bosonic algebras of lower rank which may allow to switch between
the orthosymplectic and unitary series.
We also do not consider explicitly the $\alg{u}(1)$ factors
which are present in the unitary superalgebras $\alg{su}(N|M)$,
see \cite{Hosomichi:2008jd} for more details.}
Globally, the alternating chain of nodes can be either open or closed.
If the odd parts of the Lie superalgebras coincide as well,
the supersymmetry enhances to $\superN\geq 5$.
This is equivalent to a closed quiver of length 2, see \figref{fig:structquiver}.

For $\superN\geq 5$ supersymmetry the level $\superN$
merely depends on which particular basic Lie superalgebra
the model is based upon. The cases are summarized as follows
\[\label{eq:superalgclass}
\left.\begin{array}{c}
\alg{osp}(n|2m)\\\alg{d}(2,1;\alpha)\\\alg{g}(3)\\\alg{f}(4)
\end{array}\right\}
\superN=5,\qquad
\left.\begin{array}{c}
\alg{sl}(n|m)\\\alg{osp}(2|2m)
\end{array}\right\}
\superN=6,\qquad
\alg{psl}(2|2)\colon
\superN=8.
\]
The representation of the even part on the odd part distinguishes
the three types of superalgebras:
An irreducible representation leads to $\superN=5$ supersymmetry.
If it can be reduced into two conjugate representations
supersymmetry enhances to $\superN=6$. If furthermore the
two representations are isomorphic we obtain $\superN=8$ supersymmetry.

This classification can be translated into a classification of
continuous automorphisms of the superalgebras.
The $\alg{psl}(2|2)$ superalgebra is the only
superalgebra with an $\grp{Sp}(1)$ outer automorphism.
For $\alg{sl}(n|m)$ and $\alg{osp}(2|2m)$ there exist
$\grp{U}(1)$ automorphisms whose action coincides with
the $\alg{gl}(1)$ and $\alg{so}(2)$ parts of the superalgebra, respectively.
The remaining basic superalgebras have no continuous outer automorphisms.%
\footnote{One can thus identify the three types of superalgebras
in \protect\eqref{eq:superalgclass}
with the fields $\Reals,\Complex,\Quats$, respectively.}
The classification in terms of automorphisms is natural
when one views $\superN$-extended supersymmetry
from the point of view of $\superN=4$ supersymmetry:
When breaking $\grp{OSp}(\superN|4,\Reals)$ to manifest $\grp{OSp}(4|4,\Reals)$ notation
there must be an additional $\grp{SO}(\superN-4)$ flavor symmetry.
For $\superN=5$ this requires no automorphism while
for $\superN=6$ the required automorphism is $\grp{SO}(2)\simeq\grp{U}(1)$.
Finally, for $\superN=8$ we need an $\grp{SO}(4)\simeq\grp{Sp}(1)\times\grp{Sp}(1)$
automorphism. Each of the two $\alg{psl}(2|2)$ superalgebras provides
one copy of $\grp{Sp}(1)$. Here one could also consider a manifest
$\grp{OSp}(5|4,\Reals)$ notation where the single $\alg{psl}(2|2)$
superalgebra provides the $\grp{SO}(\superN-5)=\grp{SO}(3)\simeq\grp{Sp}(1)$ automorphism.

Let us now turn to the massive case,
which was investigated in \cite{Hosomichi:2008jd, Hosomichi:2008jb} (see
 \cite{Gomis:2008cv,Song:2008bi, Honma:2008un,Gomis:2008vc,Bergshoeff:2008ta} for
 related work particularly in the context of massive M2-brane theories).
There appears to be a one-to-one correspondence
between the massive and conformal $\superN\geq 4$ supersymmetric Chern--Simons theories;
for each conformal model there is a mass deformation with the same amount
of supersymmetry and for each massive model there is a conformal limit.
The only additional parameter in the massive models is one overall
mass scale $m$. The classification in terms of superalgebras remains the same.
This result is somewhat curious because the massive models preserve less
bosonic and only half of the fermionic symmetry
and might, in principle, be less restrictive.
We define the general mass-deformed $\superN=4$ Chern--Simons theory
in \appref{app:susyinteract}.

In the massive case, the bosonic spacetime symmetry reduces to the Poincar\'e
group $\grp{SL}(2,\Reals)\ltimes\Reals^3$. Supersymmetry will be specified
by some supergroup $\grp{G}$ which enters the full super-Poincar\'e algebra as
$\grp{SL}(2,\Reals)\ltimes \grp{G}\ltimes \Reals^3$.
This means that the Lorentz algebra $\grp{SL}(2,\Reals)$
acts as an automorphism on $\grp{G}$ and that the
algebra of supercharges closes onto the momenta $\Reals^3$.
For $\superN=4$ supersymmetry the internal bosonic symmetry is $\grp{SO}(4)$.
It joins with the supersymmetry generators into the supergroup
$\grp{G}=\grp{PSU}(2|2)$.
Altogether the mass-deformed $\superN=4$ super-Poincar\'e group is \eqref{eq:TheGroup}
\[\label{eq:TheGroupAgain}
\grp{SL}(2,\Reals)\ltimes\grp{PSU}(2|2)\ltimes\Reals^3.
\]
It is one of the exceptional cases of
super-Poincar\'e algebras discussed in \cite{Nahm:1977tg}.
It is exceptional, because the supercharges close
not only onto the momentum generators,
but also onto the internal symmetries.
This type of closure of spacetime supersymmetry
is otherwise known only from the superconformal cases.
Here it requires the introduction of a mass scale $m$
to give the relation $\acomm{\gen{Q}}{\gen{Q}}\sim \gen{P}+m\gen{R}$
a consistent dimension.

For $\superN\geq 5$ supersymmetry the supergroup $\grp{G}$ splits into
two pieces, $\grp{G}=\grp{G}\indup{A}\times\grp{G}\indup{B}$ with $\grp{G}\indup{A}=\grp{PSU}(2|2)$
and $\grp{G}\indup{B}$ a supergroup of odd dimension $2(\superN-4)$.
The Lorentz group $\grp{SL}(2,\Reals)$ must act on $\grp{G}\indup{B}$
as an automorphism and $\grp{G}\indup{B}$ must close onto the momenta $\Reals^3$.
The three cases are given by $\grp{G}\indup{B}=\grp{G}_{\superN-4}$ with
\[
\grp{G}_{1}=\Reals^{0|2},
\qquad
\grp{G}_{2}=\grp{U}(1)\ltimes\grp{PSU}(1|1)^2\ltimes \grp{U}(1),
\qquad
\grp{G}_{4}=\grp{PSU}(2|2).
\]
We shall discuss the associated superalgebras in more detail below.
Here the role of the automorphisms of the superalgebra defining the field
content is even more evident:
They serve as the even part of $\grp{G}_{\superN-4}$.
For $\superN=6$ the two $\grp{U}(1)$ automorphisms
appear in $\grp{G}_2$ while for $\superN=8$
the two $\grp{Sp}(1)$ automorphism are
equivalent to the two $\grp{SU}(2)$ factors in $\grp{G}_4$.%
\footnote{It is curious to observe that
mass-deformed $\superN=8$ Chern--Simons theory is, in some sense,
constructed upon four copies of the superalgebra $\grp{PSU}(2|2)$.}
The supergroup $\grp{G}_2$ resembles $\grp{G}_4$
in that the mass appears in the anticommutation
relation $\acomm{\gen{Q}}{\gen{Q}}\sim \gen{P}+m\gen{R}$.
However, this particular $\grp{U}(1)$ generator will
commute with the remaining algebra unlike what happens in $\grp{G}_4$.
Conversely, the supergroup $\grp{G}_1$ is almost trivial
and the mass does not appear there.

Our main concern in this paper will be the case of $\superN=4$ supersymmetry
where the super group is a single copy of $\grp{G}_4$.
Let us nevertheless close this part
with some remarks on $\superN\leq 4$. There appears to be the possibility of
combining any two of the superalgebras $\grp{G}_{1,2,4}$. For instance
there could in principle be a massive $\superN=4$ supersymmetric model
which preserves only $\grp{G}_2\times\grp{G}_2$ instead of $\grp{G}_4$.
In fact the former is a subgroup of the latter and thus one can expect
it to be less constraining. In the massless limit, however, both types of
modes would result in the same supergroup $\grp{OSp}(4|4,\Reals)$.

\subsection{Mass-Deformed \texorpdfstring{$\superN=4$}{N=4} Super-Poincar\'e Algebra}
\label{sec:superalg}

The mass-deformed $D=3$, $\superN=4$ super-Poincar\'e algebra of \eqref{eq:TheGroupAgain}
consists of the bosonic Poincar\'e generators
$\gen{L}_{\alpha\beta}=\gen{L}_{\beta\alpha}$,
$\gen{P}_{\alpha\beta}=\gen{P}_{\beta\alpha}$,
the internal $\alg{su}(2)\oplus\alg{su}(2)$ generators $\gen{R}_{ab}=\gen{R}_{ba}$,
$\gen{\dot R}_{\dot a\dot b}=\gen{\dot R}_{\dot b\dot a}$
and eight supercharges $\gen{Q}_{\alpha b\dot c}$.
The Lorentz and internal algebra is
specified by its action on spinor indices
($\state{\mathcal{X}_{\ldots}}$ denotes any state with the
indicated indices)
\<\label{eq:indextrans}
\gen{L}_{\alpha\beta}\state{\mathcal{X}_{\gamma}}\eq
\half\varepsilon_{\beta\gamma}\state{\mathcal{X}_{\alpha}}
+\half\varepsilon_{\alpha\gamma}\state{\mathcal{X}_{\beta}},
\nln
\gen{R}_{ab}\state{\mathcal{X}_{c}}\eq
\ihalf\varepsilon_{bc}\state{\mathcal{X}_a}
+\ihalf\varepsilon_{ac}\state{\mathcal{X}_b},
\nln
\gen{\dot R}_{\dot a\dot b}\state{\mathcal{X}_{\dot c}}\eq
\ihalf\varepsilon_{\dot b\dot c}\state{\mathcal{X}_{\dot a}}
+\ihalf\varepsilon_{\dot a\dot c}\state{\mathcal{X}_{\dot b}}.
\>
It remains to specify the anticommutator of supercharges
\[\label{eq:susy}
\acomm{\gen{Q}_{\alpha b\dot c}}{\gen{Q}_{\delta e\dot f}}=
\varepsilon_{be}\varepsilon_{\dot c\dot f}\gen{P}_{\alpha\delta}
-2m\varepsilon_{\alpha\delta}\varepsilon_{\dot c\dot f}\gen{R}_{be}
+2m\varepsilon_{\alpha\delta}\varepsilon_{be}\gen{\dot R}_{\dot c\dot f}.
\]
In addition to the standard momentum generator $\gen{P}_{\alpha\delta}$
it contains the internal rotation generators
$\gen{R}_{be}$ and $\gen{\dot R}_{\dot c\dot f}$
which otherwise only appear in superconformal algebras.
These dimensionless generators are multiplied by a common mass $m$
for the correct mass dimension.
The constant $m$ is physical and it sets the mass scale of this model.
Due to its appearance in the supersymmetry algebra it is protected from running.

For completeness we shall write the reality conditions
of the generators in the relevant real form of the supersymmetry algebra.
For the $\alg{sl}(2)$ Lorentz and $\alg{su}(2)\oplus\alg{su}(2)$ internal rotations
we require
\[\label{eq:RealLRR}
(\gen{L}_{\alpha\beta})^\ast=\gen{L}_{\alpha\beta},
\qquad
(\gen{R}_{ab})^\ast=\varepsilon^{ac}\varepsilon^{bd}\gen{R}_{cd},
\qquad
(\gen{\dot R}_{\dot a\dot b})^\ast=\varepsilon^{\dot a\dot c}\varepsilon^{\dot b\dot d}\gen{\dot R}_{\dot c\dot d}.
\]
The supersymmetry and momentum generators obey
\[\label{eq:RealQP}
(\gen{Q}_{\alpha b\dot c})^\ast=-\varepsilon^{bd}\varepsilon^{\dot c\dot e}\gen{Q}_{\alpha d\dot e},
\qquad
(\gen{P}_{\alpha\beta})^\ast=\gen{P}_{\alpha\beta}.
\]

\subsection{\texorpdfstring{$\superN=4$}{N=4} Asymptotic Particle Representation}
\label{eq:N4rep}

We can now turn to the transformation properties of the asymptotic particles.
These particles can belong to any $D=3$ quantum field theory whose
spacetime symmetry is the above mass deformed $\superN=4$ super-Poincar\'e algebra.
We assume that the particles are on shell,
gauge-invariant and do not interact.
In particular, this means that the action of the supercharges
is linear (i.e.\ the symmetry is not spontaneously broken)
and that the supersymmetry algebra
closes exactly without additional gauge terms
or terms proportional to the equations of motion.
In particular it applies to the mass-deformed $\superN=4$ Chern--Simons theories
outlined in \appref{app:susyinteract} at arbitrary coupling.
At weak coupling it is furthermore safe to identify
the asymptotic particles with the fields.

For a fixed time-like momentum $p_{\alpha\beta}=p_\mu\sigma^\mu{}_{\alpha\beta}$, the stabilizer
of the mass-deformed super-Poincar\'e algebra
is $\alg{u}(2|2)$.
The smallest non-trivial particle multiplet
thus corresponds to the (anti) fundamental representation
of $\alg{u}(2|2)$ consisting of two bosons and two fermions.
On shell these can be identified with the (twisted) hypermultiplets
of massive $\superN=4$ super Chern--Simons theory:
\[
\mbox{(untwisted) hypermultiplet: }
\state{\phi_{a}},
\state{\psi_{\dot a}},
\qquad
\mbox{twisted hypermultiplet: }
\state{\twist\phi_{\dot a}},
\state{\twist\psi_{a}}.
\]
These both transform under $\alg{su}(2)\oplus\alg{su}(2)$
according to the general rule \eqref{eq:indextrans} but we note
that the roles of the two different $\alg{su}(2)$ indices are switched
in the twisted case relative to the untwisted case.
The most general representation of the supercharges
on the hypermultiplets
compatible with $\alg{su}(2)\oplus\alg{su}(2)$ symmetry
is given by
\[\label{eq:susyrep}
\begin{array}[b]{rclcrcl}
\gen{Q}_{\alpha b\dot c}\state{\phi_{d}}
\eq\varepsilon_{bd}u_\alpha\state{\psi_{\dot c}},
&\quad&
\gen{Q}_{\alpha b\dot c} \state{\twist\phi_{\dot d}}
\eq\varepsilon_{\dot c\dot d}v_\alpha\state{\twist\psi_{b}},
\\[3pt]
\gen{Q}_{\alpha b\dot c} \state{\psi_{\dot d}}
\eq\varepsilon_{\dot c\dot d} v_\alpha\state{\phi_{b}},
&\quad&
\gen{Q}_{\alpha b\dot c} \state{\twist\psi_{d}}
\eq\varepsilon_{bd}u_\alpha\state{\twist\phi_{\dot c}}.
\end{array}
\]
Closure of the supersymmetry algebra \eqref{eq:susy}
implies the constraint
\[\label{eq:PolSpinor}
v_\alpha u_\beta=
-p_{\alpha\beta}
-im\varepsilon_{\alpha\beta}.
\]
Note that $(u_\alpha,v_\alpha)$ and $(v_\alpha,u_\alpha)$
are effectively the incoming/outgoing
polarizations for the massive spinors $\psi$ and $\twist\psi$,
cf.\ the oscillator representation of free
fermions in \eqref{eqn:Mode_exp}.

The mass of the asymptotic particles is
constrained by the atypicality condition
of the fundamental representation of $\alg{u}(2|2)$
to equal the mass $m$ appearing in the supersymmetry algebra \eqref{eq:susy}.
In particular, this implies that the mass of
the hypermultiplets cannot run in this model.%
\footnote{Depending on the renormalization scheme
the bare and physical masses can differ by a finite amount.}

Let us investigate the relation \eqref{eq:PolSpinor} in some more detail.
It implies that the particle momentum $p_{\alpha\beta}=p_\mu\sigma^\mu{}_{\alpha\beta}$
is a function of the spinors $u_\alpha$ and $v_\alpha$.
Therefore the representation of the stabilizer is specified
through a pair of spinors $(u_\alpha, v_\alpha)$ obeying the constraint
\[
\label{eq:PolMass}
\varepsilon^{\alpha\beta}v_\alpha u_\beta=-2im.
\]
The solutions of this constraint form the three-dimensional group manifold $\grp{SL}(2)$.
Conversely, the mass shell in three dimensions is merely the two-dimensional
hyperbolic space $H^2=\grp{SL}(2)/\grp{U}(1)$. Thus representations
of the little supersymmetry algebra carry
one additional $\grp{U}(1)$ label as compared to the bosonic
subalgebra.
This label can be adjusted by changing phases of the spinors
\[
u_\alpha\to e^{+i\alpha}u_\alpha,\qquad
v_\alpha\to e^{-i\alpha}v_\alpha,
\]
which does not change the relations \eqref{eq:PolSpinor,eq:PolMass}.
In \eqref{eq:susyrep} it can be seen to determine the relative phase
between the bosons and fermions.
The $\grp{U}(1)$ degree of freedom turns out to be inessential
and it can in principle be fixed by restricting
to a particular choice of phase $u(p),v(p)$ for each momentum $p$,
e.g.\ \eqref{eq:uvofp}.
This is possible because the mass shell is topologically trivial
and there is no global obstruction in choosing a $\grp{U}(1)$ element
at each point of the $\grp{SL}(2)/\grp{U}(1)$.
Nevertheless, it is not always advisable to do this for two reasons:
The spinors $u(p),v(p)$ are not covariant under Lorentz
transformations, they are merely covariant
up to phase.
Secondly, it is sometimes convenient to complexify momenta.
This however leads to a non-trivial topology
of the above $\grp{U}(1)$ fibration
and there is no globally consistent choice $u(p),v(p)$.
In particular, this leads to potential sign ambiguities
if one tries to define the spinors $u(p),v(p)$ for
the two mass shells with positive and negative energies
with a single analytic formula.
Therefore we prefer to specify representations through
the spinors $u,v$.
However, in particular if the sign of the particle energy $p_0$ is well-known,
it is safe to specify the spinors through the particle momentum $p$
as in \eqref{eq:uvofp}.

Finally we would like to discuss unitarity conditions of the
representation.
According to \eqref{eq:RealQP} and \eqref{eq:susyrep}
hermiticity of the supersymmetry generators implies
the unitarity condition
\[\label{eq:RepUni}
u_\alpha^\ast=+v_\alpha.
\]
This also leads to a real momentum $p_\mu$
according to \eqref{eq:PolSpinor}.
Moreover, the energy $p_{0}$ is positive definite
as usual in supersymmetry algebras.
Conversely, particle multiplets with negative energy obey
\[\label{eq:RepGUni}
u_\alpha^\ast=-v_\alpha
\]
and they transform in a graded unitary representation
where all bosonic generators are hermitian and the
supercharges are anti-hermitian.

The two types of representations discussed above are just the
simplest non-trivial representations of the mass-deformed
$\superN=4$ super-Poincar\'e group \eqref{eq:TheGroupAgain}.
The representation theory follows closely the one
of $\alg{su}(2|2)$, cf.\ \cite{Beisert:2006qh} and references therein:
There are short/atypical representations
$\langle k,l\rangle$ of dimension $4(k+1)(l+1)+4kl$.
The corresponding particles have an algebraically fixed mass $m=k+l+1$.
The fundamental representations discussed above are the special case
$\langle 0,0\rangle$.
Additionally there are long representations $\{k,l\}$
of dimension $16(k+1)(l+1)$.
Their mass is unconstrained.%
\footnote{If one picks the particular value $m=k+l+2$, however,
the representation reduces to $\langle k+1,l\rangle$
and $\langle k,l+1\rangle$.
In other words, one can combine two particle multiplets
$\langle k+1,l\rangle$ and $\langle k,l+1\rangle$
to form a long multiplet whose mass
is henceforth unconstrained in close analogy to the Higgs mechanism.}
It would be interesting to study the spectrum of composite states
in supersymmetric Chern--Simons theories.

\subsection{\texorpdfstring{$\superN=5,6,8$}{N=5,6,8} Supersymmetry and Multiplets}

Let us also discuss the algebras and representations for
$\superN=5,6,8$ supersymmetry.
This is applicable to Chern--Simons theories with
coinciding representations for untwisted and twisted matter
(see \appref{app:susyinteract})
which have manifest $\superN=5,6,8$ supersymmetry.
In general the higher supersymmetries anticommute
with the $\superN=4$ supercharges
and thus they have to transform between
untwisted and twisted $\superN=4$ hypermultiplets.

In the simplest $\superN=5$ case
there are two additional supercharges $\gen{\tilde Q}_\alpha$.
Their algebra $\alg{g}_1$ closes onto the momentum generators
\[
\acomm{\gen{\tilde Q}_{\alpha}}{\gen{\tilde Q}_{\delta}}=
2\gen{P}_{\alpha\delta}.
\]
The additional supercharges $\gen{Q}_\alpha$
must act like
\[
\begin{array}[b]{rclcrcl}
\gen{\tilde Q}_{\alpha}\state{\phi_b}
\eq +v_\alpha(p)\state{\twist\psi_b},
&\quad&
\gen{\tilde Q}_{\alpha} \state{\psi_{\dot b}}
\eq -v_\alpha(p)\state{\twist\phi_{\dot b}},
\\[3pt]
\gen{\tilde Q}_{\alpha} \state{\twist\psi_b}
\eq -u_\alpha(p)\state{\phi_b},
&\quad&
\gen{\tilde Q}_{\alpha} \state{\twist\phi_{\dot b}}
\eq +u_\alpha(p)\state{\psi_{\dot b}}.
\end{array}
\]

For $\superN=6$ supersymmetry the additional algebra
$\alg{g}_2$ consists of four supersymmetries $\gen{\tilde Q}^\pm_\alpha$
and two bosonic generators $\gen{\tilde B}$ and $\gen{\tilde C}$.
Their non-trivial commutation relations are
\[
\comm{\gen{\tilde B}}{\gen{\tilde Q}^\pm_\alpha}= \pm\gen{\tilde Q}^\pm _\alpha,
\qquad
\acomm{\gen{\tilde Q}^+_{\alpha}}{\gen{\tilde Q}^-_{\beta}}=
\gen{P}_{\alpha\beta}
-im\varepsilon_{\alpha\beta}\gen{\tilde C}.
\]
There are two types of multiplets
with opposite eigenvalue of the central charge $\gen{\tilde C}$.
For $\gen{\tilde C}\simeq -1$ the action of the supercharges
reads
\[
\begin{array}[b]{rclcrcl}
\gen{\tilde Q}_{\alpha}^+\state{\phi_{b-}}
\eq +v_\alpha(p)\state{\twist\psi_{b-}},
&\quad&
\gen{\tilde Q}_{\alpha}^+\state{\psi_{\dot b-}}
\eq -v_\alpha(p)\state{\twist\phi_{\dot b-}},
\\[3pt]
\gen{\tilde Q}_{\alpha}^-\state{\twist\psi_{b-}}
\eq -u_\alpha(p)\state{\phi_{b-}},
&\quad&
\gen{\tilde Q}_{\alpha}^- \state{\twist\phi_{\dot b-}}
\eq +u_\alpha(p)\state{\psi_{\dot b-}},
\\[3pt]
\gen{\tilde Q}_{\alpha}^+\state{\twist\psi_{b-}}
\eq 0,
&\quad&
\gen{\tilde Q}_{\alpha}^+\state{\twist\phi_{\dot b-}}
\eq 0,
\\[3pt]
\gen{\tilde Q}_{\alpha}^-\state{\phi_{b-}}
\eq 0,
&\quad&
\gen{\tilde Q}_{\alpha}^-\state{\psi_{\dot b-}}
\eq 0.
\end{array}
\]
The action on the conjugate multiplet with
$\gen{\tilde C}\simeq +1$ reads
\[
\begin{array}[b]{rclcrcl}
\gen{\tilde Q}_{\alpha}^-\state{\phi_{b+}}
\eq +v_\alpha(p)\state{\twist\psi_{b+}},
&\quad&
\gen{\tilde Q}_{\alpha}^-\state{\psi_{\dot b+}}
\eq -v_\alpha(p)\state{\twist\phi_{\dot b+}},
\\[3pt]
\gen{\tilde Q}_{\alpha}^+ \state{\twist\psi_{b+}}
\eq -u_\alpha(p)\state{\phi_{b+}},
&\quad&
\gen{\tilde Q}_{\alpha}^+ \state{\twist\phi_{\dot b+}}
\eq +u_\alpha(p)\state{\psi_{\dot b+}},
\\[3pt]
\gen{\tilde Q}_{\alpha}^- \state{\twist\psi_{b+}}
\eq 0,
&\quad&
\gen{\tilde Q}_{\alpha}^- \state{\twist\phi_{\dot b+}}
\eq 0,
\\[3pt]
\gen{\tilde Q}_{\alpha}^+\state{\phi_{b+}}
\eq 0,
&\quad&
\gen{\tilde Q}_{\alpha}^+\state{\psi_{\dot b+}}
\eq 0.
\end{array}
\]
In supersymmetric Chern--Simons theories the splitting
into the $\gen{\tilde C}\simeq \pm 1$ multiplets
originates from the structure of the superalgebra
which defines the field content, cf.\ \secref{sec:susy}.

Finally, for $\superN=8$ supersymmetry
there is a complete copy the the $\superN=4$ superalgebra $\alg{g}_4$
consisting of the generators
$\gen{\tilde R}_{\tilde a\tilde b}$,
$\gen{\hat R}_{\hat a\hat b}$ and
$\gen{\tilde Q}_{\alpha \tilde b\hat c}$.
The $\alg{su}(2)\oplus\alg{su}(2)$ generators act on spinor indices
as usual
\<
\gen{\tilde R}_{\tilde a\tilde b}\state{X_{\tilde c}}\eq
\ihalf\varepsilon_{\tilde b\tilde c}\state{X_{\tilde a}}
+\ihalf\varepsilon_{\tilde a\tilde c}\state{X_{\tilde b}},
\nln
\gen{\hat R}_{\hat a\hat b}\state{X_{\hat c}}\eq
\ihalf\varepsilon_{\hat b\hat c}\state{X_{\hat a}}
+\ihalf\varepsilon_{\hat a\hat c}\state{X_{\hat b}}.
\>
Furthermore the supercharges obey the same relation as above
\[
\acomm{\gen{\tilde Q}_{\alpha \tilde b\hat c}}{\gen{\tilde Q}_{\delta \tilde e\hat f}}=
\varepsilon_{\tilde b\tilde e}\varepsilon_{\hat c\hat f}\gen{P}_{\alpha\delta}
-2m\varepsilon_{\alpha\delta}\varepsilon_{\hat c\hat f}\gen{\tilde R}_{\tilde b\tilde e}
+2m\varepsilon_{\alpha\delta}\varepsilon_{\tilde b\tilde e}\gen{\hat R}_{\hat c\hat f}.
\]
The representation on the fields is much like the one discussed in
\secref{eq:N4rep}
\[
\begin{array}[b]{rclcrcl}
\gen{\tilde Q}_{\alpha \tilde b\hat c} \state{\phi_{d\hat e}}
\eq+\varepsilon_{\hat c\hat e}v_\alpha(p)\state{\twist\psi_{d\tilde b}},
&\quad&
\gen{\tilde Q}_{\alpha \tilde b\hat c} \state{\psi_{\dot d\hat e}}
\eq-\varepsilon_{\hat c\hat e} v_\alpha(p)\state{\twist\phi_{\dot d\tilde b}},
\\[3pt]
\gen{\tilde Q}_{\alpha \tilde b\hat c} \state{\twist\psi_{d\tilde e}}
\eq+\varepsilon_{\tilde b\tilde e}u_\alpha(p)\state{\phi_{d\hat c}},
&\quad&
\gen{\tilde Q}_{\alpha \tilde b\hat c}\state{\twist\phi_{\dot d\tilde e}}
\eq-\varepsilon_{\tilde b\tilde e}u_\alpha(p)\state{\psi_{\dot d\hat c}}.
\end{array}
\]
Again, the additional indices $\hat e$ and $\tilde e$ on the
untwisted and twisted multiplets, respectively, originate from
the structure of the defining superalgebra.

Note that $\grp{U}(N)\times\grp{U}(N)$ $\superN=6$ Chern--Simons models
at levels $k=1$ or $k=2$ are expected
to have $\superN=8$ enhanced supersymmetry \cite{Aharony:2008ug}.
This may appear impossible considering that the
$\grp{Sp}(1)$ automorphism required for $\superN=8$ supersymmetry cannot
act on a single $\superN=6$ particle representation.
Here one has to bear in mind that the points $k=1,2$ are strongly coupled.
Particles can bind to disorder operators of rank $k+k$ to effectively
double the set of fundamental particles \cite{Aharony:2008ug}.
On these pairs the $\grp{Sp}(1)$ automorphism can act.

\begin{figure}
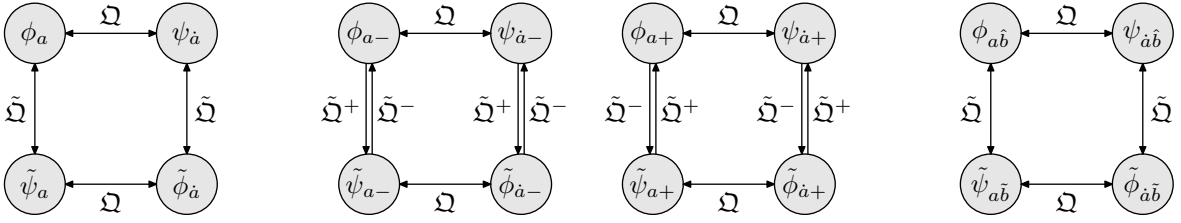
\centering
\includegraphics{FigSusyN5.mps}\hfill
\includegraphics{FigSusyN6A.mps}\quad
\includegraphics{FigSusyN6B.mps}\hfill
\includegraphics{FigSusyN8.mps}

\caption{The action of the extended $\superN=5,6,8$ supersymmetries
(left, middle two, right) on one-particle asymptotic states.}
\label{fig:SusyAction}
\end{figure}

To conclude, we summarize the action of the $\superN=5,6,8$ supersymmetry generators
$\gen{Q}$ and $\gen{\tilde Q}$
on the untwisted and twisted hypermultiplets
in \figref{fig:SusyAction}.

\section{Scattering Amplitudes from Supersymmetry}
\label{sec:SusyScat}

In this section we shall derive the form of scattering amplitudes
by means of supersymmetry and compare these predictions to
field theory calculations.

\subsection{Pure Scattering}

We will now set up the amplitudes
for a scattering process of four hypermultiplets.
The processes described here account for scattering of purely untwisted hypermultiplets
in models with $\mathcal{N}\geq 4$ supersymmetry.
We assume that all four particle momenta $p_k$, $k=1,2,3,4$,
are incoming and on shell.
The polarization spinors $(u_k,v_k)$
are on-shell according to \eqref{eq:PolSpinor}.

The amplitudes will be represented by an operator $\bra{\scat}$
acting on four-particle states and returning the corresponding amplitude%
\footnote{Alternatively we may give an invariant four-particle
state or an invariant two-to-two scattering operator,
cf.\ \protect\secref{sec:2to2}.}
\[
\braket{\scat}{1234}=A_{1234}.
\]
The most general ansatz for the scattering matrix elements
with manifest $\alg{su}(2)\oplus\alg{su}(2)$ symmetry reads
\<\label{eq:SMat}
\braket{\scat}{\phi_a\phi_b\phi_c\phi_d} \eq
\bigbrk{+\half (A+B)\varepsilon_{ad}\varepsilon_{bc}
+\half (A-B)\varepsilon_{ac}\varepsilon_{bd} }\,\delta^3(p_1+p_2+p_3+p_4),
\nln
\braket{\scat}{\psi_{\dot a}\psi_{\dot b}\psi_{\dot c}\psi_{\dot d}} \eq
\bigbrk{+\half (D+E)\varepsilon_{\dot a\dot d}\varepsilon_{\dot b\dot c}
+\half (D-E)\varepsilon_{\dot a\dot c}\varepsilon_{\dot b\dot d}}\,\delta^3(p_1+p_2+p_3+p_4),
\nln
\braket{\scat}{\phi_{a}\psi_{\dot b}\phi_{c}\psi_{\dot d}} \eq
-G\varepsilon_{ac}\varepsilon_{\dot b\dot d}\, \delta^3(p_1+p_2+p_3+p_4),
\nln
\braket{\scat}{\psi_{\dot a}\phi_{b}\psi_{\dot c}\phi_d} \eq
-L\varepsilon_{\dot a\dot c}\varepsilon_{bd}\, \delta^3(p_1+p_2+p_3+p_4),
\nln
\braket{\scat}{\phi_a\phi_b\psi_{\dot c}\psi_{\dot d}} \eq
-\half C\varepsilon_{ab}\varepsilon_{\dot c\dot d}\, \delta^3(p_1+p_2+p_3+p_4),
\nln
\braket{\scat}{\phi_{a}\psi_{\dot b}\psi_{\dot c}\phi_d} \eq
-H\varepsilon_{ad}\varepsilon_{\dot b\dot c}\, \delta^3(p_1+p_2+p_3+p_4),
\nln
\braket{\scat}{\psi_{\dot a}\psi_{\dot b}\phi_{c}\phi_{d}} \eq
-\half F\varepsilon_{\dot a\dot b}\varepsilon_{cd}\, \delta^3(p_1+p_2+p_3+p_4),
\nln
\braket{\scat}{\psi_{\dot a}\phi_{b}\phi_{c}\psi_{\dot d}} \eq
-K\varepsilon_{\dot a\dot d}\varepsilon_{bc}\, \delta^3(p_1+p_2+p_3+p_4).
\>
At this stage there are 10 independent matrix elements
$A,\ldots,L$ of the scattering amplitude.

We would now like to impose invariance under supersymmetry on
the amplitude leading to further constraints on the
matrix elements.
This is conveniently done by imposing
invariance conditions of the sort $\bra{\scat}\gen{Q}_{11}\state{\phi_1\phi_2\phi_2\psi_2}=0$.
From these we obtain altogether 32 constraints which are
collected in the following 16 spinor-valued equations
\[\label{eq:SMatConstr}
\begin{array}[b]{rclcrcl}
0\eq Av_3+Hu_2+Lu_1,
&\quad&
0\eq Bv_3+Cu_4+Hu_2-Lu_1,
\\[3pt]
0\eq Av_4+Gu_2+Ku_1,
&&
0\eq Bv_4-Cu_3-Gu_2+Ku_1,
\\[3pt]
0\eq Av_2-Hu_3-Gu_4,
&&
0\eq Bv_2-Fu_1-Hu_3+Gu_4,
\\[3pt]
0\eq Av_1-Lu_3-Ku_4,
&&
0\eq Bv_1+Fu_2+Lu_3-Ku_4,
\\[3pt]
0\eq Du_3+Gv_1-Kv_2,
&&
0\eq Eu_3-Fv_4-Gv_1-Kv_2,
\\[3pt]
0\eq Du_4-Hv_1+Lv_2,
&&
0\eq Eu_4-Fu_3-Hv_1-Lv_2,
\\[3pt]
0\eq Du_1-Gv_3+Hv_4,
&&
0\eq Eu_1-Cv_2+Gv_3+Hv_4,
\\[3pt]
0\eq Du_2+Kv_3-Lv_4,
&&
0\eq Eu_2+Cv_1+Kv_3+Lv_4.
\end{array}\]
In order to extract the matrix elements $A,\ldots,L$ it
is convenient to contract the equations
with spinors appearing in the equation
using the antisymmetric tensor $\varepsilon^{\alpha\beta}$.
To avoid clutter we introduce an antisymmetric scalar product for spinors
\[\label{eq:twistor}
\tprod{u}{v}=\varepsilon^{\alpha\beta}u_\alpha v_\beta.
\]
In analogy to the scattering amplitudes in
four-dimensional Yang--Mills theory using the spinor-helicity
formalism we shall call this product
a twistor bracket.
Moreover, we shall use the short notation
\[\label{eq:twistorshort}
\tprods{k}{l}:=\tprod{u_k}{u_l},\quad
\tprods{\bar k}{l}:=\tprod{v_k}{u_l},\quad
\tprods{\bar k}{\bar l}:=\tprod{v_k}{v_l}.
\]
Somewhat remarkably, all constraints can be solved simultaneously
on the ten matrix elements leaving just one overall
factor $T$ for the amplitude.
The solution reads
\[\label{eq:SMatEl}
\begin{array}[b]{rclcrclcrcl}
A\eq \displaystyle
T\,,
&&
D\eq\displaystyle
-T\,\frac{\tprods{\bar 3}{\bar 4}}{\tprods{1}{2}}\,,
&&
G\eq \displaystyle
+T\,\frac{\tprods{\bar 4}{1}}{\tprods{1}{2}}\,,
\\[10pt]
\half (A+B)\eq\displaystyle
-T\,\frac{\tprods{\bar 3}{1}\tprods{2}{4}}{\tprods{1}{2}\tprods{\bar 3}{4}}\,,
&&
\half (D+E)\eq\displaystyle
+T\,\frac{\tprods{\bar 3}{1}\tprods{\bar 1}{\bar 3}}{\tprods{1}{2}\tprods{\bar 3}{4}}\,,
&&
H\eq \displaystyle
+T\,\frac{\tprods{\bar 3}{1}}{\tprods{1}{2}}\,,
\\[10pt]
\half (A-B)\eq\displaystyle
+T\,\frac{\tprods{1}{4}\tprods{\bar 3}{2}}{\tprods{1}{2}\tprods{\bar 3}{4}}\,,
&&
\half (D-E)\eq\displaystyle
-T\,\frac{\tprods{\bar 2}{\bar 3}\tprods{\bar 3}{2}}{\tprods{1}{2}\tprods{\bar 3}{4}}\,,
&&
K\eq \displaystyle
-T\,\frac{\tprods{\bar 4}{2}}{\tprods{1}{2}}\,,
\\[10pt]
\half C\eq\displaystyle
-T\,\frac{\tprods{\bar 3}{1}\tprods{\bar 3}{2}}{\tprods{1}{2}\tprods{\bar 3}{4}}\,,
&&
\half F\eq\displaystyle
+T\,\frac{\tprods{\bar 3}{1}\tprods{\bar 1}{4}}{\tprods{1}{2}\tprods{\bar 3}{4}}\,,
&&
L\eq \displaystyle
-T\,\frac{\tprods{\bar 3}{2}}{\tprods{1}{2}}\,.
\end{array}\]
The overall factor $T$ is at this point completely general and there are
no additional restrictions on its form in our initial ansatz. We will see later that
crossing symmetry does put additional requirements however
 these are compatible with the definition in perturbative field theory.
Clearly there are many equivalent ways of writing the matrix elements
which also explains why many of the 32 constraints are degenerate.
A useful set of identities for the four scattering particles
with $p_1+p_2+p_3+p_4=0$ is given by
\[\label{eq:twistoridentity}
\frac{\tprods{l}{m}}{\tprods{k}{n}}
=\frac{\tprods{\bar k}{l}}{\tprods{\bar m}{n}}
=\frac{\tprods{\bar k}{\bar n}}{\tprods{\bar l}{\bar m}}\,,
\qquad
\tprods{\bar k}{k}=-2im,
\qquad
\set{k,l,m,n}=\set{1,2,3,4}.
\]
Additionally there is a cyclic identity which holds for any
four two-component spinors
\[\label{eq:twistorcyclic}
0=
 \tprod{a}{b}\tprod{c}{d}
+\tprod{b}{c}\tprod{a}{d}
+\tprod{c}{a}\tprod{b}{d}.
\]
There are three simple relations among the matrix
elements which can be checked using the above identities
\<\label{eq:SMatRel}
0\eq AD+GL-HK,
\nln
0\eq AD-BE+CF,
\nln
0\eq (A-B)(D-E)-CF+4GL.
\>
The remaining seven matrix elements
are independent: they can be adjusted
freely by choosing one overall factor,
one fermion phase
for each leg
($u_k,v_k\mapsto e^{\pm i\alpha_k}u_k,v_k$)
and the two Mandelstam invariants
$s=(p_1+p_2)^2$ and $t=(p_1+p_4)^2$.
Note that by a Lorentz transformation one
can change only three of the fermion phases.
Thus if one uses a particular choice
of spinor polarization as a function of momenta,
e.g.\ \eqref{eq:uvofp}, then only
six elements are independent.

\subsection{Mixed Scattering}
\label{sec:mixmat}

Next we will consider scattering of mixed matter fields.
Twisted hypermultiplets transform under supersymmetry
equivalently to untwisted multiplets, however
with the statistics of the on-shell particles flipped.
To obtain the twisted scattering amplitudes
we can thus simply replace a $(\phi_a,\psi_{\dot a})$ by
a $(\twist\psi_{a},\twist\phi_{\dot a})$
and insert the appropriate signs due to the change of statistics.

For correct overall statistics and parity of the internal symmetry,
we can only twist multiplets in pairs.
First we twist particles 3 and 4 of the four-particle scattering amplitude
\eqref{eq:SMat}. A suitable ansatz for the mixed scattering amplitude is
\<\label{eq:SMatT}
\braket{\scat}{\phi_a\phi_b\twist\psi_c\twist\psi_d} \eq
\bigbrk{+\half (A+B)\varepsilon_{ad}\varepsilon_{bc}
+\half (A-B)\varepsilon_{ac}\varepsilon_{bd}}\,\delta^3(p_1+p_2+p_3+p_4),
\nln
\braket{\scat}{\psi_{\dot a}\psi_{\dot b}\twist\phi_{\dot c}\twist\phi_{\dot d}} \eq
\bigbrk{-\half (D+E)\varepsilon_{\dot a\dot d}\varepsilon_{\dot b\dot c}
-\half (D-E)\varepsilon_{\dot a\dot c}\varepsilon_{\dot b\dot d}}\,\delta^3(p_1+p_2+p_3+p_4),
\nln
\braket{\scat}{\phi_{a}\psi_{\dot b}\twist\psi_{c}\twist\phi_{\dot d}} \eq
+G\varepsilon_{ac}\varepsilon_{\dot b\dot d}\,\delta^3(p_1+p_2+p_3+p_4),
\nln
\braket{\scat}{\psi_{\dot a}\phi_{b}\twist\phi_{\dot c}\twist\psi_d} \eq
-L\varepsilon_{\dot a\dot c}\varepsilon_{bd}\,\delta^3(p_1+p_2+p_3+p_4),
\nln
\braket{\scat}{\phi_a\phi_b\twist\phi_{\dot c}\twist\phi_{\dot d}} \eq
+\half C\varepsilon_{ab}\varepsilon_{\dot c\dot d}\,\delta^3(p_1+p_2+p_3+p_4),
\nln
\braket{\scat}{\phi_{a}\psi_{\dot b}\twist\phi_{\dot c}\twist\psi_d} \eq
-H\varepsilon_{ad}\varepsilon_{\dot b\dot c}\,\delta^3(p_1+p_2+p_3+p_4),
\nln
\braket{\scat}{\psi_{\dot a}\psi_{\dot b}\twist\psi_{c}\twist\psi_{d}} \eq
-\half F\varepsilon_{\dot a\dot b}\varepsilon_{cd}\,\delta^3(p_1+p_2+p_3+p_4),
\nln
\braket{\scat}{\psi_{\dot a}\phi_{b}\twist\psi_{c}\twist\phi_{\dot d}} \eq
+K\varepsilon_{\dot a\dot d}\varepsilon_{bc}\,\delta^3(p_1+p_2+p_3+p_4).
\>
The constraints due to supersymmetry turn out to be exactly the same as in \eqref{eq:SMatConstr}:
Due to the change of statistics of particles 3 and 4, we have to flip
the signs of all instances of $u_4,v_4$. Furthermore, the signs of
the matrix elements $C,D,E,G,K$ in \eqref{eq:SMatT}
have been flipped with respect to those in \eqref{eq:SMat}.
Altogether the sign flips cancel out and the solution
\eqref{eq:SMatEl} applies to the mixed scattering matrix as well.
Note that for the amplitudes $A,\ldots,L$
we can use a different prefactor $T$ which will be denoted
by $T_{12\twist{3}\twist{4}}$. In general a particle index $\twist{k}$ will
indicate a twisted hypermultiplet.

Finally, let us state the result for the scattering matrix of four
twisted multiplets
\<\label{eq:SMatTT}
\braket{\scat}{\twist\psi_a\twist\psi_b\twist\psi_c\twist\psi_d} \eq
\bigbrk{+\half (A+B)\varepsilon_{ad}\varepsilon_{bc}
+\half (A-B)\varepsilon_{ac}\varepsilon_{bd}}\,\delta^3(p_1+p_2+p_3+p_4),
\nln
\braket{\scat}{\twist\phi_{\dot a}\twist\phi_{\dot b}\twist\phi_{\dot c}\twist\phi_{\dot d}} \eq
\bigbrk{+\half (D+E)\varepsilon_{\dot a\dot d}\varepsilon_{\dot b\dot c}
+\half (D-E)\varepsilon_{\dot a\dot c}\varepsilon_{\dot b\dot d}}\,\delta^3(p_1+p_2+p_3+p_4),
\nln
\braket{\scat}{\twist\psi_{a}\twist\phi_{\dot b}\twist\psi_{c}\twist\phi_{\dot d}} \eq
-G\varepsilon_{ac}\varepsilon_{\dot b\dot d}\,\delta^3(p_1+p_2+p_3+p_4),
\nln
\braket{\scat}{\twist\phi_{\dot a}\twist\psi_{b}\twist\phi_{\dot c}\twist\psi_d} \eq
-L\varepsilon_{\dot a\dot c}\varepsilon_{bd}\,\delta^3(p_1+p_2+p_3+p_4),
\nln
\braket{\scat}{\twist\psi_a\twist\psi_b\twist\phi_{\dot c}\twist\phi_{\dot d}} \eq
+\half C\varepsilon_{ab}\varepsilon_{\dot c\dot d}\,\delta^3(p_1+p_2+p_3+p_4),
\nln
\braket{\scat}{\twist\psi_{a}\twist\phi_{\dot b}\twist\phi_{\dot c}\twist\psi_d} \eq
+H\varepsilon_{ad}\varepsilon_{\dot b\dot c}\,\delta^3(p_1+p_2+p_3+p_4),
\nln
\braket{\scat}{\twist\phi_{\dot a}\twist\phi_{\dot b}\twist\psi_{c}\twist\psi_{d}} \eq
+\half F\varepsilon_{\dot a\dot b}\varepsilon_{cd}\,\delta^3(p_1+p_2+p_3+p_4),
\nln
\braket{\scat}{\twist\phi_{\dot a}\twist\psi_{b}\twist\psi_{c}\twist\phi_{\dot d}} \eq
+K\varepsilon_{\dot a\dot d}\varepsilon_{bc}\,\delta^3(p_1+p_2+p_3+p_4).
\>
Here the signs of $H,K,C,F$ have been flipped with
respect to \eqref{eq:SMat}. Flipping as well the signs
of $u_2,v_2,u_4,v_4$ results in the same
set of constraints \eqref{eq:SMatConstr} whose
solution is given by \eqref{eq:SMatEl}.
The prefactor for this scattering process
will be denoted by $T_{\twist{1}\twist{2}\twist{3}\twist{4}}$.

\subsection{Scattering with \texorpdfstring{$\mathcal{N}>4$}{N>4} Supersymmetry}
\label{sec:higher_susy}

Let us now consider the additional constraints that follow if we extend the supersymmetry to
$\superN=5$. We have new invariance conditions of the type
$\langle \scat | \twist{\gen{Q}}_\alpha |\phi_1\phi_1\phi_2\twist{\psi}_2\rangle$  which
in principle give sixteen constraints on the eight independent
matrix structures
\[\begin{array}[b]{lclclcl}
T_{1234}, & & T_{12\twist3\twist4},& & T_{1\twist23\twist4},& & T_{1\twist2\twist34},\\
T_{\twist 1\twist 234}, & & T_{\twist12\twist34},& & T_{\twist123\twist4},& & T_{\twist1\twist2\twist3\twist4}.
\end{array}\]
However most of the constraints are redundant and there are only six which are independent.
We choose these to be
\<
\mathrel{}
+\tprods{\bar1 }{\bar 2}T_{1\twist23\twist4}+\tprods{\bar 1}{\bar 3}T_{12\twist3\twist4}-\tprods{\bar 1}{4}T_{1234}\eq0,\nn\\
-\tprods{\bar1 }{\bar 2}T_{\twist 123 \twist4}+\tprods{\bar 2}{\bar 3}T_{12\twist3\twist4}-\tprods{\bar 2}{4}T_{1234}\eq0,\nn\\
+\tprods{\bar1 }{\bar 2}T_{ 1\twist2\twist3 4}-\tprods{\bar 1}{ 3}T_{1234}-\tprods{\bar 1}{\bar 4}T_{12\twist3\twist4}\eq0,\nn\\
-\tprods{\bar1 }{ 2}T_{\twist 1\twist23 4}+\tprods{\bar 3}{ 2}T_{1\twist2\twist34}+\tprods{\bar 4}{2}T_{1\twist23\twist4}\eq0,\nn\\
+\tprods{\bar2 }{ 1}T_{\twist 1\twist23 4}+\tprods{\bar 3}{ 1}T_{\twist 12\twist34}+\tprods{\bar 4}{1}T_{\twist123\twist4}\eq0,\nn\\
+\tprods{1 }{2}T_{\twist 12\twist3 4}-\tprods{1}{ 3}T_{\twist 1\twist234}+\tprods{\bar 4}{1}T_{\twist1\twist2\twist3\twist4}\eq0.
\>
We can thus express all scattering elements in terms of  $T_{1234}$ and $T_{12\twist3\twist4}$
\<
\label{eq:pieq}
T_{\twist 1\twist 2\twist 3\twist 4}
\eq
+2\,\frac{\tprods{1}{\bar 1}}{\tprods{\bar 1}{\bar 2}}\, T_{1 2\twist3\twist4}
-\frac{\tprods{3}{4}}{ \tprods{\bar 1}{\bar 2} }\,T_{1234}
,\nn \\
T_{\twist 1\twist 234} \eq
+\frac{\tprods{\bar 3}{\bar 4}}{\tprods{\bar 1}{\bar 2}} \ T_{12 \twist 3  \twist 4}
,\nn \\
T_{1\twist23\twist4}\eq
-\frac{\tprods{\bar1}{\bar3}}{\tprods{\bar1}{\bar2}}\,T_{12\twist 3\twist 4}
+\frac{\tprods{\bar1}{4}}{\tprods{\bar1}{\bar2}}\,T_{1234}
,\nn\\
T_{1\twist2\twist34}\eq
+\frac{\tprods{\bar1}{\bar4}}{\tprods{\bar1}{\bar2}}\,T_{12 \twist 3  \twist 4}
+\frac{\tprods{\bar1}{3}}{\tprods{\bar1}{\bar2}}\,T_{1234}
,\nn\\
T_{\twist12\twist34}\eq
-\frac{\tprods{\bar2}{\bar4}}{\tprods{\bar1}{\bar2}}\,T_{12 \twist 3  \twist 4}
-\frac{\tprods{\bar2}{3}}{\tprods{\bar1}{\bar2}}\,T_{1234}
,\nn\\
 T_{\twist123\twist4}\eq
+\frac{\tprods{\bar2}{\bar3}}{\tprods{\bar1}{\bar2}}\,T_{12 \twist 3  \twist 4}
-\frac{\tprods{\bar2}{4}}{\tprods{\bar1}{\bar2}}\,T_{1234}
.
\>

There are further, similar,  relations if we consider the $\superN=6$ algebra,
however the multiplet structure is slightly more complicated.
In the case of $\superN=8$
the algebra is composed of two copies of $\alg{psu}(2|2)$
and the scattering matrix takes a simple tensor product form.
Let us consider in a little more detail the $\superN=8$ case.
There the fields have an additional $\alg{su}(2)$ index $\hat a$ or $\tilde a$
(in Chern--Simons models
these can originate from the defining superalgebra as explained in \secref{sec:susy}):
$\phi_{a\hat a}$,
$\psi_{\dot a\hat a}$,
$\twist\psi_{a\twist a}$,
$\twist\phi_{\dot a \twist a}$.
The $\superN=4$ prefactors obtain the corresponding indices, e.g.\
\[
\braket{\scat}{\phi_{1\hat a\vphantom{\hat b}}\phi_{1\hat b}\phi_{2\hat c\vphantom{\hat b}}\phi_{2\hat d}}=
T_{1234,\hat a\hat b\hat c\hat d}\,\delta^3(p_1+p_2+p_3+p_4).
\]
These are all related by the additional $\alg{su}(2)$'s and so we can pick a single
representative $T^{\superN=8}$ in each sector in terms of which we can express all other elements.
Let us define it such that the respective $\alg{su}(2)$ indices take values
$1,2,1,2$, for example
\[
T_{12 3 4}^{\superN=8}=T_{1234,\hat1\hat2\hat1\hat2},
\qquad
T_{12\twist 3\twist 4}^{\superN=8}=T_{12\twist3\twist4,\hat1\hat2\tilde1\tilde2}.
\]
As before we consider the constraints following from invariance conditions such as
\[
\bra{\scat}\twist{\gen{Q}}_{\alpha\twist e\hat f}
\ket{\phi_{1\hat a\vphantom{\hat b}}\phi_{1\hat b\vphantom{\hat b}}\phi_{2\hat c\vphantom{\hat b}}\twist{\psi}_{2\twist d\vphantom{\hat b}}}=0.
\]
and in addition to \eqref{eq:pieq}
we can find a relation between $T^{\superN=8}_{12\twist 3\twist 4}$ and $T^{\superN=8}_{1234}$
\[
\label{eq:ne8eq}
T_{12\twist 3\twist 4}^{\superN=8} =\frac{\tprods{\bar 1}{ 4}}{  \tprods{\bar 1}{\bar 3} }\ T_{1234}^{\superN=8}.
\]
%

\subsection{Crossing Symmetry}
\label{sec:cross}

First consider the exchange
of particles $1\leftrightarrow 2$.
In the scattering matrix \eqref{eq:SMat,eq:SMatEl}
it corresponds to the following exchange of elements
\[
\begin{array}[b]{rclcrclcrclcrclcrcl}
A\earel{\leftrightarrow}+A,&&
B\earel{\leftrightarrow}-B,&&
C\earel{\leftrightarrow}-C,&&
G\earel{\leftrightarrow}-K,&&
H\earel{\leftrightarrow}-L,
\\[3pt]
D\earel{\leftrightarrow}-D,&&
E\earel{\leftrightarrow}+E,&&
F\earel{\leftrightarrow}+F,&&
K\earel{\leftrightarrow}-G,&&
L\earel{\leftrightarrow}-H.
\end{array}
\]
It is straight-forward to verify that the whole
scattering matrix is symmetric under the interchange
of particles if the prefactor obeys $S_{2134}=S_{1234}$.
Similarly, the exchange $3\leftrightarrow 4$
leads to the following map of matrix elements
\[
\begin{array}[b]{rclcrclcrclcrclcrcl}
A\earel{\leftrightarrow}+A,&&
B\earel{\leftrightarrow}-B,&&
C\earel{\leftrightarrow}+C,&&
G\earel{\leftrightarrow}+H,&&
H\earel{\leftrightarrow}+G,
\\[3pt]
D\earel{\leftrightarrow}-D,&&
E\earel{\leftrightarrow}+E,&&
F\earel{\leftrightarrow}-F,&&
K\earel{\leftrightarrow}+L,&&
L\earel{\leftrightarrow}+K.
\end{array}
\]
Again this leaves the scattering matrix invariant provided that
$S_{1243}=S_{1234}$.

As a third type of crossing we consider the cyclic permutation $1\to 2\to 3\to 4\to 1$.
It turns out to reshuffle the matrix elements in a more elaborate fashion
\[\begin{array}[b]{rclcrcl}
A_{2341}\earel{\to} -\half (A_{1234}-B_{1234}),
&&
B_{2341}\earel{\to} +\half (3A_{1234}+B_{1234}),
\\[3pt]
D_{2341}\earel{\to} +\half (D_{1234}-E_{1234}),
&&
E_{2341}\earel{\to} -\half (3D_{1234}+E_{1234}),
\\[3pt]
G_{2341}\earel{\to} +L_{1234},
&&
L_{2341}\earel{\to} -G_{1234},
\\[3pt]
C_{2341}\earel{\to} +2K_{1234},
&&
K_{2341}\earel{\to} +\half F_{1234},
\\[3pt]
F_{2341}\earel{\to} -2H_{1234},
&&
H_{2341}\earel{\to} -\half C_{1234}.
\end{array}
\]
Cyclic crossing symmetry on all the matrix elements is achieved by demanding
\[\label{eq:cross1234}
1\to 2\to 3\to 4\to 1\colon\quad
T_{2341}=-T_{1234}\frac{\tprods{2}{3}\tprods{\bar 4}{1}}{\tprods{1}{2}\tprods{\bar 4}{3}}\,.
\]

Finally, consider the twisted scattering amplitudes \eqref{eq:SMatT,eq:SMatTT}.
These have the same crossing relations up to some signs due to statistics.
A summary of the crossing relations is provided in the following table:
\[\label{eq:cross}
\begin{array}[b]{rcl|rcl|rcl}
\multicolumn{3}{c|}{1\leftrightarrow 2}
&
\multicolumn{3}{c|}{3\leftrightarrow 4}
&
\multicolumn{3}{c}{1\to 2\to 3\to 4\to 1}
\\\hline
T_{2134}\eq +T_{1234}
&
T_{1243}\eq +T_{1234}
&
T_{2341}\eq \displaystyle-T_{1234}\frac{\vphantom{\hat 1}\tprods{2}{3}\tprods{\bar 4}{1}}{\tprods{1}{2}\tprods{\bar 4}{3}}
\\
T_{21\twist3\twist4}\eq +T_{12\twist3\twist4}
&
T_{12\twist4\twist3}\eq -T_{12\twist3\twist4}
&
T_{2\twist3\twist41}\eq \displaystyle-T_{12\twist3\twist4}\frac{\vphantom{\hat 1}\tprods{2}{3}\tprods{\bar 4}{1}}{\tprods{1}{2}\tprods{\bar 4}{3}}
\\
T_{\twist2\twist1\twist3\twist4}\eq -T_{\twist1\twist2\twist3\twist4}
&
T_{\twist1\twist2\twist4\twist3}\eq -T_{\twist1\twist2\twist3\twist4}
&
T_{\twist2\twist3\twist4\twist1}\eq \displaystyle+T_{\twist1\twist2\twist3\twist4}\frac{\vphantom{\hat 1}\tprods{2}{3}\tprods{\bar 4}{1}}{\tprods{1}{2}\tprods{\bar 4}{3}}
\end{array}
\]
Note that the cyclic crossing relation for the mixed hypermultiplets
maps between scattering matrices with different hypermultiplet assignments.

\subsection{Two-to-Two Particle Scattering}
\label{sec:2to2}

The scattering amplitude $\bra{\scat_{1234}}$ in \eqref{eq:SMat}
is written such that all four particles are on an equal footing.
For various purposes, however, it is convenient to write
the scattering amplitude as an operator $\scat{}_{12}^{43}$
acting on two-particle states and returning
(a linear combination of) two-particle states.

To convert between the two pictures, let us first introduce a
two-particle state $\state{\singlet}$ which is invariant under
the super-Poincar\'e algebra.
In particular, its total momentum must be zero, $p_2=-p_1$.
This implies the following relation for the polarization spinors
\[
u_2=ie^{+i\alpha}v_1,\qquad
v_2=ie^{-i\alpha}u_1,
\]
with some free parameter $\alpha$ representing the
relative polarization between the spinors.
It is then straight-forward to confirm that the following
composite state is annihilated by all generators%
\footnote{This expression implies the use of polarization spinors $u(p),v(p)$
with definite phase for a given momentum $p$, see
the discussion in \protect\secref{eq:N4rep}.
If we consider $u_{1,2},v_{1,2}$ as the fundamental degrees
of freedom, we should choose the integral
$\int d^2u_1\,d^2v_1\,d^2u_2\,d^2v_2\,\delta(\tprod{v_1}{u_1}+2im)\,\delta^3(p_1+p_2)\ldots$\,.
}
\[\state{\singlet}=
\int d^3p\, 2\pi\delta(p^2+m^2)
\lrbrk{
\varepsilon^{ab}\state{\phi_a\phi_b}
+ie^{i\alpha}\varepsilon^{\dot a\dot b}\state{\psi_{\dot a}\psi_{\dot b}}
}.
\]
This state is invariant under the full
super-Poincar\'e algebra.
Note that without the integration over the mass shell it would only be
invariant under supercharges and internal rotations which form
an ideal of the algebra.
The momenta $p_{1,2}$ of the individual particles break Lorentz invariance.
We have inserted a normalization factor of $2\pi$
corresponding to the imaginary part of
a propagator $2\Im (p^2+m^2-i\epsilon)^{-1}=2\pi\delta(p^2+m^2)$.

Now we can define the two-to-two scattering operator $\scat{}_{12}^{43}$ as
\[
\scat{}_{12}^{43}
\state{\mathcal{X}_1\mathcal{X}_2}
=
\half\braket{\scat_{12\bar3\bar4}}{\mathcal{X}_1\mathcal{X}_2\singlet_{\bar44}\singlet_{\bar33}},
\]
which is invariant by construction.
The factor of $1/2$ is a symmetry factor
to account for two identical outgoing particle multiplets;
it is compensated by the phase space integrals
in the S-matrix which count each state twice
modulo permutation.
More explicitly,
using the action \eqref{eq:SMat}, the operator takes the form
\<\label{eq:2to2T}
\scat\state{\phi_a\phi_b} \eq
2\pi^2\int d^3p\, \delta(p^2_3+m^2)\, \delta(p^2_4+m^2)
\bigbrk{A\state{\phi_{(a}\phi_{b)}}
+B\state{\phi_{[a}\phi_{b]}}
+\half C\varepsilon_{ab}\varepsilon^{\dot c\dot d} \state{\psi_{\dot c}\psi_{\dot d}}},
\nln
\scat\state{\psi_{\dot a}\psi_{\dot b}} \eq
2\pi^2\int d^3p\, \delta(p^2_3+m^2)\, \delta(p^2_4+m^2)
\bigbrk{D\state{\psi_{(\dot a}\psi_{\dot b)}}
+E\state{\psi_{[\dot a}\psi_{\dot b]}}
+\half F\varepsilon_{\dot a\dot b}\varepsilon^{cd} \state{\phi_{c}\phi_{d}}},
\nln
\scat\state{\phi_{a}\psi_{\dot b}} \eq
2\pi^2\int d^3p\, \delta(p^2_3+m^2)\, \delta(p^2_4+m^2)
\bigbrk{G\state{\psi_{\dot b}\phi_{a}}
+H\state{\phi_{a}\psi_{\dot b}}},
\nln
\scat\state{\psi_{\dot a}\phi_{b}} \eq
2\pi^2\int d^3p\, \delta(p^2_3+m^2)\, \delta(p^2_4+m^2)
\bigbrk{K\state{\psi_{\dot a}\phi_{b}}
+L\state{\phi_{b}\psi_{\dot a}}}.
\>
Here we have labeled the two outgoing particles as $4$ and $3$.
We have also fixed the above phase to $\alpha=\half\pi$.
In other words, the polarization spinors between particles
$3,4$ and their conjugates $\bar3,\bar4$ are related by
\[
\label{eq:conjspin}
u_{\bar k}=+v_k,
\qquad
v_{\bar k}=-u_k
\]
and the matrix elements
$A,\ldots,L$ equal those in
\eqref{eq:SMatEl} with indices $A_{12\bar3\bar4},\ldots L_{12\bar3\bar4}$.

\subsection{Worldsheet Scattering Matrix for AdS/CFT and Integrability}
\label{sec:ads5}

The extended $\alg{psu}(2|2)$ algebra plays an important role
in the investigation of integrability in
the planar AdS/CFT correspondence between strings
on $AdS_5\times S^5$ and $\superN=4$ super Yang--Mills theory \cite{Beisert:2005tm,Hofman:2006xt,Arutyunov:2006ak}.
It also appears analogously in the recently discussed duality between strings
on $AdS_4\times\Complex\Projspc^3$ and $\superN=6$ Chern--Simons theory
\cite{Aharony:2008ug,Minahan:2008hf,Gaiotto:2008cg,Gromov:2008qe,Ahn:2008aa,Bak:2008cp}.
The algebra serves as the symmetry in a light cone gauge of string theory
or equivalently in a ferromagnetic excitation picture of gauge theory spin chains.

For the two body scattering of the $\superN=4$ super Yang--Mills spin chain
each of the excitations can be one of 16 flavors and so the scattering is described
by a  $16^2 \times16^2 $ matrix. In \cite{Beisert:2005tm} it was shown that the symmetries
determine the matrix structure uniquely and so it is determined up to an overall phase.
The same result holds for the scattering of worldsheet excitations on the string
worldsheet \cite{Hofman:2006xt,Klose:2006zd,Arutyunov:2006yd}.

Hofman and Maldacena, \cite{Hofman:2006xt} (see also \cite{Lin:2005nh}),
pointed out that the constraints imposed
by the $\alg{psu}(2|2)$ algebra on the spin chain were exactly those that a four
particle scattering amplitude in $2+1$-dimensions in a theory with the same
super-algebra would have. They further
pointed out that the ``dynamic'' nature of the spin chain scattering, whereby
the length of the chain changes in certain scattering processes, is related
to the non-Lorentz invariance of the $2+1$ scattering matrix. Under an overall
rotation it picks up a phase due to the fermions spin.

While the matrix structure of the scattering amplitude is
identical between the spin chain/worldsheet theory and the $2+1$ Lorentz
invariant theories the kinematics are quite distinct which can be seen in the difference
between the overall two- and three-dimensional momentum delta functions.
In two dimensions the scattering momenta can not change in magnitude and at
most can be exchanged between particles. In three dimensions the final state phase
space is larger and includes the relative angle between the two particles.

The matrix elements $A,\ldots,L$ are related in terms of
the spinors $u,v$ from the supersymmetry representation \eqref{eq:susyrep}.
In \cite{Beisert:2005tm,Beisert:2006qh} the supersymmetry representation
is specified in terms of the parameters $a,b,c,d$ instead.
We thus have to relate these sets of parameters first:
This is easily achieved by
\[
u=\sqrt{2im}\matr{c}{+a\\-c},\quad
v=\sqrt{2im}\matr{c}{-b\\+d}.
\]
The constraint $\tprod{v}{u}=-2im$ is equivalent to $ad-bc=1$.
This simple choice leads to the following incoming momentum components%
\footnote{%
Unfortunately, it turns out that $p_0$ and $p_2$ are purely imaginary for
physical magnons of the worldsheet theory.
For magnons of the mirror worldsheet theory
\cite{Arutyunov:2007tc}, however,
all momentum components are real.
In fact one could choose for incoming particles
$u=\sqrt{im}(+a+ic,-c-ia)$,
$v=\sqrt{im}(-b-id,+d+ib)$
which leads to exactly the same S-matrix elements.
In that case $p_0$ and $p_2$ are interchanged and multiplied by $i$
thus making them real.
For the alternative choice,
the intermediate expressions are somewhat cluttered
and hence we shall stick to the above unphysical choice.}
\<
p_0-p_1\eq -2ig \ m\alpha (1-x^+/x^-),
\nln
p_0+p_1\eq -2ig\ m\alpha^{-1}(1-x^-/x^+),
\nln
p_2\eq -2g \ m(x^+-x^-)+im.
\>
Note that the parameters $a,b,c,d$ associated to the magnons
depend on the ordering of magnons, see \cite{Beisert:2006qh}
for details. The correct assignment is
\<\label{eq:uvmagnon}
u_1\eq\sqrt{2igm}\,\gamma_1\matr{c}{1\\-i\alpha^{-1}/x^+_1},
\nln
v_1\eq \sqrt{-2igm}\,\gamma_1^{-1}(x^+_1-x^-_1)\matr{c}{i\alpha/x^-_1\\1},
\nln
u_2\eq \sqrt{2igm}\,\gamma_2\xi_1\matr{c}{1\\-i\alpha^{-1}x^-_1/x^+_1x^+_2 },
\nln
v_2\eq \sqrt{-2igm}\,\gamma_2^{-1}\xi^{-1}_1(x^+_2-x^-_2)\matr{c}{i\alpha x^+_1/x^-_1x^-_2\\1},
\nln
u_3=-v_{\bar 3}\eq \sqrt{2igm}\,\gamma_1\xi_2\matr{c}{1\\-i\alpha^{-1}x^-_2/x^+_1x^+_2},
\nln
v_3=+u_{\bar 3}\eq \sqrt{-2igm}\,\gamma_1^{-1}\xi^{-1}_2(x^+_1-x^-_1)\matr{c}{i\alpha x^+_2/x^-_1x^-_2\\1},
\nln
u_4=-v_{\bar 4}\eq \sqrt{2igm}\,\gamma_2\matr{c}{1\\-i\alpha^{-1}/x^+_2},
\nln
v_4=+u_{\bar 4}\eq \sqrt{-2igm}\,\gamma_2^{-1}(x^+_2-x^-_2)\matr{c}{i\alpha/x^-_2\\1}.
\>
Note that the energies are related as $(E_3,E_4)=(E_1,E_2)$.
Thus the prefactor $T$ should contain a factor of $\delta(E_3-E_1)$.
Let us thus compute the contribution from the delta functions%
\footnote{We will consistently drop the contribution
from diagonal scattering where $(p_1,p_2)=(p_3,p_4)$.}
\[
2\pi^2\int d^3p\, \delta(p_3^2+m^2)\delta(p_4^2+m^2)\delta(E_3-E_1)
=\frac{\pi^2}{2|p_{1x}p_{2y}-p_{1y}p_{2x}|}\,.
\]
We will thus need a compensating factor $\mathnormal{\Delta}_{12}$ for the prefactor
\[
\mathnormal{\Delta}_{12}=\frac{2}{\pi^2}\,(p_{1x}p_{2y}-p_{1y}p_{2x})
=
\frac{4ig^2m^2}{\pi^2}\lrbrk{1-\frac{x^+_1}{x^-_1}}\lrbrk{1-\frac{x^+_2}{x^-_2}}
\lrbrk{1-\frac{x^-_1x^-_2}{x^+_1x^+_2}}.
\]
Substituting the representation spinors \eqref{eq:uvmagnon} into the matrix elements
\eqref{eq:SMatEl} one recovers the S-matrix
presented in \cite{Beisert:2006qh}
provided that the prefactors of the S-matrices are
related as follows
\[
T_{12\bar3\bar4}=\delta(E_1-E_3)\mathnormal{\Delta}_{12} S^0_{12}\,\frac{x^+_2-x^-_1}{x^-_2-x^+_1}\,.
\]
The matrix elements $A,\ldots,L$ have been normalized such that
they can be compared directly to the results of \cite{Beisert:2005tm,Beisert:2006qh}.
A priori the magnon S-matrix depends on nine parameters,
$S^0,x_1,x_2,g,\alpha,\gamma_1,\gamma_2,\xi_1,\xi_2$.
As the matrix elements in \eqref{eq:SMatEl}
have only seven degrees of freedom, there must be
two directions in the nine-dimensional
parameter space along which the S-matrix is invariant.
From \eqref{eq:uvmagnon} one can easily infer that
the parameters $\gamma_1,\gamma_2,\xi_1,\xi_2$
correspond to phases of the fermion spinors,
see \cite{Hofman:2006xt}. In the integrable system these
parameters and the phase factor $S^0$ are usually fixed
leaving only three degrees of freedom $x_1,x_2,g$.
In that sense, the integrable S-matrix is merely a
special case of the most general spacetime S-matrix.

Let us now consider the crossing symmetry
of the magnon S-matrix \cite{Janik:2006dc}.
and compare it to the crossing studied in \secref{sec:cross}.
Crossing of the magnon S-matrix corresponds to
interchanging particles 2 and 4.
From iterating \eqref{eq:cross} we can derive
the $2\leftrightarrow 4$ crossing relation
\[
T_{1432}=\frac{\tprods{1}{4}}{\tprods{1}{2}}\,
\frac{\tprods{\bar 3}{2}}{\tprods{\bar 3}{4}}\,T_{1234}
\quad\mbox{or}\quad
T_{1\bar4\bar32}=\frac{\tprods{1}{\bar4}}{\tprods{1}{2}}\,
\frac{\tprods{3}{2}}{\tprods{3}{\bar 4}}\,T_{12\bar 3\bar 4}.
\]
This combination of spinors takes the following form in $x^\pm$ notation
\[
\frac{\tprods{1}{\bar4}}{\tprods{1}{2}}\,\frac{\tprods{3}{2}}{\tprods{3}{\bar 4}}
=
\frac{x^-_2-x^-_1}{x^+_2-x^-_1}\,\frac{1/x^-_2-x^+_1}{1/x^+_2-x^+_1}\,.
\]
We furthermore have to relate the crossed prefactor $T_{1\bar4\bar32}$
to the crossed prefactor $S^0_{1\bar 2}$ and also fix the
parameters $\xi_k$ %
\footnote{We do not understand the appearance of the inverse in the formula,
but it is necessary to make the below crossing relation work.}
\[
T_{1\bar4\bar32}=\delta(E_1-E_3)\mathnormal{\Delta}_{12}
\lrbrk{S^0_{1\bar 2}\,\frac{1/x^+_2-x^-_1}{1/x^-_2-x^+_1}}^{-1},
\qquad
\xi_k^2=\frac{x^+_k}{x^-_k}\,.
\]
We then recover the crossing relation \cite{Janik:2006dc,Beisert:2006qh}
\[
1=\frac{S^0_{12}S^0_{1\bar 2}}{\xi_1^2}\,
\frac{x^+_{\bar2}-x^-_1}{x^-_{\bar2}-x^-_1}\,
\frac{1/x^+_{\bar2}-x^+_1}{1/x^-_{\bar2}-x^+_1}\,.
\]

Given this formal equivalence, it is natural to ask what role
the known integrable structures of the spin chain
might play in the $2+1$ dimensional supersymmetric Chern--Simons theory. 
However due to the different kinematical structure (see also \cite{Hofman:2006xt}),
which is imposed by the delta-function prefactor in the $1+1$ dimensional model,
the Chern--Simons scattering matrix does not satisfy the Yang--Baxter equation.

\subsection{Six-Particle Scattering}
\label{sec:scattersix}

Let us briefly comment on scattering of more than four particles.
In fact, scattering must always involve an
even number of external physical particles due to charge conservation:
All particles transform as a doublet
of one of the two internal $\alg{su}(2)$ symmetries.
In other words they form a doublet of the diagonal $\alg{su}(2)$.
A singlet of this $\alg{su}(2)$ can only be composed from
an even number of doublets.

The next non-trivial case is thus six external particles.
Altogether there are $4^6=4096$ components most
of which are zero due to charge conservation.
Taking into account the $\alg{su}(2)\times\alg{su}(2)$
internal symmetry, there are 70 remaining invariant structures.
Finally, supersymmetry relates most of them and there are
only two invariant structures leading to two prefactors \cite{Beisert:2006qh},
see also \cite{Puletti:2007hq}.
These can, for instance, be obtained from the purely bosonic
and purely fermionic scattering processes
\[
\braket{\scat}{\phi_1\phi_1\phi_1\phi_2\phi_2\phi_2},\qquad
\braket{\scat}{\psi_1\psi_1\psi_1\psi_2\psi_2\psi_2}.
\]

In this work we will not need higher-particle scattering
because it contributes to unitarity relations starting
from three loops. As there are no physical gluon states
all scattering amplitudes must involve an even number
of external particles. For two-to-two scattering this implies that in the
unitarity relations we must have four internal, cut legs which implies at least
three loop momenta. Nevertheless, it would be interesting to
see whether one can set up recursion relations similar to
those obtained for $\superN=4$ SYM \cite{Cachazo:2004kj, Britto:2004ap, Britto:2005fq}
or find related generating functions for amplitudes
\cite{Nair:1988bq,Georgiou:2004by, Bianchi:2008pu, Elvang:2008na, Drummond:2008cr}.
The form of the Chern--Simons matter scattering amplitudes is actually
quite similar to those of $\superN=4$ SYM but again one must be careful to
take into account the significant differences due to the three-dimensional
kinematics.

\section{Color Structures}
\label{sec:color_structures}

Feynman diagrams in a gauge theory consist of two parts,
spacetime functions and color structures.
Here we shall discuss the color structures
relevant to four-point scattering in $\superN\geq 4$ Chern--Simons theories
at tree level and at one loop.
The scattering amplitudes will then be written as linear combinations
of these color structures multiplied by space-time functions.
In fact, it usually suffices to compute so-called color-ordered amplitudes.
However, because the gauge group decomposes
into multiple factors potentially of different rank
it is worth the effort of analyzing the structures in detail.
It will turn out that the numerical factors from the color structure
crucially depend on the colors of the external legs.
In this section we will be quite general and the discussion, at least
initially,  is valid for all the
theories characterized by the various super-Lie  algebras enumerated in \secref{sec:susy}
including the exceptional superalgebras. Naturally when the discussion turns to
the planar limit we implicitly restrict ourselves to the theories with a large
$N$ limit.

\subsection{Color Graphs}

\begin{figure}\centering
\includegraphics{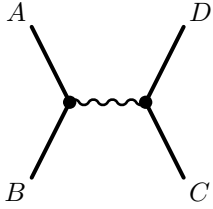}
\caption{Sample Feynman graph and color structure.}
\label{fig:sampstruct}
\end{figure}

We start by introducing a graphical notation
which will be very useful to classify color structures.
Consider, for example, the scattering of two particles
by exchange of a gluon,
cf.\ the Feynman graph in \figref{fig:sampstruct}.
The Feynman rules associate a color factor
to each vertex (e.g.\ $T_{MAB}$) and to each line (e.g.\ $K^{MN}$).
In gauge theories the color structure of
\figref{fig:sampstruct} is typically
\[\label{eq:sampstruct}
T_{MAB}K^{MN}T_{NCD}.
\]
We shall use the same graph \figref{fig:sampstruct}
to denote this color structure.
Nevertheless, the correspondence between Feynman graphs
and color structures is not one-to-one:
There can be vertices with a composite color structure,
e.g.\ a single vertex can be of the form \eqref{eq:sampstruct}.
Therefore several Feynman diagrams will have one and the same color structure.
Furthermore, different color structures are often related by some identities.

We now set up the specific rules
for a generic $\superN=4$ Chern--Simons model,
cf.\ \appref{app:susyinteract} for a brief summary.
There are three types of fields:
untwisted matter, twisted matter and gluons.
We shall use solid, double and wiggly lines to distinguish between them,
see \figref{fig:structlines}.
The associated color factors are
$L^{AB}$, $\twist L^{\twist A\twist B}$ and $K^{MN}$, respectively.
There are also three types of vertices:
They connect a gluon to two untwisted fields, two twisted fields
or to two further gluons.
The vertices are depicted in \figref{fig:structvert}
and they correspond to the structures
$M_{MAB}=K_{MN}M^N_{AB}$,
$\twist M_{M\twist A\twist B}=K_{MN}\twist M^{N}_{\twist A\twist B}$
and $F_{MPQ}=K_{MN}F^N_{PQ}$, respectively.
All the terms in the Lagrangian in \appref{app:susyinteract.action}
have a graphical representation using the above lines and 3-vertices.

\begin{figure}\centering
\includegraphics{FigColorLineA.mps}\qquad
\includegraphics{FigColorLineB.mps}\qquad
\includegraphics{FigColorLineC.mps}
\caption{Color lines:
$L^{AB}$,
$\twist L^{\twist A\twist B}$,
$K^{MN}$.
}
\label{fig:structlines}
\end{figure}

\begin{figure}\centering
\includegraphics{FigColorVertexA.mps}\qquad
\includegraphics{FigColorVertexB.mps}\qquad
\includegraphics{FigColorVertexC.mps}
\caption{Color vertices:
$M_{MAB}$, $\twist M_{M\twist A\twist B}$, $F_{MPQ}$.}
\label{fig:structvert}
\end{figure}

\begin{figure}
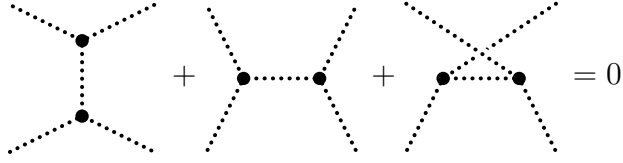
\centering
$\includegraphicsbox{FigColorJacobiA.mps}+
\includegraphicsbox{FigColorJacobiB.mps}+
\includegraphicsbox{FigColorJacobiC.mps}=0$
\caption{Jacobi identities for color structures.}
\label{fig:structjacobi}
\end{figure}

In a gauge theory the vertices are structure constants of the
gauge group. They therefore obey a host of identities,
e.g.\ Jacobi identities.
In our case there are five Jacobi identities,
see \appref{sec:susyinteract.def}.
They all have the same form and are summarized
graphically in \figref{fig:structjacobi}.
Note that the Jacobi identity only exists if
all involved vertices exist: There is no
Jacobi identity for a gluon line joining a
twisted with an untwisted vertex!
Here our main interest is the enumeration
of distinct structures and not their precise prefactors.
For instance, we shall not always pay close attention to signs.

\subsection{Tree Level}
\label{sec:StructTree}

\begin{figure}\centering
\includegraphics{FigColorTreeUA.mps}\qquad
\includegraphics{FigColorTreeUB.mps}\qquad
\includegraphics{FigColorTreeUC.mps}
\caption{Untwisted tree graphs:
$\cstree_{12,34}$,
$\cstree_{13,24}$,
$\cstree_{14,23}$.}
\label{fig:structtreeun}
\end{figure}

Let us first consider a scattering process of
four untwisted matter fields.
At tree level we need two 3-vertices
to connect the four external lines.
There are three ways in which this can be done,
see \figref{fig:structtreeun}.
We shall denote the structures by
\[
\cstree_{AB,CD}=M_{MAB}K^{MN}M_{NCD}.
\]
They have the obvious eight-fold symmetries
$\cstree_{AB,CD}=\cstree_{BA,CD}=\cstree_{CD,AB}$.
Furthermore, the Jacobi identity in \figref{fig:structjacobi}
relates these three structures
\[\label{eq:structjactree}
\cstree_{AB,CD}+\cstree_{AC,DB}+\cstree_{AD,BC}=0.
\]
Now any scattering amplitude at tree level can be written as
\[
T=
T_s\cstree_{12,34}+
T_t\cstree_{14,23}+
T_u\cstree_{13,24}.
\]
The Jacobi identity \eqref{eq:structjactree}, however,
states that the basis $\cstree_{12,34},\cstree_{13,24},\cstree_{14,23}$
is over-complete. Thus for any $\delta$ the amplitude is equivalent to
\[
T=
(T_s+\delta)\cstree_{12,34}+
(T_t+\delta)\cstree_{14,23}+
(T_u+\delta)\cstree_{13,24}.
\]
We can use this freedom to remove one of the coefficients,
for example $\delta=-T_s$ simplifies the amplitude to
\[
T=
T'_t\cstree_{14,23}+
T'_u\cstree_{13,24}.
\]

Next we consider scattering amplitudes between two
untwisted and two twisted fields. They can be written using the symbol
\[
\cstree_{AB,\twist C\twist D}=M_{MAB}K^{MN}\twist M_{N\twist C\twist D}.
\]
There are six permutations for the function:
$\cstree_{12,\twist 3\twist 4}$,
$\cstree_{13,\twist 2\twist 4}$,
$\cstree_{14,\twist 2\twist 3}$,
$\cstree_{23,\twist 1\twist 4}$,
$\cstree_{24,\twist 1\twist 3}$,
$\cstree_{34,\twist 1\twist 2}$,
see \figref{fig:structtreemix}.
In this case there are no Jacobi identities because
there is no vertex to connect untwisted and twisted fields directly.

\begin{figure}\centering
\includegraphics{FigColorTreeMA.mps}\quad
\includegraphics{FigColorTreeMB.mps}\quad
\includegraphics{FigColorTreeMC.mps}\quad
\includegraphics{FigColorTreeMD.mps}\quad
\includegraphics{FigColorTreeME.mps}\quad
\includegraphics{FigColorTreeMF.mps}
\caption{Mixed tree graphs:
$\cstree_{12,\twist3\twist4}$,
$\cstree_{12,\twist3\twist4}$,
$\cstree_{13,\twist2\twist4}$,
$\cstree_{\twist1\twist2,34}$,
$\cstree_{\twist1\twist2,34}$,
$\cstree_{\twist1\twist3,24}$.}
\label{fig:structtreemix}
\end{figure}

Finally, the amplitudes for four twisted fields are
analogous to the untwisted fields discussed above.
Altogether there are $2+6+2$ color structures for
four-particle scattering at tree level.

\subsection{One Loop}
\label{sec:structloop}

\begin{figure}
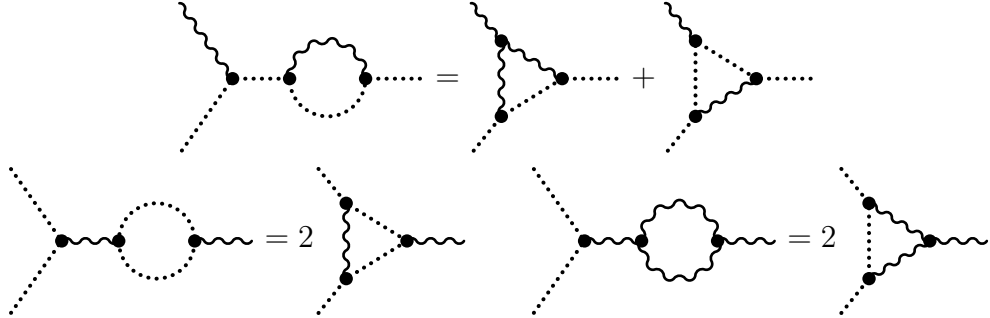
\centering
$\includegraphicsbox{FigColorBubTrigE.mps}
=
\includegraphicsbox{FigColorBubTrigA.mps}+
\includegraphicsbox{FigColorBubTrigB.mps}$

$\includegraphicsbox{FigColorBubTrigF.mps}
=
2\includegraphicsbox{FigColorBubTrigC.mps}$\qquad
$\includegraphicsbox{FigColorBubTrigG.mps}
=
2\includegraphicsbox{FigColorBubTrigD.mps}$

\caption{Bubble-triangle relations.}
\label{fig:structbubtrig}
\end{figure}

For four particle scattering at the one-loop level
there must be four 3-vertices which can be connected in various ways.
It is obvious that the graph has one internal loop
which permits a rough classification:
The loop can have two sides (bubble), three sides (triangle)
or four sides (box). A bubble can be understood to dress a line while
a triangle dresses a vertex.
Bubbles and vertices are in fact closely related,
see \figref{fig:structbubtrig}
Consider a bubble connected to a vertex by a line.
Applying the Jacobi identity (\figref{fig:structjacobi})
to the connecting line will move the loop onto the vertex.
Consider instead a triangle with two sides of equal kind
(in our model triangles always have this property).
Applying the Jacobi identity (\figref{fig:structjacobi}) to the
third side will move the loop onto the line at the opposite side.
We can thus convert freely between bubbles and triangles.
The only exception where this is not possible is for
configurations of mixed particles which lack a
Jacobi identity.

\begin{figure}
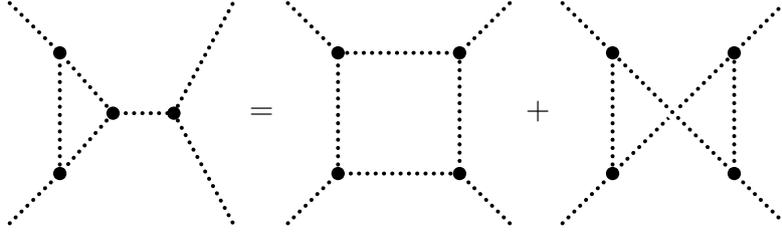
\centering
$\includegraphicsbox{FigColorTrigBoxC.mps}
=
\includegraphicsbox{FigColorTrigBoxA.mps}+
\includegraphicsbox{FigColorTrigBoxB.mps}$
\caption{Triangle-box relation.}
\label{fig:structtrigbox}
\end{figure}

A similar relation holds between triangles and boxes.
Consider two adjacent vertices, one of them being dressed by a loop,
see \figref{fig:structtrigbox}.
Then apply the Jacobi identity (\figref{fig:structjacobi})
to the connecting line. This yields two boxes.
The boxes are distinct and thus this conversion is a one-way procedure:
Triangles can be converted to boxes (unless they contain mixed particles),
but not vice versa.

Our strategy for enumerating independent
one-loop structures for four-particle
scattering is clear. We should convert
bubbles to triangles and triangles to boxes as far as possible.

\begin{figure}
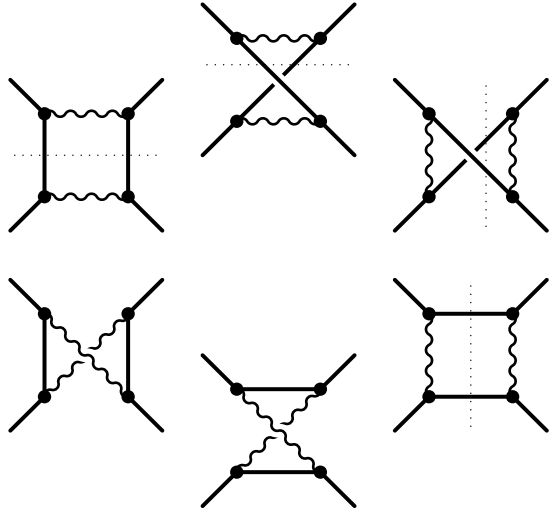
\centering
\includegraphics{FigColorBoxUA.mps}\quad
\raisebox{1cm}{\includegraphics{FigColorBoxUB.mps}}\quad
\includegraphics{FigColorBoxUC.mps}\vspace{0.5cm}\par

\includegraphics{FigColorBoxUF.mps}\quad
\raisebox{-1cm}{\includegraphics{FigColorBoxUE.mps}}\quad
\includegraphics{FigColorBoxUD.mps}
\caption{Purely untwisted scattering at one loop.
Clockwise from top left:
$\csbox_{14,23}$,
$\csbox_{13,24}$,
$\csbox_{13,42}$,
$\csbox_{12,43}$,
$\csbox_{12,34}$,
$\csbox_{14,32}$.
The horizontal and vertical dotted lines
indicate possible unitarity cuts in the
$s$- and $t$-channels, respectively, to be discussed in
\protect\secref{sec:UniUntwist}.
Note that gluon lines cannot be cut in Chern--Simons theories.}
\label{fig:structlooppure}
\end{figure}

We start with only untwisted particles.
Clearly the color structures can be converted
to boxes by the above procedure.
A box has the underlying structure
\[\label{eq:structbox}
\csbox_{AB,CD}=M_{MAE}M_{NCF}K^{MN}K^{PQ}L^{EG}L^{FH}M_{PGB}M_{QHD}.
\]
It has a fourfold symmetry $\csbox_{AB,CD}=\csbox_{BA,DC}=\csbox_{CD,AB}$.
In total there are six boxes, all of the same structure,
but with a permutation of the external legs, see \figref{fig:structlooppure}.
The Jacobi identity relates all six of these, but only at the expense of
triangles which have been eliminated earlier. It turns out that the basis
of six boxes is minimal.

\begin{figure}
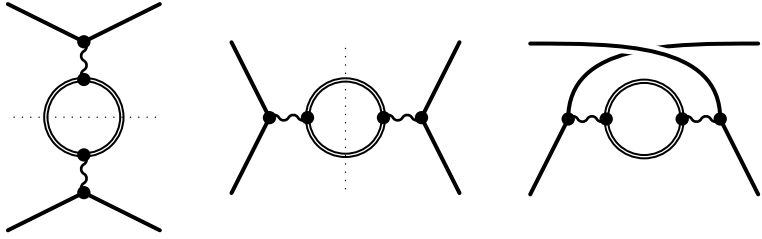
\centering
\includegraphicsbox{FigColorBubbleUTA.mps}\qquad
\includegraphicsbox{FigColorBubbleUTB.mps}\qquad
\includegraphicsbox{FigColorBubbleUTC.mps}
\caption{Untwisted scattering with internal twisted loop:
$\csbubt_{12,34}$,
$\csbubt_{14,23}$,
$\csbubt_{13,24}$.
The horizontal and vertical dotted lines
indicate possible unitarity cuts in the
$s$- and $t$-channels, respectively, to be discussed in
\protect\secref{sec:UniMix}.
}
\label{fig:structlooptwisted}
\end{figure}

For four external untwisted fields there is also the option to
have twisted particles run in the internal loop. In this case the
loop must be a bubble dressing the central gluon line.
Jacobi identities are ineffective here and thus there are
three structures denoted by
$\csbubt_{12,34}$, $\csbubt_{14,23}$, $\csbubt_{13,24}$
(see \figref{fig:structlooptwisted}).
Bubble graphs are defined by
\<\label{eq:structbubble}
\csbubu_{AB,CD}\eq
M_{MAB}K^{MP}M_{PEF}
L^{EG}L^{FH}
M_{QGH}K^{QN}M_{NCD},
\nln
\csbubt_{AB,CD}\eq
M_{MAB}K^{MP}\twist M_{P\twist E\twist F}
\twist L^{\twist E\twist G}\twist L^{\twist F\twist H}
\twist M_{Q\twist G\twist H}K^{QN}M_{NCD},
\>
they have the same eightfold symmetry as tree graphs
$\cstree_{AB,CD}$, but they do not obey an
identity like \eqref{eq:structjactree}.
Note that the untwisted bubble can be expanded
to a sum of boxes
\[\label{eq:structbubbleexpand}
\csbubu_{AB,CD}=2\csbox_{AD,BC}+2\csbox_{AC,BD}.
\]

Finally, we must enumerate the structures for mixed four-particle scattering
at one loop. It turns out that some graphs can be promoted to a box
while others cannot. The latter ones can however be brought to the form
of a bubble dressing the central gluon line. There are four diagrams
$\csbox_{14,\twist2\twist3}$,
$\csbox_{14,\twist3\twist2}$,
$\csbubu_{14,\twist2\twist3}$,
$\csbubt_{14,\twist2\twist3}$
for each assignment of the untwisted and twisted particles to the external legs,
see \figref{fig:structmixed}, giving a total of 24.
The structures are analogous to those in \eqref{eq:structbox,eq:structbubble}
but with some structure constants replaced by twisted ones.

\begin{figure}\centering
\includegraphicsbox{FigColorBoxMA.mps}\qquad
\includegraphicsbox{FigColorBoxMB.mps}\qquad
\includegraphicsbox{FigColorBubbleMA.mps}\qquad
\includegraphicsbox{FigColorBubbleMB.mps}
\caption{Mixed scattering at one loop with
untwisted particles $1,4$ and twisted particles $2,3$:
$\csbox_{14,\twist2\twist3}$,
$\csbox_{14,\twist3\twist2}$,
$\csbubu_{14,\twist2\twist3}$,
$\csbubt_{14,\twist2\twist3}$.
}
\label{fig:structmixed}
\end{figure}

Before we close this part, it is useful to mention that
the one-loop structures can be understood as squares of tree structures.
For example the box can be written as an iterated tree
\[
\csbox_{AB,CD}=\cstree_{AE,CF}L^{EG}L^{FH}\cstree_{GB,HD},
\]
or $\cstree_{16,25}\cstree_{64,53}=\csbox_{14,23}$ for short.
For the three basic tree color structures
$\cstree_{12,34}$, $\cstree_{14,23}$ and $\cstree_{13,24}$
we can set up a convenient multiplication table:
\[\label{eq:structtreemult}
\begin{array}[b]{c|ccc}
\cdot&\cstree_{56,34}&\cstree_{53,64}&\cstree_{63,54}\\\hline
\cstree_{12,56}&\csbubu_{12,34}&-\half \csbubu_{12,34}&-\half \csbubu_{12,34}\\
\cstree_{16,25}&-\half \csbubu_{12,34}&\csbox_{14,23}&\csbox_{13,24}\\
\cstree_{15,26}&-\half \csbubu_{12,34}&\csbox_{13,24}&\csbox_{14,23}\\
\end{array}
\]

\subsection{Planar Limit and Color Ordering}

It is often convenient to consider gauge groups
with a very large rank where the class of
planar Feynman diagrams contributes dominantly.
At the one-loop level it is in fact often sufficient
to just compute the planar Feynman diagrams and
all the non-planar corrections follow
by completion of the color structures.
Color ordering for scattering amplitudes
is also based intrinsically on the availability of many colors.
Here we discuss the large-$N$ behavior of the color structures
discussed above.

A prototypical $\superN=4$ supersymmetric Chern--Simons model
with mixed hypermultiplets is a quiver theory with
$\grp{U}(N_k)$ gauge groups, see \figref{fig:structquiver} on
page \pageref{fig:structquiver}.
The gauge fields belong to the adjoint of the $\grp{U}(N_k)$
while the matter fields are bi-fundamentals
connecting two adjacent gauge group factors.
Let us for definiteness assume that untwisted matter connects
$\grp{U}(N_{2k-1})$ to $\grp{U}(N_{2k})$
and twisted matter connects
$\grp{U}(N_{2k})$ to $\grp{U}(N_{2k+1})$, cf.\ \figref{fig:structquiver}.
In the planar limit all the $N_k$ are taken to be proportional
to some large number $N$.

\begin{figure}\centering
\includegraphicsbox{FigColorOrderA.mps}\qquad
\includegraphicsbox{FigColorOrderB.mps}\qquad
\includegraphicsbox{FigColorOrderC.mps}

\caption{Three color ordering structures
$\Tr(\bar 12)(\bar 21)(\bar 12)(\bar 21)$ for $1234$,
$\Tr(\bar 12)(\bar 21)(\bar 10)(\bar 01)$ for $12\twist3\twist4$,
$\Tr(\bar 12)(\bar 23)(\bar 32)(\bar 21)$ for $1\twist2\twist34$}
\label{fig:structcolororder}
\end{figure}

Now we shall consider color ordering of the legs in a scattering graph:
Each leg is assigned a pair of fundamental color indices
$(\bar k,k\pm 1)$. Two adjacent legs have a common but
mutually conjugate color index $\ldots,k)(\bar k,\ldots$.
A sample color ordering structure for four untwisted fields
is thus $\Tr(\bar 12)(\bar 21)(\bar 12)(\bar 21)$.
A similar color ordering structure for two untwisted and two
twisted fields would be $\Tr(\bar 12)(\bar 21)(\bar 10)(\bar 01)$,
see \figref{fig:structcolororder}.

\begin{figure}\centering
\parbox{2.1cm}{\centering\includegraphicsbox{FigColorDLineA.mps}\\[0.1cm]
\includegraphicsbox{FigColorDLineB.mps}\\[0.3cm]$\ldots$}\qquad
\parbox{2.1cm}{\centering\includegraphicsbox{FigColorDLineC.mps}\\[0.1cm]
\includegraphicsbox{FigColorDLineD.mps}\\[0.3cm]$\ldots$}\qquad
\parbox{2.7cm}{\centering\includegraphicsbox{FigColorDLineE.mps} $-$\\[0.1cm]
\includegraphicsbox{FigColorDLineF.mps} $+$\\[0.1cm]
\includegraphicsbox{FigColorDLineG.mps} $-$\\[0.3cm]
$\ldots$\phantom{ $-$}}

\caption{Double line notation for
untwisted, twisted and gluon color lines,
cf.\ \protect\figref{fig:structlines}.
Gluon lines have associated sign factors.}
\label{fig:structordline}
\end{figure}

\begin{figure}\centering
\parbox{2cm}{\centering\includegraphicsbox{FigColorDVertexA.mps}\\[0.1cm]
\includegraphicsbox{FigColorDVertexB.mps}\\[0.3cm]$\ldots$}\qquad
\parbox{2cm}{\centering\includegraphicsbox{FigColorDVertexC.mps}\\[0.1cm]
\includegraphicsbox{FigColorDVertexD.mps}\\[0.3cm]$\ldots$}\qquad
\parbox{2.5cm}{\centering\includegraphicsbox{FigColorDVertexE.mps} $+$\\[0.1cm]
\includegraphicsbox{FigColorDVertexF.mps} $-$\\[0.3cm]$\ldots$ \smash{\phantom{$+$}}}

\caption{Double line notation for vertices, cf.\ \protect\figref{fig:structvert}.
Pure gluon vertices have associated sign factors.}
\label{fig:structordvert}
\end{figure}

\begin{figure}
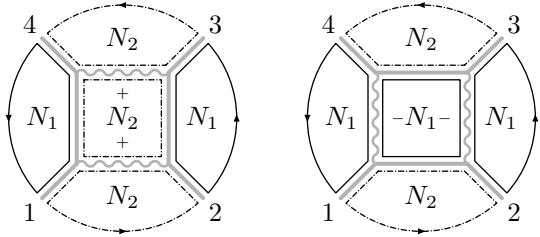
\centering
\includegraphicsbox{FigColorDBoxE.mps}\qquad
\includegraphicsbox{FigColorDBoxF.mps}

\caption{Two box graphs $\csbox_{14,23}$ and $\csbox_{12,43}$
with different assignment of internal gluons
contracted with the color structure
$\Tr(\bar 12)(\bar 21)(\bar 12)(\bar 21)$
in the planar limit.}
\label{fig:structboxorder}
\end{figure}

A color ordering structure can be applied to a color graph
in order to yield a polynomial in the ranks $N_k$.
It is straight-forward to evaluate
the polynomial when the graph is represented in a double line notation:
The lines of a color structure (see \figref{fig:structlines}) are
thickened to a ribbon and the two sides of the ribbon are attributed
a certain color $N_k$, see \figref{fig:structordline}.
The color indices must be adjacent for matter lines
and equal for gauge lines as explained above.
A vertex connects sides of equal color in
two possible ways, see \figref{fig:structordvert}.
Each closed loop of color $k$ then contributes
one power of $N_k$ to the monomial associated to the ribbon graph.
A (possibly incomplete) set of rules to
determine the sign of a ribbon graph is as follows:
Signs originate from gluon lines with odd color (\figref{fig:structordline})
as well as from one out of two pure gluon vertices (\figref{fig:structordvert}).
Note that we will not be careful about some overall signs of color graphs
when they are not related in some way.
Sample ribbon graphs are provided in \figref{fig:structboxorder}.
It is obvious that the large-$N$ asymptotics follows
from the planar structure of the graph.
However, the precise distribution of the $N_k\sim N$ factors
is not as easily recognized. In the example in \figref{fig:structboxorder}
the two different orientations of the box
lead to two different leading-$N$ contributions, $N_1^2N_2^3$ vs.\ $N_1^3N_2^2$.
Let us therefore evaluate the color structures
discussed above which appear in the field theory calculation at one loop.

\paragraph{Pure Scattering.}

For scattering of four untwisted particles
we shall always take the standard color ordering
of $N_1^{-2}N_2^{-2}\Tr(\bar 12)(\bar 21)(\bar 12)(\bar 21)$,
cf.\ \figref{fig:structcolororder}, to evaluate color structures.
The prefactor cancels the color factors which originate
from the color ordering structure itself
and they make the large-$N$ expansion more transparent.
For the tree graphs in \figref{fig:structtreeun}
we obtain the following exact results
\[
\cstree_{12,34}\to -1+\frac{1}{N_3N_4}\,,\qquad
\cstree_{14,23}\to +1-\frac{1}{N_3N_4}\,,\qquad
\cstree_{13,24}\to 0.
\]
The large-$N$ asymptotics agrees with the planar structure
of the underlying graphs, the first two are planar while the third one
is non-planar.
Also the Jacobi identity \eqref{eq:structjactree} is fulfilled.

\begin{figure}
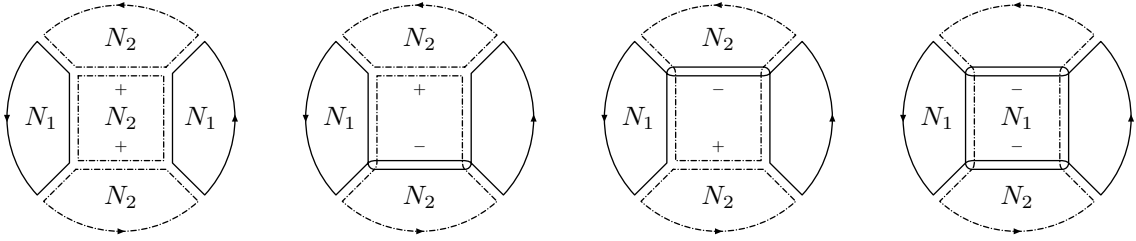
\centering
\includegraphicsbox{FigColorDBoxA.mps}\qquad
\includegraphicsbox{FigColorDBoxB.mps}\qquad
\includegraphicsbox{FigColorDBoxC.mps}\qquad
\includegraphicsbox{FigColorDBoxD.mps}

\caption{Full evaluation of the box graph $\csbox_{14,23}$
contracted with the color structure
$N_1^{-2}N_2^{-2}\Tr(\bar 12)(\bar 21)(\bar 12)(\bar 21)$:
$N_2-N_1^{-1}-N_1^{-1}+N_2^{-1}$.}
\label{fig:structboxexpand}
\end{figure}

Next we evaluate the box graphs in \figref{fig:structlooppure},
see \figref{fig:structboxexpand} for an explicit example,
\[\begin{array}[b]{rclcrclcrcl}
\csbox_{14,23}\earel{\to}\displaystyle N_2-\frac{2}{N_1}+\frac{1}{N_2}\,,&&
\csbox_{14,32}\earel{\to}\displaystyle -\frac{2}{N_1}+\frac{2}{N_2}\,,&&
\csbox_{13,24}\earel{\to}\displaystyle 0\,,
\\[2ex]
\csbox_{12,43}\earel{\to}\displaystyle N_1-\frac{2}{N_2}+\frac{1}{N_1}\,,&&
\csbox_{12,34}\earel{\to}\displaystyle +\frac{2}{N_1}-\frac{2}{N_2}\,,&&
\csbox_{13,42}\earel{\to}\displaystyle 0\,.
\end{array}
\]
The bubble graphs in \figref{fig:structlooptwisted} yield similar expressions
\[\label{eq:structordbub}
\begin{array}[b]{rclcrclcrcl}
\csbubu_{12,34}\earel{\to}\displaystyle 2N_2-\frac{4}{N_1}+\frac{2}{N_2}\,,&&
\csbubu_{14,23}\earel{\to}\displaystyle 2N_1-\frac{4}{N_2}+\frac{2}{N_1}\,,&&
\csbubu_{13,24}\earel{\to}\displaystyle 0\,.
\\[2ex]
\csbubt_{12,34}\earel{\to}\displaystyle 2N_0+\frac{2N_3}{N_1N_2}\,,&&
\csbubt_{14,23}\earel{\to}\displaystyle 2N_3+\frac{2N_0}{N_1N_2}\,,&&
\csbubt_{13,24}\earel{\to}\displaystyle 0\,.
\end{array}
\]
The untwisted bubbles are related to the boxes via \eqref{eq:structbubbleexpand}
and the above expressions obey the rule.
Note that we have evaluated the twisted bubbles
under the assumption of many gauge group factors in
the $\superN=4$ quiver diagram (\figref{fig:structquiver}).
For $\superN=5,6,8$ supersymmetric models
twisted and untwisted representations are the same
and thus the bubbles in \eqref{eq:structordbub} must be
the same as well
\[\label{eq:structordbubext}
\csbubu=\csbubt.
\]
The expressions \eqref{eq:structordbub} for $N_3\to N_1$ and $N_0\to N_2$
do not reflect the equality because the above assumptions
for evaluating the twisted bubble $\csbubt$ do not apply
in a closed quiver of length two (see \figref{fig:structquiver}
on page \pageref{fig:structquiver}).
Instead we must set $\csbubt\to\csbubu$ for $\superN=5,6,8$.

\paragraph{Mixed Scattering.}

Our standard color ordering for mixed scattering
of the type $12\twist3\twist4$
will be $N_0^{-1}N_1^{-2}N_2^{-1}\Tr(\bar 12)(\bar 21)(\bar 12)(\bar 21)$,
cf.\ \figref{fig:structcolororder}.
The single tree diagram for this assignment of twisted legs
evaluates to
\[
\cstree_{12,\twist3\twist4}\to -1.
\]
The one-loop graphs can be found in \figref{fig:structmixed},
they yield
\[
\csbox_{12,\twist3\twist4}\to N_1,
\qquad
\csbox_{12,\twist4\twist3}\to \frac{1}{N_1}\,,
\qquad
\csbubu_{12,\twist3\twist4}\to 2N_2-\frac{2}{N_1}\,,
\qquad
\csbubt_{12,\twist3\twist4}\to 2N_0-\frac{2}{N_1}\,.
\]

Finally let us consider another assignment of twisted legs
$12\twist3\twist4$
which will become useful later.
The standard color ordering
will be $N_1^{-1}N_2^{-2}N_3^{-1}\Tr(\bar 12)(\bar 23)(\bar 32)(\bar 21)$,
cf.\ \figref{fig:structcolororder}.
The color ordered tree graph reads
\[
\cstree_{14,\twist2\twist3}\to +1.
\]
while the color ordered loop amplitudes in \figref{fig:structmixed}
yield
\[
\csbox_{14,\twist2\twist3}\to N_2,
\qquad
\csbox_{14,\twist3\twist2}\to \frac{1}{N_2}\,,
\qquad
\csbubu_{14,\twist2\twist3}\to 2N_1-\frac{2}{N_2}\,,
\qquad
\csbubt_{14,\twist2\twist3}\to 2N_3-\frac{2}{N_2}\,.
\]

For purely twisted scattering we use the ordering
$N_2^{-2}N_3^{-2}\Tr(\bar 32)(\bar 23)(\bar 32)(\bar 23)$,
but the results will be analogous to those of purely
untwisted scattering discussed above.

\section{Four-Particle Scattering in Field Theory}
\label{sec:fieldtheory}

In this section we compare the results for the four particle scattering matrix,
obtained in \secref{sec:SusyScat} using the supersymmetry algebra,
with the predictions of perturbative field theory computations at the tree and one-loop level.

In what is to follow, we shall start with the $\mathcal{N}=4$ theory without twisted hypermultiplets,
we shall then build on that by including twisted hypermultiplets but without imposing any particular
conditions on the representations under which the two matter multiplets transform.
Thus the solutions we obtain will hold for theories with $\mathcal{N} = 4,5,6$ and $8$ supersymmetries.
We will relegate some of the details regarding the field theory conventions
to the appendices where one can find, for example,
the explicit expression for the action \eqref{eqn:CSLag}
and the oscillator expansion of the fields \eqref{eqn:Mode_exp}.

We can define, as is usual, two particle in and out states for the scalars and
fermions in terms of their free oscillator expressions:
\[
\state{B,p_2;A,p_1}\indup{in} =\sqrt{ 2 E_1}\sqrt{2 E_2}\,
C^{\dagger}_{B}(p_2)\,C^{\dagger}_{A}(p_1)
\state{0, t=-\infty}
\]
where $A, B$ denote, here collectively, all the particle labels
and $C^{\dagger}_{A,B}$ are the corresponding positive energy creation operators.
One can then define the $\smat$-matrix elements between in and out states as usual which in
turn, in
the notation of (\Secref{sec:2to2}), defines the operator $\scat{}_{12}^{43}$ via the usual relation
$\smat=1 + i \scat$. Thus,
\[
{}\indup{out}\langle p_4, p_3|p_2,p_1\rangle\indup{in}=
\delta_{1(3}\delta_{4)2}
+ i \delta\left(p_1+p_2-p_3-p_4\right)\ T_{12\bar 3\bar 4}~.
\]
However to connect with the four-particle scattering matrix used in \eqref{eq:SMat},
where all the momenta are on an equal footing
and we have negative energy particles, we must rather consider four point correlation functions.
As is standard, we identify the one-particle
 irreducible four point functions with all momenta incoming,  $\Gamma(p_1,p_2,p_3,p_4)$ with
$-i \ T_{1234}$.

In doing this we must be slightly careful regarding the definition of our asymptotic states. At tree-level
we will simply include an addition factor $\sqrt{4\pi/k}$ per external field but at higher loops
we must include the non-trivial field renormalization that occurs.

\subsection{Pure Amplitudes at Tree Level}
\label{sec:QFTTree}

Let us initially consider the element of $\scat$ given by
\[
A = i \bigvev{
\phi^{A}_{1}(p_1)\,
\phi^{B}_{1}(p_2)\,
\phi^{C}_{2}(p_3)\,
\phi^{D}_{2}(p_4)
}
\]
where $A$ denotes the total contribution for untwisted scalar to scalar scattering,
transforming in the symmetric representation of $\alg{su}(2)$
and without any implicit requirements of color ordering on the indices $A,B,C,D$.

\begin{figure}[ht]\centering
\includegraphics[scale=0.4]{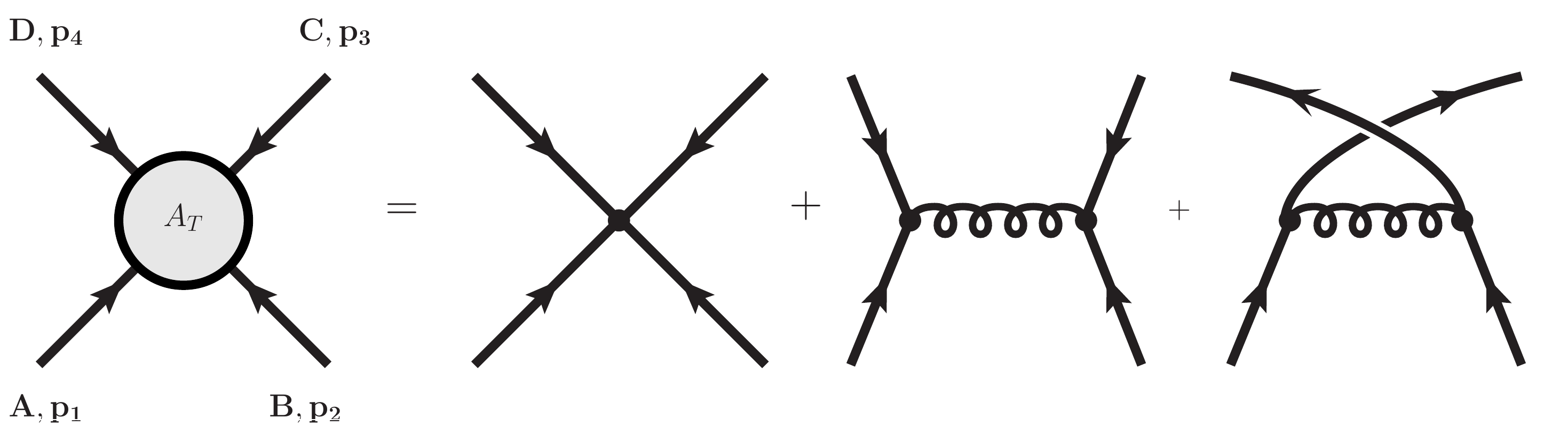}
\caption{Diagrams for the $A$ element at tree level.}
\label{fig:Atree}
\end{figure}

Explicit evaluation of the complete,
non-color ordered matrix element $A$
follows from the Feynman diagrams (\Figref{fig:Atree})
and, in terms of a simplified coupling
\[
\label{eq: coupling}
g=\frac{4\pi}{k}\,,
\]
gives the expression
\[\label{eq:Atree}
A = g\ K_{MN}\lrbrk{M^{M}_{AD}M^{N}_{BC}A_t + M^{M}_{AC}M^{N}_{BD} A_u},
\]
with the color stripped amplitudes $A_t,A_u$
\[
A_t=2i\left(im - \frac{\varepsilon_{\mu\nu\rho} p_1^\mu p_3^\nu p_4^\rho}{\left(p_1+p_4\right)^2}\right),
\qquad
A_u=2i\left(im - \frac{\varepsilon_{\mu\nu\rho} p_1^\mu p_4^\nu p_3^\rho}{\left(p_1+p_3\right)^2}\right).
\]
There are equivalent expressions for $A_t$
 expressed in terms of the three dimensional Lorentz
 invariants $s,t,u$ or the spinors, which themselves are related by,
 \<
\label{eqn:identities_1}
s=(p_1+p_2)^2 \eq- \tprods{1}{\bar 2}\tprods{\bar 1}{2}
=\tprods{1}{\bar 1}^2-\tprods{1}{2}\tprods{\bar 1}{\bar 2},
\nln
t=(p_1+p_4)^2 \eq-\tprods{1}{\bar 4}\tprods{\bar 1}{4}
=\tprods{1}{\bar 1}^2-\tprods{1}{4}\tprods{\bar 1}{\bar 4},
\nln
u=(p_1+p_3)^2 \eq -\tprods{1}{\bar 3}\tprods{\bar 1}{3}
=\tprods{1}{\bar 1}^2-\tprods{1}{3}\tprods{\bar 1}{\bar 3},
\nln
\sqrt{-stu}\eq\pm\tprods{1}{\bar 2}\tprods{2}{\bar 3}\tprods{3}{\bar 1}.
\>
In terms of these variables we can write
\footnote{In the left-most expression there is a sign ambiguity due to the square root.
In checking such properties as crossing it
is therefore always preferable to use the spinor formulation.}
%
\[
A_t = i  \frac{\tprods{1}{2}\tprods{\bar{2}}{\bar 4}}{\tprods{\bar 4}{1}}
= i  \left(2im - \frac{\sqrt{-stu}}{t}\right),
\]
The $A_t$ term in the above expression captures the contribution from one ordering of the contact term indices
and the $t$-channel gluon exchange, while the second term $A_u$ is the remaining part of
the contact term and the $u$-channel  gluon exchange. Due to the choice of $\alg{su}(2)$ indices
there is no $s$-channel gluon exchange diagram.
In obtaining  this answer we have used the fact that the Chern--Simons propagator is given by
\[
\bigvev{A^{\mu}_{M}(p)A^{\nu}_{N}(q)} =  g \ K_{MN}\frac{\varepsilon^{\mu \nu \rho}p_\rho}{2 p^2}\,\delta^3(p+q).
\]
As discussed in \Secref{sec:StructTree} the symmetries of $K_{MN}$ and $M^M_{AB}$ in conjunction
with the  fundamental identity imply that it is possible to write any tree level amplitude involving only untwisted
hypermultiplets as
\[
A^{(0)}=
A_t\cstree_{14,23}+
A_u\cstree_{13,24}.
\]
Note that \eqref{eq:Atree} is already of this form.
For amplitudes where this is not automatically the case,
we can eliminate the $s$-channel color structure using the fundamental identity
\[
K_{MN}\left(M^{M}_{AC}M^{N}_{BD} + M^{M}_{CB}M^{N}_{AD}+ M^{M}_{BA}M^{N}_{CD}\right) =0.
\]
The color ordered contribution to the amplitude is taken to be simply
the first ($t$-channel) term of the above expression, after we strip away the color factor: $A_t$.
The second ($u$-channel) term is related to the first term by crossing.

The undetermined prefactor $T$ appearing in the matrix elements \eqref{eq:SMatEl}
simply equals the $A$ element.
Thus the color ordered normalization factor $T_t$ is nothing but the
$t$-channel amplitude for scalars transforming in the symmetric representation of $\alg{su}(2)$
\[
T_t=A_t  =  i  \frac{\tprods{1}{2}\tprods{\bar{2}}{\bar 4}}{\tprods{\bar 4}{1}}\,.
\]
Once this factor has been set, as above,
we can check the other matrix elements
all of which will have the same prefactor:
\[
g\ K_{MN}M^{M}_{AD}M^{N}_{BC}= g\ \cstree_{14,23}
\]
and for which the tree level  perturbative computations yield:
\[\begin{array}[b]{rclrclrcl}
A\eq \displaystyle+
i \frac{\tprods{1}{2}\tprods{\bar 2}{\bar 4}}{\tprods{\bar 4}{1}}\,,
&
D\eq\displaystyle
- i \frac{\tprods{\bar 2}{\bar 4}\tprods{\bar 3}{\bar 4}}{\tprods{\bar 4}{1}}\,,
&
G\eq \displaystyle
+i\tprods{\bar 2}{\bar 4}\,,
\\[2ex]
\half (A+B)\eq\displaystyle
-i \frac{\tprods{\bar 3}{1}\tprods{2}{4}\tprods{\bar 2}{\bar 4}}{\tprods{\bar 4}{1}\tprods{\bar 3}{4}}\,,
&
\half (D+E)\eq\displaystyle
-i\frac{\tprods{\bar 3}{1}\tprods{\bar 1}{\bar 3}\tprods{\bar 2}{\bar 4}}{\tprods{\bar 4}{1}\tprods{\bar 3}{4}}\,,
&
H\eq \displaystyle
+i\frac{\tprods{\bar 3}{1}\tprods{\bar 2}{\bar 4}}{\tprods{\bar 4}{1}}\,,
\\[2ex]
\half (A-B)\eq\displaystyle
-i \frac{\tprods{1}{4}\tprods{\bar 1}{\bar 3}}{\tprods{\bar 3}{4}}\,,
&
\half (D-E)\eq\displaystyle
+i \frac{\tprods{\bar 1}{\bar 3}\tprods{\bar 2}{\bar 3}}{\tprods{\bar 3}{4}}\,,
&
K\eq \displaystyle
-i \frac{\tprods{\bar 4}{2}\tprods{\bar 2}{\bar 4}}{\tprods{\bar 4}{1}}\,,
\\[2ex]
\half C\eq\displaystyle
-i \frac{\tprods{\bar 3}{1}\tprods{\bar 1}{\bar 3}}{\tprods{\bar 3}{4}}\,,
&
\half F\eq\displaystyle
+i \frac{\tprods{\bar 2}{4}\tprods{\bar 1}{\bar 3}}{\tprods{\bar 3}{4}}\,,
&
L\eq \displaystyle
+i \tprods{\bar 1}{\bar 3}\,.
\end{array}\]
These results are in manifest agreement with the predictions from the supersymmetry algebra
in \eqref{eq:SMatEl}.
We can now write the complete tree-level untwisted-untwisted scattering prefactor
\[
\label{eq:uuuu_tree}
T^{(0)}_{1234}=
\cstree_{14,23} \left(i g \frac{\tprods{1}{2}\tprods{\bar 2}{\bar 4}}{\tprods{\bar 4}{1}}\right)
+\cstree_{13,24} \left(i g \frac{\tprods{1}{2}\tprods{\bar 2}{\bar 3}}{\tprods{\bar 3}{1}}\right).
\]

Since the twistor brackets satisfy a host of non-linear identities
such as \eqref{eq:twistoridentity,eq:twistorcyclic}
there is no canonical way of representing the results of the perturbative computations
however the above seems to be particularly simple.
We note the following identities have been used to compute
the scattering amplitudes involving four fermions
i.e.\ the matrix elements $D$ and $E$:
\<
\left(u_{\bar 4}\gamma_\mu v_{1}\right)\left(u_{\bar 3}\gamma_\nu v_2\right)\frac{ \varepsilon^{\mu \nu\rho}(p_1+p_4)_\rho}{t} \eq 2\, \frac{\tprods{\bar 2}{\bar 4}\tprods{\bar 3}{\bar 4}}{\tprods{1}{\bar 4}}\,,\nln
\left(u_{\bar 3}\gamma_\mu u_{\bar 4}\right)\left(v_1\gamma_\nu v_2\right)\frac{ \varepsilon^{\mu \nu\rho}(p_1+p_2)_\rho}{s} \eq 2\left(\frac{\tprods{1}{2}\tprods{\bar{2}}{\bar 4}}{\tprods{1}{\bar 4}}\right)\left(\frac{\tprods{\bar 3}{ 2}\tprods{\bar 2}{\bar 3}}{\tprods{1}{2}\tprods{\bar 3}{4}}\right).
\>
On the l.h.s.,
we have the contributions of gluon exchanges
between two fermions as they would usually appear in a (tree level) perturbative computation.
The r.h.s.\ are the relevant twistorial expressions.

Given the explicit expressions for the matrix elements it is straightforward to check that they
satisfy the crossing relations described in \Secref{sec:cross}. For example the invariance
under exchange $3 \leftrightarrow 4$ is immediate from \eqref{eq:uuuu_tree} and using
the identities \eqref{eq:twistoridentity} one can see that it is invariant under exchange
of $1\leftrightarrow2$.
To further see that the prefactor transforms as
\[
T^{(0)}_{2341}=-\frac{\tprods{2}{3}\tprods{\bar 4}{1}}{\tprods{1}{2}\tprods{\bar4}{3}}\ T^{(0)}_{1234}
\]
under $1\rightarrow2\rightarrow3\rightarrow4\rightarrow1$ we need to use \eqref{eq:twistorcyclic} and
the Jacobi identities relating the tree level color structures.

\subsection{Mixed Amplitudes at Tree Level}
\label{sec:QFTTreeMix}

The  `twisted-untwisted' multiplet scattering is much the same
and the overall prefactor $T$,
corresponding to the $A$ element, is
\[
A=T=\braket{\scat}{\phi_{(a}{\phi}_{b)}{\twist \psi}_{(c}\twist\psi_{d)} }.
\]
In this case is the color ordered amplitude is defined to be the coefficient
of the single mixed color structure at tree level
\[
 g\ K_{MN}M^{M}_{BA}\twist{M}^{N}_{DC}= g\ \cstree_{12,\twist3\twist4}.
\]
We have the following perturbative tree-level results:
\[\begin{array}[b]{rclrclrcl}
A\eq \displaystyle-i \tprods{3}{4}\,,
&
D\eq\displaystyle +i \tprods{\bar 1}{\bar 2} \,,
&
G\eq \displaystyle
-i \tprods{3}{\bar 2} \,,
\\[2ex]
\half (A+B)\eq\displaystyle
+i \frac{\tprods{\bar2}{4}\tprods{2}{4}}{\tprods{\bar3}{ 4}}
\,,
&
\half (D+E)\eq\displaystyle
-i \frac{\tprods{\bar 2}{4} \tprods{\bar 1}{\bar 3}}{\tprods{\bar 3}{4}}
\,,
&
H\eq \displaystyle
- i \tprods{\bar 2}{4}\,,
\\[2ex]
\half (A-B)\eq\displaystyle
+i \frac{\tprods{\bar1}{4}\tprods{1}{4}}{\tprods{\bar3}{ 4}}
\,,
&
\half (D-E)\eq\displaystyle
-i \frac{\tprods{\bar 2}{\bar3}\tprods{\bar 1}{4}}{\tprods{\bar 3}{4}}
\,,
&
K\eq \displaystyle
-i \tprods{3}{\bar 1}\,,
\\[2ex]
\half C\eq\displaystyle
+i\frac{\tprods{\bar 2}{4}\tprods{\bar3}{2}}{\tprods{\bar3}{4}}\,,
&
\half F\eq\displaystyle
-i   \frac{\tprods{\bar 2}{4}\tprods{\bar1}{4}}{\tprods{\bar3}{4}}\,,
&
L\eq \displaystyle
-i\tprods{\bar 1}{4}\,.
\end{array}\]
Once again, we note that the perturbative tree level results completely
agree with the computations based on the supersymmetry algebra \eqref{eq:SMatEl}.
In this case the complete result for tree level untwisted-twisted scattering
is given by
\[
\label{eq:uutt_tree}
T^{(0)}_{12\twist 3\twist 4} =-ig\tprods{3}{4}  \ \cstree_{12,\twist 3\twist 4} .
\]
For the sake of completeness,
we list some of the key identities
that are useful for converting the results obtained
from standard perturbation theory
to the twistorial expressions used in the expression for the scattering matrix
\<
(v_{\bar 3}\varepsilon v_{\bar4})\left[\frac{\tprods{\bar 3}{1}\tprods{2}{4}}{\tprods{1}{2}\tprods{\bar3}{ 4}} + \frac{\tprods{1}{4}\tprods{\bar3}{2}}{\tprods{1}{2}\tprods{\bar3}{ 4}}\right] \eq -2\varepsilon^{\mu\nu\sigma}
(v_{\bar 4} \sigma_\mu v_{\bar 3}) \frac{(p_1)_\nu(p_2)_\sigma}{s}\,,\nln
( v_{\bar 3}\varepsilon v_{\bar 4})\left[\frac{\tprods{\bar 3}{2}\tprods{\bar2}{\bar3}}{\tprods{1}{2}\tprods{\bar3}{ 4}} - \frac{\tprods{\bar3}{1}\tprods{\bar1}{\bar3}}{\tprods{1}{2}\tprods{\bar3}{ 4}}\right] \eq +2\varepsilon^{\mu\nu\sigma}\left(v_1\sigma_\mu v_2\right)\frac{(p_3)_\nu(p_4)_\sigma}{s}\,,\nln
2\left(v_{\bar 3} \varepsilon v_{\bar 4}\right) \left[\frac{\tprods{\bar 3}{1}\tprods{\bar1}{4}}{\tprods{1}{2}\tprods{\bar3}{ 4}}\right] \eq -\frac{\varepsilon^{\mu \nu \sigma}}{s}(p_1+p_2)_\sigma
\left(v_{\bar 4} \sigma_\mu v_{\bar 3} \right) \left(v_1 \sigma_\nu v_2\right)\,,\nln
\left(v_{\bar 3} \varepsilon v_{\bar 4} \right) \left[\frac{\tprods{\bar 3}{1}\tprods{\bar3}{2}}{\tprods{1}{2}\tprods{\bar3}{ 4}}\right]\eq  -\frac{\varepsilon^{\mu\nu\sigma}(p_3-p_4)_\mu(p_1)_\nu(p_2)_\sigma}{s}\,.
\>

As it is also useful for later considerations we record the field theory result
for scattering when particles 2 and 3 are twisted. In this case the overall
prefactor, corresponding to the $A$ element, is
\[
A=T=\braket{\scat}{\phi_{(a}{\twist \psi}_{b)}{\twist \psi}_{(c}\phi_{d)} }
=i\, \frac{\tprods{\bar 1}{2}\tprods{1}{2} }{\tprods{\bar 3}{2}}
=-i\, \frac{\tprods{\bar 4}{3}\tprods{4}{3} }{\tprods{\bar 2}{3}}\,.
 \]
 Thus we have
\[
\label{eq:uttu_tree}
T^{(0)}_{1\twist 2\twist 3 4}=
- ig \frac{\tprods{\bar 1}{2}\tprods{1}{2}}{\tprods{2}{\bar 3}}\ \cstree_{14,\twist 2\twist 3}.
\]

In the case of scattering between the twisted matter content of the theory,
we have to consider the fact that  the quartic bosonic vertex as well as
the fermion mass terms come with opposite signs as compared to the untwisted sector.
The explicit formulae for the scattering of the untwisted multiplet can be readily adapted
to the case at hand. The matrix elements for any four untwisted fields
can be taken over to their twisted counterparts by changing the sign of the mass
and replacing $u(p)$ by $v(p)$ and vice versa. Thus the matrix element $D$ is given by
\[
D
=i \frac{\tprods{\bar 1}{\bar 2}\tprods{2}{4}}{\tprods{4}{\bar 1}}
= i \left(-2im - \frac{\sqrt{-stu}}{t}\right)
\,,
\]
where the relevant color prefactor is $g \cstree_{\twist 1\twist 4,\twist 2\twist 3}$.
It is a straightforward exercise to see that
\eqref{eq:SMatEl} continues to describe the four particle scattering
matrix relevant to  the twisted sector of the theory.
In particular the overall prefactor, $T=A$, for this sector is given by
\[
T=A=-i \frac{\tprods{2}{ 4}\tprods{3}{4}}{\tprods{4}{\bar 1}}
\]
and so
\[
\label{eq:tttt_tree}
T^{(0)}_{\twist{1}\twist{2}\twist{ 3}\twist{ 4}}=
\cstree_{\twist1\twist4,\twist2\twist3} \left(-ig\,\frac{\tprods{2}{ 4}\tprods{3}{4}}{\tprods{4}{\bar 1}}\right)
+\cstree_{\twist1\twist3,\twist2\twist4}  \left(-ig\, \frac{\tprods{2}{ 3}\tprods{4}{3}}{\tprods{3}{\bar 1}}\right)
.
\]

Finally one can check that the untwisted-twisted and twisted-twisted scattering matrix elements
satisfy the crossing relations in \Secref{sec:cross}.

If one chooses the matter to be in representations such that there is extended $N=5$
supersymmetry there are additional
relations between the pure twisted-twisted, pure untwisted-untwisted
and the mixed untwisted-twisted scattering as described in \Secref{sec:higher_susy}.
It is straightforward to  check that the perturbative calculations are consistent with those additional relations.
In particular, if the untwisted and twisted multiplets transform in the same gauge representation,
supersymmetry implies that
\[
 T_{12 \twist 3  \twist 4}
 =
 \frac{ \tprods{\bar 3}{1} \tprods{\bar 2}{\bar 1} T_{\twist1\twist2\twist3\twist4} +
 \tprods{1}{2} \tprods{\bar 2}{4} T_{1234} }{\tprods{1}{2} \tprods{ \bar 2}{\bar 3}+\tprods{1}{4}\tprods{\bar4}{\bar3} }~
\]
which is indeed satisfied by the tree-level expressions above.

To check the $\superN=8$ relations one must make use of the simplifications
in the color structure that occur for the gauge group $\grp{SO}(4)$
\<
\cstree_{12,34}\eq M_{AB}^M K_{MN}M^N_{CD}
\propto \eps_{ABCD}\eps^{\hat a\hat b} \eps^{\hat c\hat d}\nn\\
\cstree_{13,24}\eq M_{AC}^M K_{MN}M^N_{CD}
\propto -\eps_{ABCD}\eps^{\hat a\hat c} \eps^{\hat b\hat d}\nn\\
\cstree_{14,23}\eq M_{AC}^M K_{MN}M^N_{CD}
\propto \eps_{ABCD}\eps^{\hat a\hat d} \eps^{\hat b\hat c}~.
\>

Using these relations one can check that the constraints \eqref{eq:ne8eq}
are satisfied.

\subsection{One-Loop Amplitudes in Pure \texorpdfstring{$\superN=4$}{N=4} SCS}
\label{sec:QFTLoop}

In this section we shall calculate the one-loop correction to the scattering matrix using standard off-shell
methods. As is well known,
it is possible using unitarity relations, to reconstruct the imaginary part of
the one-loop amplitudes from the phase space integral over products of
tree-level amplitudes (which can of course be extended to higher orders).
This provides a very efficient
method for calculating scattering amplitudes and the results that follow from
unitarity for the theories at hand, elaborated upon in detail in the next section, are in perfect agreement with the predictions following from the supersymmetry algebra. However the drawback of
these methods is the so called ``polynomial ambiguity'' whereby the cut construction can
miss contributions which are rational functions of the kinematic invariants lacking a cut.
In four dimensional super Yang--Mills all one-loop massless amplitudes are cut constructible (for
discussion see e.g.\  \cite{Bern:1996je}) as all rational terms are related to terms with cuts at
$\mathcal{O}(\epsilon^0)$ (where $\epsilon$ is the dimensional regularisation parameter). This is
true for the maximally supersymmetric $\superN=4$ case but also for $\superN=1$ theories.
However it is certainly not true in general and non-supersymmetric Yang--Mills theories
are not one-loop cut constructable. Thus while
it is reasonable to expect the one-loop amplitudes in the supersymmetric
Chern--Simons matter theories to be cut-constructible it is by no means guaranteed.
The off-shell methods provide a check that the cuts are indeed capturing all the amplitude
and that there is no rational piece unrelated to a logarithm.
In principle one could also fix any rational function by working to sufficiently high order in $\epsilon$
however this can be  involved and the direct calculation is quite feasible for the two to two scattering.


We will consider the matrix element $G$ in \eqref{eq:SMat}
\[
G \ \delta^3(p_1+p_2+p_3+p_4)= -\braket{\scat}{\phi_1\psi_{\dot 1}\psi_{\dot 2}\phi_2}
\]
as it
involves the fewest number of Feynman diagrams at the one-loop level. We initially consider
the contribution with only gluons or untwisted matter running in loops and
then separately add the contribution from twisted matter. Let us now make a few remarks about
the color structure; following an examination of  interactions following from the action
\eqref{eqn:CSLag} we can use the Jacobi identities on the vertices to see that,
in the notation of \secref{sec:structloop}, only box-like structures occur.
The color structures that appear in the $s$-channel diagrams are:
\[
\csbox_{14,23}\ \ \mbox{and}\  \ \csbox_{13,24}\ ~ ,
\]
in the $t$- and $u$-channels we have, respectively,
\[
\csbox_{13,42},\csbox_{12,43}
\qquad
\mbox{and}
\qquad
\csbox_{12,34},\csbox_{14,32}.
\]
Thus we see that all box-like color structures enumerated in \secref{sec:structloop} can
appear. However the
coefficients of the different structures are all related by crossing and
so we need only calculate a single coefficient.
We will thus focus on the $s$-channel
contribution to the color-ordered amplitude which occurs with the
prefactor $\csbox_{14,23}$.
\begin{figure}[ht]
\begin{center}
\includegraphics[scale=0.4]{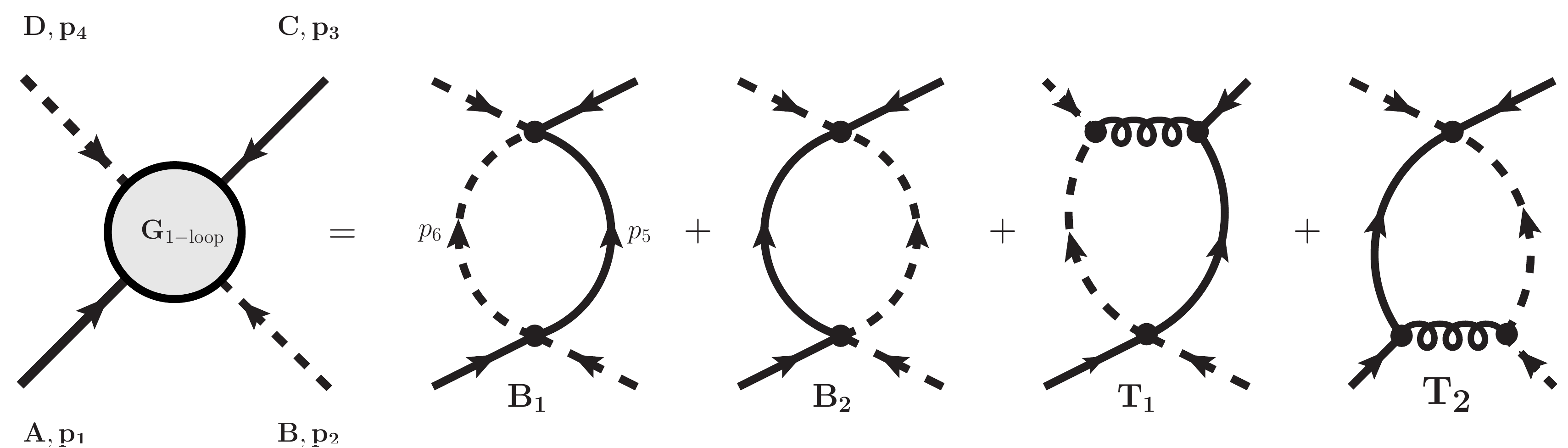}
\caption{Diagrams for the $G$ element at one-loop.}
\label{fig:G1loop}
\end{center}
\end{figure}

The one-loop correction to $G$ involves four Feynman diagrams (\Figref{fig:G1loop}).
Two of these diagrams correspond to `bubbles'
with a scalar and a fermionic propagator forming a closed loop
and two are `triangles', where a gluon exchange between the
intermediate scalar and fermion fields completes the loop.
The complete answer is given by
\[
G^{(1)}
=i \  \csbox_{14,23}  \int d^3 \ell \left(B_1+B_2+T_1+T_2\right)
\]
The integrands for the bubble diagrams with all momenta incoming are:
\<
B_1 \eq- i\frac{v_2\varepsilon(ip_6 + m\varepsilon)\varepsilon v_4}{2(p_6^2+m^2)(p_5^2+m^2)}\,,\nln
B_2 \eq- i\frac{v_2\varepsilon(ip_5 + m\varepsilon)\varepsilon v_4}{2(p_6^2+m^2)(p_5^2+m^2)}\,.
\>
 The integrands for the triangles are:
\<
T_1 \eq \frac{i}{D}\left(v_2\varepsilon(ip_6 + m\varepsilon)\bar{\sigma}^\rho v_4\,\varepsilon_{\rho \nu \mu}(p_6+p_4)^\nu (p_5-p_3)^\mu\right),\nln
T_2 \eq \frac{i}{D'}\left(v_2\bar{\sigma}^\rho(ip_5 + m\varepsilon)\varepsilon v_4\,\varepsilon_{\rho \nu \mu}(p_2-p_5)^\nu (p_1+p_6)^\mu\right).
\>
It is to be understood that all the spinors and $\sigma$ matrices carry lower indices and we use the shorthand notation where $(p_i)_{\alpha \beta} = (p_i)_\mu (\sigma^\mu)_{\alpha \beta}$. We denote
$\sigma$ matrices with raised indices as $\bar{\sigma}$  i.e.\ $(\bar{\sigma}^\mu)^{\alpha  \beta} = \varepsilon^{\alpha \gamma}\varepsilon^{\beta \delta}(\sigma^\mu)_{\gamma \delta}$. We also have
introduced
\<
D  \eq2 (p_6^2+m^2)(p_5^2+m^2)(p_6+p_4)^2\ \  \mbox{and}\nln
 D'  \eq 2 (p_6^2+m^2)(p_5^2+m^2)(p_5-p_2)^2
 \>
to denote the triangle denominators. There is of course only one independent loop momentum as $p_5$  and $p_6$ are related by the kinematical constraints $p_6 + p_5 = p_1+p_2 = -(p_3+p_4)$.
We now note the following identities:
\<
v_2\varepsilon(ip_6 + m\varepsilon)\bar{\sigma}^\rho v_4 \eq -iv_2(\varepsilon \eta ^{\rho \kappa} + \bar{\sigma} _\chi \varepsilon^{\chi \kappa \rho})v_4\ (p_6+p_2)_\kappa,\nln
v_2\bar{\sigma}^\rho (ip_5 + m\varepsilon)\varepsilon v_4 \eq-iv_2(\eta^{\rho \kappa}\varepsilon + \varepsilon^{\rho\kappa\chi}\bar{\sigma}_\chi)v_4\ (p_5-p_4)_\kappa.
\>
The use of these identities in $T_1$ and $T_2$ respectively, generates four terms for each of the triangle diagrams. Using $\bar{p}_{ij} = p_i-p_j$
and $p_{ij} = p_i +p_j$ for brevity, we have
\<
T\indup{1a} \eq -\frac{2}{D}\left(v_2\varepsilon v_4\right)\,\varepsilon(p_5,p_3,p_4-p_2),\nln
T\indup{1b} \eq +\frac{1}{D}\left(v_2\bar{\sigma}^\rho \eta_{\rho \lambda} v_4\right)\,p^2_{64}\bar{p}^\lambda_{53},\nln
T\indup{1c} \eq +\frac{1}{D}\left(v_2\bar{\sigma}^\rho \eta_{\rho \lambda} v_4\right)\,(p_5^2+m^2)p^\lambda_{64},\nln
T\indup{1d} \eq +\frac{1}{D}\left(v_2\bar{\sigma}^\rho \eta_{\rho \lambda} v_4\right)\,(\bar{p}_{42}\cdot \bar{p}_{53}p^\lambda_{64} - \bar{p}_{42}\cdot p_{64}\bar{p}^\lambda_{53}).
\>
Similarly, we obtain
\<
T\indup{2a} \eq -\frac{2}{D'}\left(v_2\varepsilon v_4\right)\,\varepsilon(p_4-p_2,p_6,p_1),\nln
T\indup{2b} \eq +\frac{1}{D'}\left(v_2\bar{\sigma}^\rho \eta_{\rho \lambda} v_4\right)\,\bar{p}^2_{52}{p}^\lambda_{16},\nln
T\indup{2c} \eq +\frac{1}{D'}\left(v_2\bar{\sigma}^\rho \eta_{\rho \lambda} v_4\right)\,(p_6^2+m^2)\bar{p}^\lambda_{52},\nln
T\indup{2d} \eq +\frac{1}{D'}\left(v_2\bar{\sigma}^\rho \eta_{\rho \lambda} v_4\right)\,(\bar{p}_{42}.\bar{p}_{25}p^\lambda_{16} - \bar{p}_{42}.p_{16}\bar{p}^\lambda_{25}).
\>

After introducing Feynman parameters to simplify the denominators, shifting the
loop momenta and dropping terms linear in the loop momenta it can be
straightforwardly seen that the terms $T\indup{1a,2a}$ and $T\indup{1d,2d}$ cancel using the relation
\[
\left(v_2 \varepsilon v_4\right)\, \varepsilon(p_4,p_3,p_2)=-v_2 \bar \sigma^{\lambda} v_4\Big[p_2\cdot \bar p_{24}\ (p_1)_{\lambda}-p_1\cdot {\bar p}_{24}\ (p_2)_{\lambda}\Big].
\]
 Further, the terms $T\indup{1b}$, $T\indup{2b}$, $B_1$ and $B_2$ combine to give
\[
\label{eq:integrand_fin}
T\indup{1b}+T\indup{2b}+B_1+B_2= +i\frac{ v_2 \varepsilon( i p_1-m\varepsilon) \varepsilon v_4}{(p_6^2+m^2)(p_5^2+m^2)}
\]
which corresponds to an $s$-channel
 massive scalar bubble integral, $\intbub(s)$,  with a coefficient proportional
to $\tprods{\bar 2}{ \bar 1}\tprods{1}{ \bar 4}$.
The massive bubble integral can be evaluated
\<
\label{eq:massbub}
\intbub(s) \eq \int d^{3} \ell \ \frac{1}{\left(\ell^2+m^2\right)\left((\ell-p_{12})^2+m^2\right)}\nn\\
                          \eq \frac{i \pi^2}{\sqrt{-s}} \ln \left(\frac{2m+\sqrt{-s}}{2m-\sqrt{-s}}\right).
\>
Finally there is the contribution from the terms $T_{1c}$
and $T_{2c}$ which can be written as
\[
T_{1c}+T_{2c}=\frac{- i m\  (v_2\varepsilon v_4)}{(\tilde \ell^2+\Delta)^2}(x-1),
\]
where we have introduced the Feynman parameter $x$ and $\Delta=(1-x)^2m^2$.
In fact the integral over the loop momenta and
Feynman parameter can be trivially done and the result is
\[
\int dx\ \int d\ell^3  \frac{1}{(\tilde \ell^2+\Delta)^2}(x-1)=  \frac{\pi^2}{ m}\,.
\]
This thus contributes to the amplitude a term
proportional to the tree level contribution
\[
\label{eq:rational_rem}
\pi^2 \tprods{\bar 2}{\bar 4}  \csbox_{14,23},
\]
however it is cancelled by identical factors
coming from the renormalization of the fermionic fields
and which contribute to the S-matrix via the LSZ reduction formula.

\begin{figure}[ht]\centering
\includegraphics[scale=0.4]{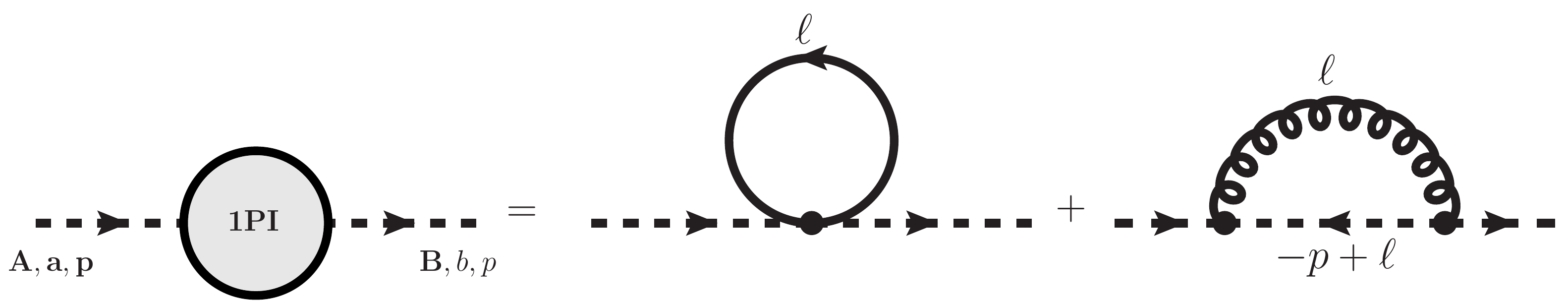}
\caption{Diagrams for fermionic propagator at one-loop.}
\label{fig:Fermion_prop}
\end{figure}

\begin{figure}[ht]\centering
\includegraphics[scale=0.35]{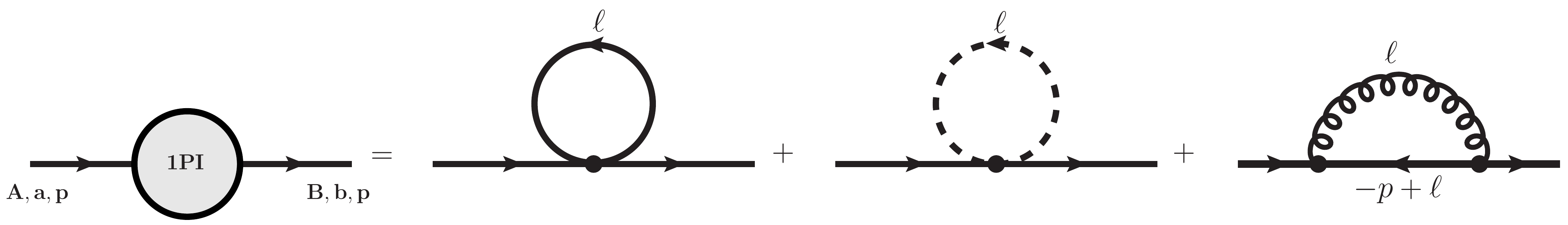}
\caption{Diagrams for bosonic propagator at one-loop.}
\label{fig:Boson_prop}
\end{figure}

We first consider the diagrams
contributing to the fermionic self-energy (\Figref{fig:Fermion_prop}):
a tadpole with scalars in the loop and a gluon correction.
There is also in principle a contribution from the twisted fields,
if they are present,
however their contribution vanishes due to color index contractions.
The tadpole diagram gives an integrand
\[
T(p)= -iK_{MN}M^M_{A{\hat A}}M^N_{{\hat B} B}L^{\hat B \hat A}\varepsilon^{\dot a\dot b}\frac{\eps}{\ell^2+m^2}
\]
and the gluon correction is
\[
I(p)= -K_{MN}M^M_{A{\hat A}}M^N_{{\hat B}B}L^{\hat B \hat A}\varepsilon^{\dot a\dot b}
\frac{\bar \sigma^\mu\left(i(\ell-p)+\varepsilon m\right)\bar \sigma^\nu \varepsilon_{\mu\rho\nu}\ell^\rho}{2\left[(\ell-p)^2+m^2\right]\ell^2}.
\]
For simplicity we strip off the color and flavor indices, then using the relations
\<
\sigma^{\lambda}\bar \sigma^{\nu}\eq-\eta^{\lambda\nu}+\sigma_{\kappa}\varepsilon\varepsilon^{\kappa\lambda\nu},
\nln
\bar \sigma^{\lambda} \sigma^{\nu}\eq-\eta^{\lambda\nu}+\bar \sigma_{\kappa}\varepsilon\varepsilon^{\kappa\lambda\nu}
\>
we can simplify the gluon contribution
\<
I(p) \eq-\frac{-i\ell\cdot (\ell-p)\ \varepsilon -m\ \ell}{\left[(\ell-p)^2+m^2\right]\ell^2}\nln
  \eq- \frac{-i\varepsilon}{(\ell-p)^2-m^2}+\frac{\ell\cdot p\ \varepsilon- m\ \ell}{\left[(\ell-p)^2+m^2\right]\ell^2}
\>
where the first term can be seen to cancel against the tadpole diagram. We can simplify
the remaining term by introducing Feynman parameter $x$ and shifting the loop-momenta.
Thus, with $\Delta=x(1-x)p^2+x m^2$, we have
\[
M^2(p)=
im\int dx\int d^3\ell\, x \, \frac{ \varepsilon(i p + p^2/m\ \varepsilon)\varepsilon }{\left[\ell^2+\Delta(p)\right]^2 }\,.
\]
Iterating these 1PI diagrams we find the correction to the propagator
\[
\frac{(i p+ m \varepsilon)}{p^2+m^2+M^2(p)}= Z\indup{f}(p)\, \frac{i p+\varepsilon m}{p^2+m^2}
 +\mbox{terms regular as } p_0\rightarrow E(p)
\]
where we see that the mass remains unchanged (the pole is not shifted as the correction is proportional to
the inverse propagator) and  that the one-loop shift in the field renormalization is $\delta Z\indup{f}(p)=i g^2\pi^2$.
 The one-loop correction to the bosonic propagator can be easily seen to be zero. The relevant
 diagrams are given in (\Figref{fig:Boson_prop}).

In this case the boson and fermion contributions exactly cancel and the gluon contribution is zero
due to the $\varepsilon$ tensor in the propagator, thus $Z\indup{b}(p)=1$. There are almost identical contributions to the matter in
the twisted hypermultiplets.

There is also a non-vanishing one-loop correction to the gluon propagator which is a known effect
in supersymmetric Chern--Simons theories (see e.g.\  \cite{Chen:1992ee,Gaiotto:2007qi})
and indeed in this case we find the same result.
The gluon self-interaction cancels against the ghost loop while the fermion and bosonic
contributions add to give a correction that is similar to the four-dimensional YM propagator.
However we should point out that as there are no physical gluon states,
there is no point in interpreting this correction as a field renormalization
entering into the scattering matrix.

It is worth here pausing to make a comment regarding the color structures that can arise
from the corrections to the $S$-matrix due to the field renormalizations. These have the
structure of bubbles on fermionic legs, labelled  respectively $(B,p_2)$ and $(D,p_4)$, attached
to tree-level diagrams
$\cstree_{14,23}$ and $\cstree_{13,24}$, for example
\[
\bigl[K_{MN}M^M_{DE}M^N{}^{EF}\bigr]\bigl[K_{PQ}M^P_{AF}M^Q_{BC}\bigr]~.
\]
However making use of the identities described in \Secref{sec:structloop} and in particular
\Figref{fig:structbubtrig} we can express these in the basis of one-loop box diagrams to find
the relevant term i.e.  the coefficient of the structure $\csbox_{14,23}$.

The one-loop contribution to the $G$ element from the field renormalization is
\[
\Delta G^{(1)}=\left( \sqrt{Z\indup{b}(p_1)Z\indup{f}(p_2) Z\indup{b}(p_3)Z\indup{f}(p_4)} -1\right) G^{(0)}= - \pi^2 \tprods{\bar 2}{\bar 4}
\]
which can be seen to cancel the contribution \eqref{eq:rational_rem}.

Thus we find that
\[
G^{(1)}=i \ \tprods{\bar 2}{\bar 1}\tprods{1}{\bar 4}\, \intbub(s).
\]
This is the complete $s$-channel contribution at one-loop, whose overall factor is
$\csbox_{14,23}$, for an $\mathcal{N}=4$ theory without twisted hypermultiplets.
There of course remain the other color ordering and the $t$-channel and $u$-channel diagrams;
the $t$-channel diagrams are
identical to those above after exchanging the external momenta while the $u$-channel contributions
are slightly more complicated.
However, as stated above they are all related to the calculated piece,
once one accounts for the appropriate color factors. Using this element we can determine
the one-loop piece of the overall factor undetermined by the symmetries
 \[
T^{(1)} =\frac{G^{(1)} }{ G^{(0)} }T^{(0)}= i \tprods{\bar 1}{\bar 2}\tprods{1}{2}\intbub(s)
 \]
 and thus
 \<
 \label{eq:uuuu_one_loop}
 T^{(1)}_{1234} =&&
 \left(\csbox_{14,23} +\csbox_{13,24} \right)\left[i \tprods{\bar 1}{\bar 2}\tprods{1}{2}\intbub(s)\right]
\nln
&+&\left(\csbox_{13,43} +\csbox_{12,43}\right)\left[i \frac{\tprods{\bar 2}{\bar 3}\tprods{\bar 1}{2}\tprods{3}{4}}{\tprods{\bar 1}{4}}\,\intbub(t)\right]
\nln
& +&\left(\csbox_{12,34} +\csbox_{14,32} \right)\left[i \frac{\tprods{\bar 2}{\bar 4}\tprods{ 1}{2}\tprods{\bar 3}{4}}{\tprods{\bar 3}{1}}\,\intbub(u)\right]
.
\>

\subsection{Mixed Amplitudes at One Loop}
\label{sec:QFTLoopMix}

\begin{figure}[ht]\centering
\includegraphics[scale=0.35]{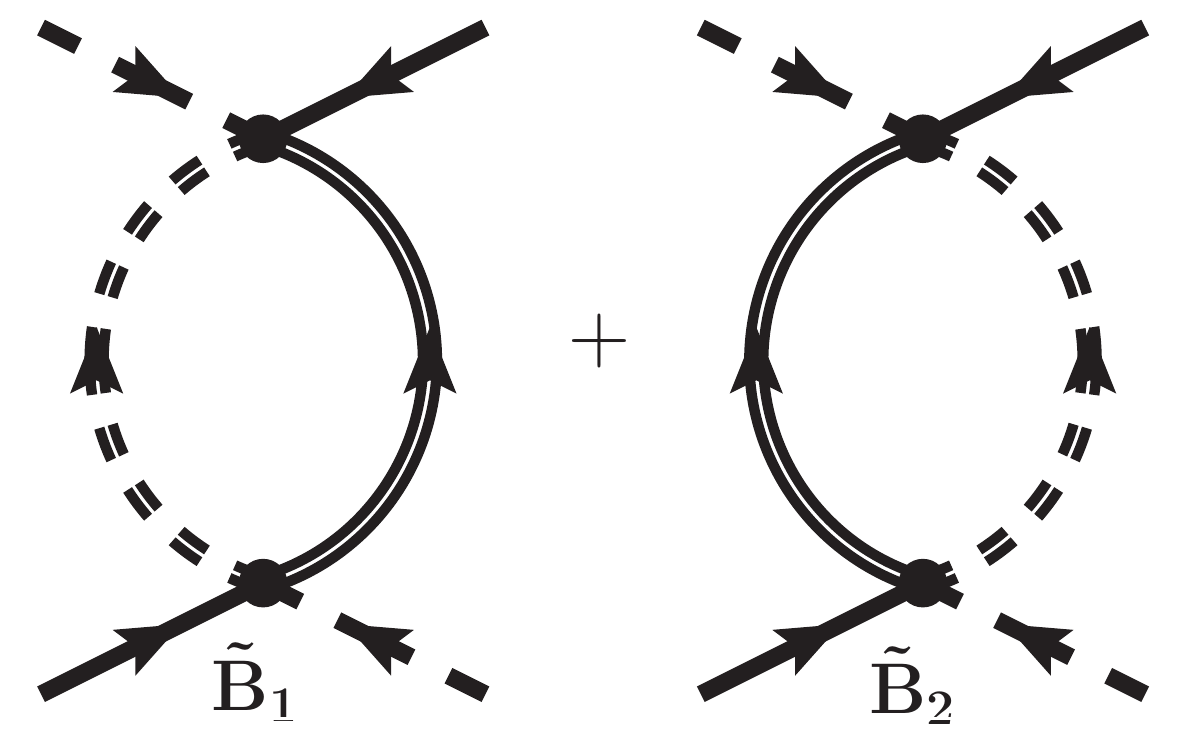}
\caption{Twisted contribution to the $G$ element at one-loop.}
\label{fig:G_1loop_twisted}
\end{figure}

We now include the contributions from the twisted hypermultiplets which give rise to two
massive bubble diagrams (\Figref{fig:G_1loop_twisted}). The color structure arises in the
$s$-channel from diagrams with twisted fields in the loop is $\csbubt_{12,34}$.
The contribution to the matrix element from these diagrams is thus
\[
G^{(1)}=i \csbubt_{12,34}\int d^3\ell \bigbrk{{\tilde B}_1+{\tilde B}_2}
\]
where the integrands for these twisted bubble diagrams are:
\<
{\tilde B}_{1} \eq -i \frac{v_2\varepsilon(ip_6 - m\varepsilon)\varepsilon v_4}{2(p_6^2+m^2)(p_5^2+m^2)}\,,\nln
{\tilde B}_{ 2} \eq  -i \frac{v_2\varepsilon(ip_5 - m\varepsilon)\varepsilon v_4}{2(p_6^2+m^2)(p_5^2+m^2)}\,.
\>
Combining these we find the same result as in  \eqref{eq:integrand_fin} but with the
opposite sign,
\[
{\tilde B}_{1}+{\tilde B}_{ 2} = -i \frac{v_2\varepsilon(ip_1 - m\varepsilon)\varepsilon v_4}{2(p_6^2+m^2)(p_5^2+m^2)}\,.
\]
Including the $t$- and $u$-channel contributions the one-loop contribution to the un\-twis\-ted-untwisted scattering matrix is
\<\label{eq:uuuu_one_loop_twint}
\Delta T^{(1)}_{1234}=& & \csbubt_{12,34}\left[-\ihalf \tprods{\bar 1}{\bar 2}\tprods{1}{2}\intbub(s)\right]\nln
&+& \csbubt_{12,43}\left[-\frac{i}{2}\, \frac{\tprods{\bar 2}{\bar 3}\tprods{\bar 1}{2}\tprods{3}{4}}{\tprods{\bar 1}{4}}\,\intbub(t)\right]\nln
&+& \csbubt_{13,24}\left[-\frac{i}{2}\, \frac{\tprods{\bar 2}{\bar 4}\tprods{ 1}{2}\tprods{\bar 3}{4}}{\tprods{\bar 3}{1}}\,\intbub(u)\right].
\>
We recall the contribution from the untwisted fields \eqref{eq:uuuu_one_loop} and note that we can rewrite the
the combination of untwisted color boxes as untwisted bubbles so that the total answer is
 \<
 \label{eq:uuuu_one_loop_all}
 T^{(1)}_{1234} =&&
 \left(\csbubu_{12,34} -\csbubt_{12,34}\right)\left[\ihalf \tprods{\bar 1}{\bar 2}\tprods{1}{2}\intbub(s)\right]
\nln
&+&\left( \csbubu_{12,43} -\csbubt_{12,43} \right)\left[\frac{i}{2} \,\frac{\tprods{\bar 2}{\bar 3}\tprods{\bar 1}{2}\tprods{3}{4}}{\tprods{\bar 1}{4}}\,\intbub(t)\right]
\nln
& +&\left( \csbubu_{13,24} -\csbubt_{13,24}  \right)\left[\frac{i}{2} \, \frac{\tprods{\bar 2}{\bar 4}\tprods{ 1}{2}\tprods{\bar 3}{4}}{\tprods{\bar 3}{1}}\,\intbub(u)\right]
.
\>

In the special case, of $\superN>4$ supersymmetry where both the twisted
and untwisted multiplets are in the same representation of the gauge group,
the color structures will be equal and so lead to a cancellation.
In this case we simply find that
\[
T^{(1)}_{1234}=0.
\]
For the other sectors there are almost identical
diagrams between untwisted-twisted hypermultiplets,
namely for the element $H_{12\twist3\twist4}$, which implies
\[
T^{(1)}_{12 \twist 3\twist 4}=0
\]
and for twisted-twisted scattering, $L_{\twist1\twist2\twist3\twist4}$, which implies
\[
T^{(1)}_{\twist 1\twist 2 \twist 3\twist 4}=0.
\]
However we will leave the more complete treatment to the substantially more efficient unitarity
methods of the next section.
Note that a vanishing one-loop contribution is obviously in agreement
with the $\superN>4$ constraints on scattering amplitudes
discussed in \secref{sec:higher_susy}.

\section{Scattering Unitarity}
\label{sec:unitarity}

The scattering matrix $\smat=1+i\scat$ in a reasonable quantum field theory
is expected to be unitary, $\smat^\dagger\smat=1$. For the
scattering amplitudes $\scat$ it implies the unitarity condition
\[\label{eq:uni}
-i(\scat-\scat^\dagger)=\scat^\dagger \scat.
\]
In this section we would like to compare the one-loop
field theory results of the previous section
with scattering unitarity. In particular, we want to
see whether the field theory results stand a chance of
being cut constructible.

\subsection{Adjoint and Multiplication}

In order to confirm unitarity for the
amplitudes derived above,
we should first understand how to
take the adjoint and how to multiply
two-to-two scattering amplitudes $\scat[T]$,
where $T$ is the overall factor as
defined in \eqref{eq:SMatEl}.

The adjoint scattering amplitude has the same structure
as the original amplitude but with different matrix elements
\[
\begin{array}[b]{rclcrcl}
A^\dagger_{12\bar3\bar4}\eq (A_{43\bar2\bar1})^\ast,
&&
D^\dagger_{12\bar3\bar4}\eq (D_{43\bar2\bar1})^\ast,
\\[3pt]
B^\dagger_{12\bar3\bar4}\eq (B_{43\bar2\bar1})^\ast,
&&
E^\dagger_{12\bar3\bar4}\eq (E_{43\bar2\bar1})^\ast,
\\[3pt]
C^\dagger_{12\bar3\bar4}\eq (F_{43\bar2\bar1})^\ast,
&&
F^\dagger_{12\bar3\bar4}\eq (C_{43\bar2\bar1})^\ast,
\\[3pt]
G^\dagger_{12\bar3\bar4}\eq (L_{43\bar2\bar1})^\ast,
&&
L^\dagger_{12\bar3\bar4}\eq (G_{43\bar2\bar1})^\ast,
\\[3pt]
H^\dagger_{12\bar3\bar4}\eq (H_{43\bar2\bar1})^\ast,
&&
K^\dagger_{12\bar3\bar4}\eq (K_{43\bar2\bar1})^\ast.
\end{array}
\]
Here the spinors $\bar1,\bar2,\bar3,\bar4$
are conjugate to $1,2,3,4$, respectively, according to
\eqref{eq:conjspin}
\[
u_{\bar k}=+v_{k},\qquad
v_{\bar k}=-u_{k}.
\]
For unitary representations \eqref{eq:RepUni},
or more generally by replacing
$u_\alpha^\ast\to v_\alpha$,
$v_\alpha^\ast\to u_\alpha$,
the adjoint matrix elements take the same form
as the original matrix elements \eqref{eq:SMatEl}.
We can thus write the adjoint scattering amplitude
as a regular scattering amplitude
\[\label{eq:adjointscat}
\scat[T]^\dagger=\scat[T^\dagger],
\]
but instead of $T$ with the prefactor $T^\dagger$ defined by
\[\label{eq:adjointpre}
T_{12\bar3\bar4}^\dagger=(T_{43\bar2\bar1})^\ast.
\]
This is because for unitary representations
the adjoint scattering matrix obeys the same
symmetries as the original one.

Iterative two-to-two particle scattering
also satisfies the transformation laws of
overall two-to-two scattering,
hence
\[\label{eq:iteratescat}
\scat[T']\,\scat[T'']=\scat[T].
\]
The following relations between the matrix elements
ensure that the product takes the expected form
\<
\frac{T''_{12\bar5\bar6}T'_{65\bar3\bar4}}{T_{12\bar3\bar4}}
\eq
\frac{A''_{12\bar5\bar6}A'_{65\bar3\bar4}}{A_{12\bar3\bar4}}
=
\frac{D''_{12\bar5\bar6}D'_{65\bar3\bar4}}{D_{12\bar3\bar4}}
\nln\eq
\frac{B''_{12\bar5\bar6}B'_{65\bar3\bar4}+C''_{12\bar5\bar6}F'_{65\bar3\bar4}}{B_{12\bar3\bar4}}
=
\frac{E''_{12\bar5\bar6}E'_{65\bar3\bar4}+F''_{12\bar5\bar6}C'_{65\bar3\bar4}}{E_{12\bar3\bar4}}
\nln\eq
\frac{B''_{12\bar5\bar6}C'_{65\bar3\bar4}+C''_{12\bar5\bar6}E'_{65\bar3\bar4}}{C_{12\bar3\bar4}}
=
\frac{E''_{12\bar5\bar6}F'_{65\bar3\bar4}+F''_{12\bar5\bar6}B'_{65\bar3\bar4}}{F_{12\bar3\bar4}}
\nln\eq
\frac{G''_{12\bar5\bar6}K'_{65\bar3\bar4}+H''_{12\bar5\bar6}G'_{65\bar3\bar4}}{G_{12\bar3\bar4}}
=
\frac{K''_{12\bar5\bar6}K'_{65\bar3\bar4}+H''_{12\bar5\bar6}G'_{65\bar3\bar4}}{K_{12\bar3\bar4}}
\nln\eq
\frac{G''_{12\bar5\bar6}L'_{65\bar3\bar4}+H''_{12\bar5\bar6}H'_{65\bar3\bar4}}{H_{12\bar3\bar4}}
=
\frac{K''_{12\bar5\bar6}L'_{65\bar3\bar4}+L''_{12\bar5\bar6}H'_{65\bar3\bar4}}{L_{12\bar3\bar4}}\,.
\>
The prefactor of the product is thus simply given by
\[\label{eq:iteratepre}
T_{12\bar3\bar4}=2\pi^2\int d^3p\, \delta(p_5^2+m^2)\,\delta(p_6^2+m^2)\,
T''_{12\bar5\bar6}\,T'_{65\bar3\bar4}.
\]

Similar relations hold for scattering matrices
involving twisted hypermultiplets
introduced in \secref{sec:mixmat}.
We will not present these in detail here, but merely
apply them where needed.

\subsection{Unitarity Relations}

\begin{figure}\centering
\includegraphics{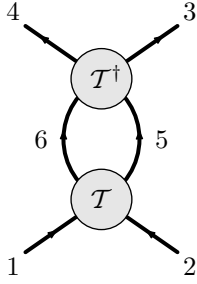}
\caption{Particle setup for the unitarity relations.}
\label{fig:UniSetup}
\end{figure}

Now we are in a position to consider unitarity for
two-to-two scattering amplitudes from field theory.
We neglect intermediate states with more than two particles.
This approximation is exact at the two-loop level because
physical particles can only be created or annihilated in pairs.
According to the above considerations,
unitarity leads to the following relation for the prefactor,
see also \figref{fig:UniSetup},
\[\label{eq:uni2to2}
-iT_{12\bar3\bar4}+i(T_{43\bar2\bar1})^\ast=
2\pi^2\int d^3p\, \delta(p_5^2+m^2)\,\delta(p_6^2+m^2)\,
T_{12\bar5\bar6}\,(T_{43\bar5\bar6})^\ast
+\order{g^4}.
\]
It is convenient to go to the center of mass frame and thus make a choice for the momenta
of the particles
\[\begin{array}[b]{rclcrcl}
p_1\eq (E,+p,0),&&
p_2\eq (E,-p,0),
\\[0.5ex]
p_3\eq (E,-p\cos\alpha,-p\sin\alpha),&&
p_4\eq (E,+p\cos\alpha,+p\sin\alpha),
\\[0.5ex]
p_5\eq (E,-p\cos\beta,-p\sin\beta),&&
p_6\eq (E,+p\cos\beta,+p\sin\beta).
\end{array}
\]
In this frame the integral over the delta
functions can be evaluated
\[
2\pi^2\int d^3p\, \delta(p_5^2+m^2)\,\delta(p_6^2+m^2)\,F_{123456}
=
\int \frac{\pi^2d\beta}{4E}\,F_{123456}.
\]

Substituting the loop expansion of the prefactor
\[
T=g T^{(0)}+g^2 T^{(1)}+g^3 T^{(2)}+\ldots,\qquad
g=\frac{4\pi}{k}\,,
\]
we obtain the unitarity relations up to two loops
\<\label{eq:uniloop}
-iT^{(0)}_{12\bar3\bar4}+i\bigbrk{T^{(0)}_{43\bar2\bar1}}^\ast\eq 0,
\nln
-iT^{(1)}_{12\bar3\bar4}+i\bigbrk{T^{(1)}_{43\bar2\bar1}}^\ast\eq
\int \frac{\pi^2d\beta}{4E}\,
T^{(0)}_{12\bar5\bar6}\,\bigbrk{T^{(0)}_{43\bar5\bar6}}^\ast
=\int \frac{\pi^2d\beta}{4E}\,
T^{(0)}_{12\bar5\bar6}\,T^{(0)}_{\bar6\bar534},
\nln
-iT^{(2)}_{12\bar3\bar4}+i\bigbrk{T^{(2)}_{43\bar2\bar1}}^\ast\eq
\int \frac{\pi^2d\beta}{4E}\,
\Bigbrk{T^{(0)}_{12\bar5\bar6}\,\bigbrk{T^{(1)}_{43\bar5\bar6}}^\ast
+T^{(1)}_{12\bar5\bar6}\,\bigbrk{T^{(0)}_{43\bar5\bar6}}^\ast}.
\>

The above relations hold for a $\superN=4$ supersymmetric model with
only one type of hypermultiplet.
If there are both types of hypermultiplets present,
the relation \eqref{eq:uni2to2} has
to be extended and supplemented by
\<\label{eq:uni2to2mix}
-i T_{12\bar3\bar4}
+i \bigbrk{T_{43\bar2\bar1}}^\ast\eq
\int \frac{\pi^2d\beta}{4E}
\Bigbrk{
T_{12\bar5\bar6}\,\bigbrk{T_{43\bar5\bar6}}^\ast
+T_{12\twist{\bar5}\twist{\bar6}}\,\bigbrk{T_{43\twist{\bar5}\twist{\bar6}}}^\ast
}
+\order{g^4},
\nln
-i T_{12\twist{\bar3}\twist{\bar4}}
+i \bigbrk{T_{\twist{4}\twist{3}\bar2\bar1}}^\ast\eq
\int \frac{\pi^2d\beta}{4E}
\Bigbrk{
T_{12\bar5\bar6}\,\bigbrk{T_{\twist{4}\twist{3}\bar5\bar6}}^\ast
+T_{12\twist{\bar5}\twist{\bar6}}\,\bigbrk{T_{\twist{4}\twist{3}\twist{\bar5}\twist{\bar6}}}^\ast
}
+\order{g^4},
\nln
-i T_{1\twist{2}\twist{\bar3}\bar4}
+i \bigbrk{T_{4\twist{3}\twist{\bar2}\bar1}}^\ast\eq
\int \frac{\pi^2d\beta}{4E}
\Bigbrk{
T_{1\twist{2}\twist{\bar5}\bar6}\,\bigbrk{T_{4\twist{3}\twist{\bar5}\bar6}}^\ast
+T_{1\twist{2}\bar5\twist{\bar6}}\,\bigbrk{T_{4\twist{3}\bar5\twist{\bar6}}}^\ast
}
+\order{g^4}.
\>
Relations among the other prefactors
can be obtained by applying discrete symmetries.
The loop expansion for all of these
is analogous to \eqref{eq:uniloop}.
As discussed in \secref{sec:higher_susy} there are
further constraints for amplitudes in
models with $\superN>4$ supersymmetry.
The above relations obey these constraints.

\subsection{Tree Level}

One can assign untwisted and twisted hypermultiplets
to the legs of two-to-two scattering amplitudes in 8 different ways.
As discussed in \secref{sec:mixmat} they all
have an equivalent matrix structure,
but the prefactors are different in general.

The tree-level prefactors with alike hypermultiplets in the in/out channels
have the following color structures, cf.\ \secref{sec:StructTree},
\[\begin{array}[b]{rclcrcl}
T^{(0)}_{12\bar3\bar4}\eq
\cstree_{14,23} t^{(0)}_{12\bar3\bar4}
+\cstree_{13,24} t^{(0)}_{12\bar4\bar3},
&&
T^{(0)}_{12\twist{\bar 3}\twist{\bar 4}}\eq
\cstree_{12,\twist3\twist4} t^{(0)}_{12\twist{\bar 3}\twist{\bar 4}},
\\[1ex]
T^{(0)}_{\twist{1}\twist{2}\bar 3\bar 4}\eq
\cstree_{\twist1\twist2,34} t^{(0)}_{\twist{1}\twist{2}\bar 3\bar 4},
 &&
T^{(0)}_{\twist{1}\twist{2}\twist{\bar 3}\twist{\bar 4}}\eq
\cstree_{\twist1\twist4,\twist2\twist3} t^{(0)}_{\twist{1}\twist{2}\twist{\bar 3}\twist{\bar 4}}
+\cstree_{\twist1\twist3,\twist2\twist4} t^{(0)}_{\twist{1}\twist{2}\twist{\bar 4}\twist{\bar 3}}
.
\end{array}\]
The coefficient functions $t$ have been evaluated
in field theory in \secref{sec:QFTTree,sec:QFTTreeMix}
\[\begin{array}[b]{rclcrcl}
t^{(0)}_{12\bar3\bar4}\eq\displaystyle
-i\frac{\tprods{1}{2}\tprods{\bar 2}{4}}{\tprods{1}{4}}\,,
&&
t^{(0)}_{12\twist{\bar 3}\twist{\bar 4}}\eq
-i\tprods{\bar 3}{\bar 4},
\\
t^{(0)}_{\twist{1}\twist{2}\bar 3\bar 4}\eq
-i\tprods{1}{2},
 &&
t^{(0)}_{\twist{1}\twist{2}\twist{\bar 3}\twist{\bar 4}}\eq\displaystyle
-i\frac{\tprods{2}{\bar 4}\tprods{\bar 3}{\bar 4}}{\tprods{\bar 4}{\bar 1}}\,.
\end{array}\]

For mixed hypermultiplets in the in/out channels the color structure
of the prefactor reads
\[\begin{array}[b]{rclcrcl}
T^{(0)}_{1\twist{2}\twist{\bar 3}\bar 4}
\eq \cstree_{14,\twist{2}\twist 3} t^{(0)}_{1\twist{2}\twist{\bar 3}\bar 4},
&&
T^{(0)}_{1\twist{2}\bar 3\twist{\bar 4}}
\eq \cstree_{13,\twist2\twist4} t^{(0)}_{1\twist{2}\twist{\bar 4}\bar 3},
\\[1ex]
T^{(0)}_{\twist{1}2\twist{\bar 3}\bar 4}
\eq \cstree_{24,\twist1\twist3} t^{(0)}_{2\twist{1}\twist{\bar 3}\bar 4},
 &&
T^{(0)}_{\twist{1}2\bar 3\twist{\bar 4}}
\eq \cstree_{24,\twist1\twist3} t^{(0)}_{2\twist{1}\twist{\bar 4}\bar 3},
\end{array}\]
with the single coefficient function
\[\label{eq:unitreemix}
t^{(0)}_{1\twist{2}\twist{\bar 3}\bar 4}=
i\frac{\tprods{\bar3}{\bar4}\tprods{\bar3}{4}}{\tprods{\bar2}{\bar3}}\,.
\]

Using the spinor identities \eqref{eq:twistoridentity},
it is straightforward to confirm the tree-level unitarity conditions
\<
\bigbrk{t^{(0)}_{43\bar2\bar1}}^\ast\eq
i\,\frac{\tprods{\bar4}{\bar3}\tprods{3}{\bar1}}{\tprods{\bar4}{\bar1}}
=t^{(0)}_{12\bar3\bar4},
\nln
\bigbrk{t^{(0)}_{43\twist{\bar 2}\twist{\bar 1}}}^\ast\eq
i\tprods{2}{1}
=t^{(0)}_{\twist{1}\twist{2}\bar 3\bar 4},
\nln
\bigbrk{t^{(0)}_{\twist{4}\twist{3}\twist{\bar 2}\twist{\bar 1}}}^\ast\eq
i\,
\frac{\tprods{\bar 3}{1}\tprods{2}{1}}{\tprods{1}{4}}=
t^{(0)}_{\twist{1}\twist{2}\twist{\bar 3}\twist{\bar 4}},
\nln
\bigbrk{t^{(0)}_{4\twist{3}\twist{\bar 2}\bar 1}}^\ast\eq
-i\frac{\tprods{2}{1}\tprods{2}{\bar 1}}{\tprods{3}{2}}
=t^{(0)}_{1\twist{2}\twist{\bar 3}\bar 4}.
\>

\subsection{Pure Matter at One Loop}
\label{sec:UniUntwist}

Next we consider one-loop unitarity
for a model with only one type of hypermultiplet.
First the color structure of the integrand in \eqref{eq:uniloop} is investigated
\[
T^{(0)}_{12\bar5\bar6}T^{(0)}_{65\bar3\bar4}
=
2\csbox_{14,23} t^{(0)}_{12\bar5\bar6} t^{(0)}_{65\bar3\bar4}
+2\csbox_{13,24} t^{(0)}_{12\bar5\bar6} t^{(0)}_{65\bar4\bar3}
\]
We have used the crossing property
$t^{(0)}_{12\bar3\bar4}=t^{(0)}_{21\bar4\bar3}$
and identified the color structure as a box
$\cstree_{16,25} \cstree_{64,53}=\csbox_{14,23}$,
cf.\ \eqref{eq:structtreemult}.
For convenience we have indicated the unitarity cuts
in \figref{fig:structlooppure} on page \pageref{fig:structlooppure}.
We evaluate and simplify the product of coefficient functions
\[\label{eq:uniintone}
\frac{\pi^2}{4E}\,t^{(0)}_{12\bar5\bar6}t_{65\bar3\bar4}^{(0)}
=-i\pi^2
\lrbrk{\frac{ie^{i\beta}}{e^{i\beta}-1}-\frac{ie^{i\beta}}{e^{i\beta}-e^{i\alpha}}}
t^{(0)}_{12\bar3\bar4}
-\frac{\pi^2p^2}{E}\,.
\]
Note that the integrand has single poles at the angles $\beta=0$ and $\beta=\alpha$.
At these points the momenta of the intermediate particles $6,5$
agree precisely with the ones of the ingoing particles $1,2$ ($\beta=0$)
or outgoing particles $4,3$ ($\beta=\alpha$). They originate from
a gluon exchange with zero momentum. Fortunately the poles
have exactly opposite residues and thus we can ignore their contribution
altogether. The unitarity condition leads to
an imaginary part which is independent of the overall scattering angle $\alpha$
\<\label{eq:uniloopres}
-i T^{(1)}_{12\bar3\bar4}
+i \bigbrk{T^{(1)}_{43\bar2\bar1}}^\ast\eq
\int \frac{\pi^2d\beta}{4E}\,T^{(0)}_{12\bar5\bar6}T_{65\bar3\bar4}^{(0)}
=-\frac{4\pi^3 p^2}{E}
\lrbrk{\csbox_{14,23}+\csbox_{13,24}}
\nln
\eq
-\frac{2\pi^3 p^2}{E}\,
\csbubu_{12,34}
=
-\frac{\pi^3 \tprods{1}{2}\tprods{\bar 1}{\bar 2}}{\sqrt{\tprods{1}{\bar2}\tprods{\bar1}{2}}}\,
\csbubu_{12,34}
.
\>
The conversion between different color structures
is due to the identity \eqref{eq:structbubbleexpand}
and the final transformation
$\sqrt{\tprods{1}{\bar2}\tprods{\bar1}{2}}=2E$
and $\tprods{1}{2}\tprods{\bar 1}{\bar 2}=4p^2$ makes the
result independent of a specific frame.
This expression is in agreement with
the field theory calculation \eqref{eq:uuuu_one_loop} in \secref{sec:QFTLoop}:
The expression \eqref{eq:uuuu_one_loop}
obeys $T^{(1)}_{12\bar3\bar4}=\bigbrk{T^{(1)}_{43\bar2\bar1}}^\ast$
except for branch cut discontinuities in the loop integrals.
The integral $\intbub(s)$ in \eqref{eq:massbub} has a branch cut with discontinuity
($\sqrt{-s}=2E$)
\[\label{eq:bubblecut}
\intbub(s-i\epsilon)-\intbub(s+i\epsilon)
=-\frac{2\pi^3}{\sqrt{-s}}\,\theta(\sqrt{-s}-2m)
=-\frac{\pi^3}{E}\,\theta(E-m).
\]
The bubble integrals $\intbub(t),\intbub(u)$
in the $t$- and $u$-channels clearly have no cuts
in the physical region. For the one-loop expression \eqref{eq:uuuu_one_loop}
\[
T^{(1)}_{12\bar3\bar4}=
\sfrac{i}{2}\tprods{1}{2}\tprods{\bar 1}{\bar 2}\,\intbub(s-i\epsilon)\,\csbubu_{12,34}
+\ldots
=2ip^2\,\intbub(s-i\epsilon)\,\csbubu_{12,34}
+\ldots
\]
we thus get full agreement with unitarity \eqref{eq:uniloopres}.
We can in fact do even better and compare the results before integrating over the
loop momenta phase space. Taking the
integrand of the Feynman diagram calculation (\Secref{sec:QFTLoop}) and putting all momenta on-shell,
one finds agreement  with \eqref{eq:uniintone}.
It can be seen that the pole terms indeed come from the
`triangle' diagrams which involve gluon exchange whereas the finite piece gets contributions
from both the `triangle' and `bubble' diagrams.

It is a curious fact that the full one-loop amplitude $T^{(1)}_{12\bar3\bar4}$
from field theory is a linear combination of massive scalar bubbles $\intbub$
without further rational parts.
The coefficients of the bubbles can be reconstructed
from unitarity in all channels.
Expanding cuts using the inverse of \eqref{eq:bubblecut},
i.e.\ the minimal replacement,
therefore yields the full one-loop amplitude from unitarity.
It gives a hint that amplitudes in $\superN=4$ Chern--Simons theories
may be cut constructible.

Finally we would like to consider the total cross section
of the scattering process of two hypermultiplets.
For that purpose we shall set $\alpha=0$ so that the in and out states
are the same. The cross section is proportional to
\[\label{eq:onecrosstotal}
2\Im T^{(1)}_{12\bar2\bar1}=
\int \frac{\pi^2d\beta}{4E}\,\bigabs{T^{(0)}_{12\bar5\bar6}}^2
=-\frac{2\pi^3 p^2}{E}\,\csbubu_{12,34}.
\]
Curiously it appears that it is negative although the integrand itself
is manifestly positive. Let us thus have a closer look at the integrand
\[\label{eq:onecrossdiff}
\frac{\pi^2}{4E}\,\bigabs{t^{(0)}_{12\bar5\bar6}}^2
=\frac{\pi^2}{4E}\,t^{(0)}_{12\bar5\bar6}t_{65\bar3\bar4}^{(0)}
=\frac{\pi^2 E}{\sin^2(\half\beta)}
-\frac{\pi^2p^2}{E}
=\pi^2 E\,\cot^2(\half\beta)
+\frac{\pi^2m^2}{E}\,.
\]
In the last form it is manifestly positive.
Due to a double pole the integral is infinite
and needs to be regularized.
A principal value prescription
(or any other contour in the complex plane)
will show that the $1/\sin^2$ term does not contribute.
The finite remainder is however negative.

Essentially we have dropped a contribution from
forward scattering where a gluon with zero momentum and zero energy
is exchanged. Thus the peculiarity can be associated to an infra-red divergence.
It is in fact very similar to the collinear divergences
encountered in Yang--Mills theories, but it is milder:
It can only appear for gluons with zero momentum whereas
for Yang--Mills it appears for all light-like gluons.
The effects are nevertheless similar.
The reason why the IR singularity
cannot directly be seen in the result
is related to the fact that in odd spacetime dimensions
there are no divergences at one loop.

\subsection{Mixed Matter at One Loop}
\label{sec:UniMix}

We now consider one-loop scattering unitarity in
a theory with both types of hypermultiplets
using the relations \eqref{eq:uni2to2mix}.

The integrands of the first integral in \eqref{eq:uni2to2mix}
have the color structures
\[
T^{(0)}_{12\bar5\bar6}T^{(0)}_{65\bar3\bar4}
=
2\csbox_{14,23} t^{(0)}_{12\bar5\bar6} t^{(0)}_{65\bar3\bar4}
+2\csbox_{13,24} t^{(0)}_{12\bar5\bar6} t^{(0)}_{65\bar4\bar3},
\qquad
T^{(0)}_{12\twist{\bar5}\twist{\bar6}}T_{\twist{6}\twist{5}\bar 3\bar 4}^{(0)}
=
\csbubt_{12,34}
t^{(0)}_{12\twist{\bar5}\twist{\bar6}}t_{\twist{6}\twist{5}\bar 3\bar 4}^{(0)},
\]
where we have used the composition of trees to loop in \eqref{eq:structtreemult},
see also \figref{fig:structlooppure,fig:structlooptwisted}.
The remaining coefficients evaluate to
\[\label{eq:uniint1}
\frac{\pi^2}{4E}\,t^{(0)}_{12\bar5\bar6}t_{65\bar3\bar4}^{(0)}
=
-i\pi^2
\lrbrk{\frac{ie^{i\beta}}{e^{i\beta}-1}-\frac{ie^{i\beta}}{e^{i\beta}-e^{i\alpha}}}
t^{(0)}_{12\bar3\bar4}
-\frac{\pi^2p^2}{E}\,,
\qquad
\frac{\pi^2}{4E}\,t^{(0)}_{12\twist{\bar5}\twist{\bar6}}t_{\twist{6}\twist{5}\bar 3\bar 4}^{(0)}
=\frac{\pi^2p^2}{E}\,.
\]
In the integral over $\beta$ the residues cancel
and only the constant pieces remain
\[\label{eq:mixedunires}
-i T^{(1)}_{12\bar3\bar4}
+i \bigbrk{T^{(1)}_{43\bar2\bar1}}^\ast=
-\frac{2\pi^3 p^2}{E}
\lrbrk{\csbubu_{12,34}-\csbubt_{12,34}}.
\]
Again this result agrees with the field theory computation
\eqref{eq:uuuu_one_loop_all}.
Furthermore the field theory result is again a linear combination
of massive scalar bubbles $\intbub$ hinting at cut constructibility.

The two integrands of the second integral in \eqref{eq:uni2to2mix}
have the following color structures, cf.\ \figref{fig:structmixed}
\[
T^{(0)}_{12\bar5\bar6}T_{65\twist{\bar 3}\twist{\bar 4}}^{(0)}
=
-\csbubu_{12,\twist3\twist4}
t^{(0)}_{12\bar5\bar6}t_{65\twist{\bar 3}\twist{\bar 4}}^{(0)}
,
\qquad
T^{(0)}_{12\twist{\bar5}\twist{\bar6}}T_{\twist{6}\twist{5}\twist{\bar3}\twist{\bar4}}^{(0)}
=
-\csbubt_{12,\twist3\twist4}
t^{(0)}_{12\twist{\bar5}\twist{\bar6}}t_{\twist{6}\twist{5}\twist{\bar3}\twist{\bar4}}^{(0)}.
\]
The coefficient functions evaluate to
\<\label{eq:uniint2}
\frac{\pi^2}{4E}\,t^{(0)}_{12\bar5\bar6}t_{65\twist{\bar 3}\twist{\bar 4}}^{(0)}
\eq
-i\pi^2
\lrbrk{
\frac{ie^{i\beta}}{e^{i\beta}-1}
-\frac{i(E+m)}{2E}
}
t^{(0)}_{12\twist{\bar 3}\twist{\bar 4}}
\,,
\nln
\frac{\pi^2}{4E}\,t^{(0)}_{12\twist{\bar5}\twist{\bar6}}t_{\twist{6}\twist{5}\twist{\bar3}\twist{\bar4}}^{(0)}
\eq
+i\pi^2\lrbrk{
\frac{ie^{i\beta}}{e^{i\beta}-e^{i\alpha}}\,
-\frac{i(E+m)}{2E}
}
t^{(0)}_{12\twist{\bar 3}\twist{\bar 4}}
.
\>
Note that again there are two poles with residues proportional
to the tree-level amplitude. Here the poles originate from
the two different terms in the integrand.
It is not entirely clear how to perform the integral
over the poles. For practical purposes,
let us assume a principal value prescription.
The unitarity integral then evaluates to
\[
-i T^{(1)}_{12\twist{\bar3}\twist{\bar4}}
+i \bigbrk{T^{(1)}_{\twist{4}\twist{3}\bar2\bar1}}^\ast
=
-\frac{i\pi^2(E+m)\tprods{\bar 3}{\bar 4}}{2E}
\lrbrk{\csbubu_{12,\twist3\twist4}-\csbubt_{12,\twist3\twist4}},
\]

In the third integrand of \eqref{eq:uni2to2mix} we find a single color structure
\[
T^{(0)}_{1\twist{2}\twist{\bar5}\bar6}T_{6\twist{5}\twist{\bar3}\bar4}^{(0)}
=
T_{1\twist{2}\bar5\twist{\bar6}}T_{\twist{6}5\twist{\bar 3}\bar 4}
=
\csbox_{14,\twist2\twist3}
t^{(0)}_{1\twist{2}\twist{\bar5}\bar6}t_{6\twist{5}\twist{\bar3}\bar4}^{(0)}.
\]
The coefficient function yields
\[\label{eq:uniint3}
\frac{\pi^2}{4E}\,t^{(0)}_{1\twist{2}\twist{\bar5}\bar6}t_{6\twist{5}\twist{\bar3}\bar4}^{(0)}
=
i\pi^2
\lrbrk{\frac{ie^{i\beta}}{e^{i\beta}-1}-\frac{ie^{i\beta}}{e^{i\beta}-e^{i\alpha}}}
t^{(0)}_{1\twist{2}\twist{\bar3}\bar4}\,.
\]
Here both poles are present and there is no constant piece. The integral
thus vanishes exactly
\[
-i T^{(1)}_{12\twist{\bar3}\twist{\bar4}}
+i \bigbrk{T^{(1)}_{\twist{4}\twist{3}\bar2\bar1}}^\ast
= 0,
\]
%

Finally we would like to mention the curious fact that
all three integrals vanish for model with $\superN=5,6,8$ extended supersymmetry
where untwisted and twisted fields are equivalent $\csbubu=\csbubt$
\[\label{eq:uniloopzero}
-i T^{(1)}_{12\bar3\bar4}
+i \bigbrk{T^{(1)}_{43\bar2\bar1}}^\ast
=-i T^{(1)}_{12\twist{\bar3}\twist{\bar4}}
+i \bigbrk{T^{(1)}_{\twist{4}\twist{3}\bar2\bar1}}^\ast
=-i T^{(1)}_{1\twist{2}\twist{\bar3}\bar4}
+i \bigbrk{T^{(1)}_{4\twist{3}\twist{\bar2}\bar1}}^\ast
=
0.
\]
It implies a remarkable feature that the scattering amplitudes
at one loop are free from unitarity cuts.
Moreover the field theory calculations in \secref{sec:QFTLoop,sec:QFTLoopMix}
suggest that the one-loop amplitudes vanish altogether
$T^{(1)}=0$. We shall discuss the further implications below.

\subsection{Two-Loop Puzzle}

The result of vanishing one-loop unitarity cuts \eqref{eq:uniloopzero}
in $\superN=5,6,8$ supersymmetric models leads to a puzzle.
The point is that the one-loop amplitudes must be rational functions
of the momenta which in (highly) supersymmetric theories
often implies that the amplitudes $T^{(1)}$ vanish altogether.
Our field theory computations in \secref{sec:QFTLoopMix}
confirm this result for our model.
In this case, however, unitarity \eqref{eq:uniloop} implies
that the two-loop amplitudes are merely rational functions.
Blindly following the argument leads to no loop corrections at all
which is hard to believe.

There are good reasons to believe that the two-loop
amplitudes from field theory are neither zero nor merely rational functions
(cf.\ the discussion in the conclusions).
This also leads to a much more realistic pattern of non-trivial corrections
at higher loop orders. However, how does this match with our observation
of vanishing one-loop contributions?
The point is perhaps that our model does suffer from IR divergences in spite
of having only massive physical particles.
The zero mode of the Chern--Simons gauge field appears to cause the IR problems
and in the above discussions we have seen several instances of such singularities.
For example there are obvious singularities at coincident momenta 
in the tree level scattering amplitudes, \eqref{eq:uuuu_tree},
due to the singular behavior of the gluon propagator 
and related subtleties in the total cross section \eqref{eq:onecrosstotal,eq:onecrossdiff}.
The singularities effectively require to regularize the model before
computing quantum corrections.

The most reliable regulator arguably is
dimensional regularization/reduction where loop integrals are performed
in a spacetime of dimension $D=3-2\epsilon$.
Our results then imply merely that
$T^{(1)}=0+\order{\epsilon}$.
The integrand of the two-loop unitarity relation
must be suppressed likewise $T^{(0)}T^{(1)}=0+\order{\epsilon}$.
However, the integral can very well produce $1/\epsilon$ divergent terms
such that the two-loop unitarity integral is finite
$\int T^{(0)}T^{(1)}=\order{\epsilon^0}$
(or even divergent).

It would be very desirable to perform a two-loop computation
in dimensional regularization
based on both field theory and unitarity,
and consequently compare the two results.

\section{Conclusions}

In this paper we have considered the spacetime S-matrices of
various supersymmetric Chern--Simons matter theories focussing on the mass
deformed $\superN\geq 4$ theories whose super-Poincar\'e group contains the
supergroup $\grp{PSU}(2|2)$.
We have presented
the tree-level and one-loop four particle amplitudes derived using both symmetry arguments
and explicit perturbative calculations.
This extended $\grp{PSU}(2|2)$ symmetry group is almost the same as that which
occurs in the light-cone gauge fixed worldsheet theory of strings
in $AdS_5\times S^5$ or, equivalently, as the group of symmetries preserved
by the ferromagnetic vacuum in the spin chain picture of
maximally supersymmetric four dimensional Yang--Mills.
As in that context, the superalgebra greatly constrains the two-to-two
S-matrix as it interrelates all elements and determines the entire matrix structure
up to an overall factor.
This leads to the intriguing observation that the two-body spacetime S-matrix of
these three-dimensional Chern--Simons matter theories is the same as the two-dimensional
spin-chain/worldsheet S-matrix that plays such a central role in the AdS/CFT correspondence.
Furthermore this S-matrix is known to be equivalent to Shastry's R-matrix for the Hubbard model \cite{Beisert:2006qh}.
In many respects this similarity is purely formal and the kinematics are obviously quite different.
For example, due to the different kinematical structure
the Chern--Simons spacetime S-matrix, unlike the two-dimensional integrable
S-matrices, does not satisfy the Yang--Baxter equation.
Nonetheless it is certainly tempting to ask how far this analogy may be extended and
whether there are structures in common with the integrable systems
if only for certain kinematical regimes.

That all four particle scattering amplitudes should be related is perhaps not surprising
for a theory with extended supersymmetry, indeed in four-dimensional
$\superN=4$ Yang--Mills super-Ward identities imply similar relations.
For the three-dimensional Chern--Simons theories with additional twisted matter
there are of course additional unrelated amplitudes though in the cases
where the supersymmetry is extended to $\superN=5,6$ or $8$
there are further relations between amplitudes.
For the general $\superN\geq 4$ case we have fixed,
by explicit calculation, the tree level contribution to
the overall factor undetermined by the global symmetries.
We have discussed in detail the color structures that occur in the perturbative calculations
as, especially beyond tree level, these calculations
are substantially simplified by separately treating the
color and kinematical contributions.
Having the color structure in hand one can focus on the color ordered amplitudes,
which in turn can be calculated efficiently using unitarity methods (whose validity we
explicitly verified at one-loop).

For generic $\superN=4$ theories we find a one-loop contribution to the overall prefactor
corresponding  to a massive bubble diagram and, interestingly, when we include
additional twisted matter we find an identical contribution
but with the opposite sign. Thus when the twisted and untwisted matter are in the same
gauge group representations, such as in the $\superN=5,6,8$ theories,
we find that all the one-loop amplitudes vanish.
It is not immediately clear what is the physical origin of this cancellation however
it is plausible to assume that in these cases there is an additional discrete symmetry
related to the exchange of twisted and untwisted matter that
explains this seeming coincidence; 
if this is so it is not unreasonable to ask whether 
this continues at higher odd orders in the perturbative expansion.
However before going to three loops there remains the question of
how to find non-trivial two-loop scattering;
naive application of the unitarity method implies the vanishing
of the two-loop amplitudes as a consequence of the vanishing of all one-loop amplitudes.
In order to solve this puzzle it would be worthwhile to carry out an explicit two-loop calculation
either using unitarity methods, but being careful to
keep higher orders in the dimensional regularization parameter $\epsilon$, or using off-shell methods.

Taking the mass deformation to zero appears to be smooth limit for all the physical
quantities we have calculated; in particular the tree level S-matrix elements written in
spinor notation
are essentially unchanged and the ansatz for the S-matrix is thus still valid
though there may additional relations between to the elements due to extra hidden symmetries.
Of course one must now deal with the additional IR-divergences that arise at one-loop
due to the massless matter fields
but for the $\superN=5,6,8$ cases the one-loop result is still vanishing.
In this limit the theories we consider are in one-to-one correspondence with $\superN\geq 4$
superconformal Chern--Simons theories  which connects our
results to the AdS$_4$/CFT$_3$ correspondence.
In this context it is obvious to ask whether one can find, for the planar $\superN=6$ theory, a similar
relationship between scattering amplitudes and Wilson loops
as was found in the $\superN=4$ SYM/$AdS_5\times S^5$ case.
On general grounds \cite{Sen:1981sd,Korchemsky:1985xj,Magnea:1990zb,Korchemsky:1993hr,
Korchemskaya:1994qp,Sterman:2002qn,Bern:2005iz} we certainly expect that the IR
asymptotics of the scattering amplitudes
should be related to the behavior of light-like Wilson loops with a cusp
which for conformal theories is related to the anomalous
dimensions of specific twist operators
\cite{Korchemsky:1988si,Korchemsky:1992xv,Bassetto:1993xd} (see
also \cite{Kruczenski:2002fb,Frolov:2002av,Kruczenski:2007cy, Makeenko:2002qe, Alday:2007mf}).
Of course for $\superN=4$ Yang--Mills the relationship between amplitudes and Wilson
loops goes beyond the IR divergent  piece to include the finite contributions.
The four particle amplitudes display an iterative structure in perturbation theory which can
be combined into an all order exponential form, the finite piece of which is
also governed by the cusp anomalous dimension \cite{Bern:2005iz,Anastasiou:2003kj}.
It would be interesting to see if the same is true for the Chern--Simons theory or indeed whether the even more general
relation between Wilson loops and MHV amplitudes, see for example \cite{Drummond:2007aua},
can be generalized. A moderate indication is that the NLO scattering amplitudes
vanish identically for $\superN>4$ SCS
and the same is true for NLO Wilson loops \cite{Henn:2008aa},
see also \cite{Drukker:2008zx,Chen:2008bp,Rey:2008bh}.
Relatedly it may be worthwhile to look for a version
of the dual conformal symmetry found in four-dimensional amplitudes \cite{Drummond:2006rz}.
Given this possible relation between scattering
amplitudes/Wilson loops/twist operators and
previous results \cite{Aharony:2008ug, Minahan:2008hf, Gromov:2008qe}
on the anomalous dimensions of twist operators in the
planar limit we might expect
that the two-loop scattering amplitudes in $\superN=6$ CS are related to
one-loop amplitudes in four-dimensional  $\superN=4$ YM. Another hint that
this may indeed be the case is the similarity of the one-loop correction to the
CS gluon propagator to the YM propagator.
Furthermore, at strong coupling the fact that the relevant, non-compact, part
of the geometry dual to the $\superN=6$ theory, \cite{Aharony:2008ug},
is similar leads one to
believe that such a relationship between the four-dimensional YM and
three dimensional CS is plausible. Certainly in the analysis of the spectrum of
spinning strings/twist operators marked similarity to the $AdS_5\times S^5$ case is
apparent and the proof of the relationship between open strings dual to Wilson loops
and spinning strings dual to twist operators via analytic continuation
\cite{Kruczenski:2007cy} goes through exactly as in the $AdS_5$ geometry.
However while the classical string solution dual to
four particle scattering amplitudes is almost identical \cite{Alday:2007hr, Alday:2007he}
it should be mentioned that the full geometry felt
by the string, particularly for the fermions, is different and it is not certain
that the arguments relating Wilson loops to scattering
amplitudes at strong coupling via (fermionic) T-duality will be valid in this theory
\cite{Alday:2007hr, Berkovits:2008ic, Beisert:2008iq}.

The particle representations that are used in the present paper
are merely the simplest representations of the symmetry algebra.
It is conceivable that the mass-deformed CS model has bound states
which transform in larger representations.
It might be interesting to determine the spectrum of such
composite particles and also compute their scattering matrices
(by means of unitarity).

Finally, in attempting to answer questions involving higher loops,
amplitudes involving larger numbers of particles or bound states
the global symmetries will naturally be less restrictive
and there will be more independent elements.
It may therefore be useful to see whether one can adapt
the methods of recursion relations \cite{Britto:2004ap, Britto:2005fq},
generalized unitarity \cite{Bern:1997sc} and the use of
complex momenta \cite{Witten:2003nn, Britto:2004nc} to the current context.

The analysis carried out in this paper raises several interesting questions about SCS theories,
their relationship to Yang--Mills theories and the AdS$_4$/CFT$_3$ correspondence.
There are of course, several issues pertaining specifically
to a better understanding scattering processes in $\mathcal{N}\geq 4$ SCS
theories that require further studies. As explained in the preceding sections,
the resolution of the puzzle regarding a two loop
contribution to the scattering matrix that is neither zero nor purely rational,
and an understanding of the relationship between scattering amplitudes and Wilson loops,
would yield vital insights into the gauge theories in question
as well as the AdS$_4$/CFT$_3$ correspondence.

Looking beyond the immediate problems and puzzles posed by this paper,
we would like to point out that possible connections between mass deformations
of Chern--Simons models and Yang--Mills theories in three spacetime dimensions,
are worth investigating in greater detail. In the case of massless Chern--Simons models,
both the maximally supersymmetric $\mathcal{N}=8$ BLG theory,
as well as the non-supersymmetric pure Chern--Simons theory are expected to
describe the strongly coupled dynamics
of $\mathcal{N}=8$ and $\mathcal{N}=0$ Yang--Mills theories respectively.
In the later case, the vacuum wave functionals of the Chern--Simons theory,
namely the Wess--Zumino--Witten model, and that of the strongly coupled
gluonic theory are known to be the same:
a fact that can be established using a gauge invariant Hamiltonian formulation
of pure Yang--Mills theory\cite{Karabali:1997wk,Karabali:1998yq}.
In the maximally supersymmetric case, the strongly coupled Yang--Mills theory
is expected to flow to the SCS theory at strong coupling due to the
standard dualities between D2 and M2 brane dynamics\cite{Seiberg:1997ax,Seiberg:1996nz}.
Direct evidence relating the Lagrangians of the two theories
via a Higgs mechanism has also been uncovered in\cite{Mukhi:2008ux,Kluson:2008nw,Ho:2008ei,Ezhuthachan:2008ch}.
However, the  relationship between mass-deformed SCS theories
and Yang--Mills theories, if any, remains unclear.
In this context, it is worth noting that, in the special case of
three spacetime dimensions, it is possible to carry out mass-deformations
of super Yang--Mills theories on $\Reals^{1,2}$  by using a non-local,
gauge invariant mass-term for the gluons,
and ordinary quadratic mass-terms for the matter fields.
Appropriately mass-deformed super Yang--Mills theories can also
be shown to be related to matrix models in
plane wave type backgrounds by the methods of dimensional
reduction \cite{Agarwal:2008rr}.
It is thus worth investigating if the interrelationships between massless SCS
and super Yang--Mills theories has a parallel in connections between
mass deformed Chern--Simons models and  massive Yang--Mills theories
of the type investigated in \cite{Agarwal:2008rr}.

On a related note, it might be interesting to investigate the role of
mass-deformed algebras in constraining the spacetime physics of
other gauge theories, such as $\mathcal{N}=8$
supersymmetric Yang--Mills theory on $\Reals\times S^2$
and other related Yang--Mills Chern--Simons theories
constructed in \cite{Pope:2003jp, Itzhaki:2005tu,Lin:2005nh}.

\paragraph{Acknowledgements.}

We would like to thank
L.\ Dixon,
N.\ Drukker,
J.\ Henn,
J.\ Plefka,
S.-J.\ Rey
and
R.\ Roiban
for many useful and enlightening discussions.

\appendix

\section{Conventions}
\label{app:conv}

\subsection{Spacetime}
\label{app:convspacetime}

We give a brief summary of spacetime index conventions used in this paper
following the conventions of HLLLP \cite{Hosomichi:2008jd} to a large extent.

\paragraph{Vectors.}

For vector indices we choose the signature of spacetime and
the antisymmetric tensor according to
\[
\eta^{\mu\nu}=\diag (-,+,+),
\qquad
\eta_{\mu\nu}=\diag (-,+,+),
\qquad
\varepsilon^{012}=+1.
\qquad
\varepsilon_{012}=-1.
\]

\paragraph{Spinors.}

We start by defining a basis of real symmetric and antisymmetric $2\times 2$ matrices:
\[
[\sigma^\mu]^{\alpha\beta}=
\matr{cc}{-&0\\0&-},
\matr{cc}{+&0\\0&-},
\matr{cc}{0&-\\-&0},
\qquad
\varepsilon^{\alpha\beta}=\matr{cc}{0&+\\-&0}.
\]
The conjugate basis with lower indices is defined by
$\sigma^\mu_{\alpha\beta}=\varepsilon_{\alpha\gamma}\varepsilon_{\beta\delta}\sigma^{\mu,\gamma\delta}$
and
$\varepsilon_{\alpha\beta}=\varepsilon_{\alpha\gamma}\varepsilon_{\beta\delta}\varepsilon^{\gamma\delta}$
\[
[\sigma^\mu]_{\alpha\beta}=
\matr{cc}{-&0\\0&-},
\matr{cc}{-&0\\0&+},
\matr{cc}{0&+\\+&0},
\qquad
\varepsilon_{\alpha\beta}=\matr{cc}{0&+\\-&0}.
\]
If we lower only one spinor index
$[\gamma^\mu]{}_\alpha{}^\beta=-\varepsilon_{\alpha\gamma}[\sigma^\mu]^{\gamma\beta}
=\varepsilon^{\beta\gamma}[\sigma^\mu]_{\alpha\gamma}$
we obtain
\[
[\gamma^\mu]{}_\alpha{}^\beta=
\matr{cc}{0&+\\-&0},
\matr{cc}{0&+\\+&0},
\matr{cc}{+&0\\0&-}=
i\sigma^2,\sigma^1,\sigma^3,
\]
where the latter three $\sigma^k$ refer to the standard Pauli matrices.
The gamma matrices obey the algebra
\[
\gamma^\mu\gamma^\nu = \eta^{\mu\nu}+\varepsilon^{\mu\nu\rho} \gamma_{\rho}.
\]

Spinors $\psi$ will usually carry a lower spinor index $\psi_\alpha$
so that one can conveniently multiply gamma matrices to their left, $\gamma_\mu\psi$.
To close off a sequence of gamma matrices from the left
one can use a spinor $\psi$ followed by $\varepsilon$ to raise
the index $\psi_\alpha\varepsilon^{\alpha\beta}$.
Barred spinors $\bar\psi=\psi^\ast\varepsilon$ have an upper index
$\bar\psi^\alpha=\psi^\ast_\beta\varepsilon^{\beta\alpha}$
and one can multiply gamma matrices to their right, $\bar\psi\gamma_\mu$.

To convert between vectors and bi-spinors we use the map
\[
p_{\alpha\beta}=p_\mu \sigma^\mu_{\alpha\beta}=
\matr{cc}{-p_0-p_1&p_2\\p_2&-p_0+p_1},
\qquad
p_\mu=-\half \sigma_\mu^{\alpha\beta}p_{\alpha\beta}.
\]

\paragraph{Conversion from HLLLP notation.}
\
For compatibility reasons we adopt the spinor conventions
used in HLLLP \cite{Hosomichi:2008jd}
with the only exception of the $\varepsilon_{\alpha\beta}$ symbol
with lower indices. The reason is that lowering both indices
of $\varepsilon^{\alpha\beta}$ with two $\varepsilon\supup{HLLLP}_{\alpha\beta}$
according to the prescription in \cite{Hosomichi:2008jd} leads to
$-\varepsilon\supup{HLLLP}_{\alpha\beta}$.
We thus choose the opposite sign for $\varepsilon_{\alpha\beta}$.
This is consistent with the fact that $\det \varepsilon=+1$:
The relative normalization of totally antisymmetric tensors $\varepsilon$
with upper and lower indices should be determined by
the determinant of the matrix to raise or lower indices (here: the same $\varepsilon$).

In order to avoid sign confusions we will always raise or
lower spinor indices explicitly by means of $\varepsilon$ tensors.
All the symbols will have a definite position of spinor indices;
we commonly use lower indices which are contracted
by $\varepsilon^{\alpha\beta}$.

Thus, the conversion from HLLLP \cite{Hosomichi:2008jd} to our notation consists of
the following two replacements
\[
\psi^\alpha\indup{HLLLP} = \varepsilon ^{\alpha \beta }\psi_\beta, \qquad
\varepsilon\supup{HLLLP}_{\alpha\beta}  = -\varepsilon_{\alpha \beta}.
\]

\subsection{Polarization Spinors}
\label{app:convpol}

Consider now the Dirac equation
\[
\brk{\gamma^\mu\partial_\mu -m}\psi=0.
\]
It is solved by
$\psi=\exp(+ip_\mu x^\mu)u(+p)$
and
$\psi=\exp(-ip_\mu x^\mu)u(-p)$
with the polarization spinors
\[\label{eq:uvofp}
u(p)=
\frac{1}{\sqrt{p_0-p_1}}
\matr{c}{
  p_2-im \\
  p_1-p_0
},
\qquad
v(p)=
\frac{1}{\sqrt{p_0-p_1}}
\matr{c}{
  p_2+im \\
  p_1-p_0
}.
\]
These are normalized such that
\[
v_\alpha(p)\,u_\beta(p)=-p_{\alpha\beta}-im\varepsilon_{\alpha\beta}.
\]
Obviously, the Dirac equation with opposite mass
\[
\brk{\gamma^\mu\partial_\mu +m}\twist\psi=0
\]
has the solutions
$\twist\psi= \exp(+ip_\mu x^\mu)v(+p)$
and
$\twist\psi = \exp(-ip_\mu x^\mu)v(-p)$
with $u$ replaced by $v$ interchanged.
By construction, it is also clear that for inverted momentum one obtains
\[
u(-p)=i \sign (p_0)\, v(p),\qquad
v(-p)=i \sign (p_0)\, u(p).
\]
Finally, let us note that the two polarization spinors
are related complex conjugation
\[\label{eq:uvconj}
u(p)^\ast=-i\, u(-p^\ast)
,\qquad
v(p)^\ast=
-i\, v(-p^\ast)
.
\]
%
\subsection{Completeness Relations and Conversion}

We list two completeness relations for symmetrized bispinors
\[
\sigma^{\alpha\beta}_\mu
\sigma^\mu_{\gamma\delta}
=
-\delta^\alpha_\gamma\delta^\beta_\delta
-\delta^\alpha_\delta\delta^\beta_\gamma,
\qquad
\varepsilon^{\alpha\beta}
\varepsilon_{\gamma\delta}
=
\delta^\alpha_\gamma\delta^\beta_\delta
-\delta^\alpha_\delta\delta^\beta_\gamma.
\]
These can be used to convert between vectors
and symmetric bispinors
\[
a_{\alpha\beta} = \sigma^\mu_{\alpha\beta} a_\mu,
\qquad
a_\mu=-\half \sigma_\mu^{\alpha\beta} a_{\alpha\beta}.
\]
Furthermore in three dimensions vectors and two-forms are equivalent
\[
a_\rho=\half \varepsilon_{\mu\nu\rho}a^{\mu\nu},
\qquad
a_{\mu\nu}=-\varepsilon_{\mu\nu\rho}a^\rho.
\]
Thus we can also convert directly between symmetric bispinors and two-forms
\[
a_{\alpha\beta}=-\half \sigma^\mu_{\alpha\gamma}\varepsilon^{\gamma\delta}\sigma^\nu_{\delta\beta} a_{\mu\nu} ,
\qquad
a_{\mu\nu}=-\half \sigma_\mu^{\alpha\gamma}\varepsilon_{\gamma\delta}\sigma_\nu^{\delta\beta} a_{\alpha\beta}.
\]


\section{The \texorpdfstring{$\superN=4$ Chern--Simons}{N=4 Chern-Simons}  Model}
\label{app:susyinteract}

In this appendix we define the $\superN=4$ supersymmetric Chern--Simons model
and give a summary of its symmetries.

\subsection{Definitions}
\label{sec:susyinteract.def}

We start by listing the basic fields, symbols and indices that
appear in the model.

\paragraph{Types of Indices.}
%
\begin{itemize}
\item
$M,N,P,\ldots$: gauge adjoint indices,
\item
$A,B,C,\ldots$: gauge untwisted representation indices,
\item
$\twist A,\twist B,\twist C,\ldots$: gauge twisted representation indices,
\item
$\alpha,\beta,\gamma,\ldots$: spacetime spinor indices (cf.\ \appref{app:convspacetime}),
\item
$\mu,\nu,\rho,\ldots$: spacetime vector indices (cf.\ \appref{app:convspacetime}),
\item
$a,b,c,\ldots$: flavor indices of first $\alg{su}(2)$,
\item
$\dot a,\dot b,\dot c,\ldots$: flavor indices of second $\alg{su}(2)$,
\item
$\tilde a,\tilde b,\tilde c,\ldots$: flavor indices of third $\alg{su}(2)$,
\item
$\hat a,\hat b,\hat c,\ldots$: flavor indices of fourth $\alg{su}(2)$.
\end{itemize}

\paragraph{Gauge Invariant Symbols.}


\begin{itemize}
\item
gauge algebra structure constants $F^{M}_{NP}=-F^M_{PN}$,

\item
gauge algebra (Cartan--Killing) metric $K_{MN}=K_{NM}$,

\item
untwisted (twisted) representation $T^{A}_{MB}$ ($\twist T^{\twist A}_{M\twist B}$),

\item
untwisted (twisted) moments $M^{M}_{AB}=M^M_{BA}$ ($\twist M^{M}_{\twist A\twist B}=\twist M^{M}_{\twist B\twist A}$),

\item
untwisted (twisted) metric $L_{AB}=-L_{BA}$ ($\twist L_{\twist A\twist B}=-\twist L_{\twist B\twist A}$).

\end{itemize}
The untwisted structure constants
$F^M_{NP},T^A_{MB},M^M_{AB}$
obey the Jacobi identities of a Lie superalgebra
\<
0\eq F^M_{RN}F^R_{PQ}+F^M_{RP}F^R_{QN}+F^M_{RQ}F^R_{NP},
\nln
0\eq T^A_{CM}T^C_{NB}-T^A_{CN}T^C_{MB}+T^A_{PB}F^P_{MN},
\nln
0\eq F^M_{NP}M^P_{AB}+M^M_{AC}T^C_{NB}+M^M_{BC}T^C_{NA},
\nln
0\eq T^A_{MB}M^M_{CD}+T^A_{MC}M^M_{DB}+T^A_{MD}M^M_{BC}.
\>
Furthermore the compatibility of structure constants
and metric implies the relation
\[
L_{AC}T^C_{MB}=K_{MN}M^N_{AB}.
\]
The twisted constants
$F^M_{NP},\twist T^{\twist A}_{M\twist B},\twist M^M_{\twist A\twist B},\twist L_{\twist A\twist B}$
obey the same Lie superalgebra relations.
The Lie superalgebra for the twisted sector need not
be isomorphic to the one for the ``untwisted'' sector;
the even parts defined through $F^M_{NP}$ must be isomorphic,
but the odd parts
$\twist T^{\twist A}_{M\twist B},\twist M^M_{\twist A\twist B},\twist L_{\twist A\twist B}$
can differ.

\paragraph{Fields and Combinations.}

The most general non-abelian model is based upon the following five types of fields:
\begin{itemize}
\item
gauge field $\mathcal{A}_{\alpha\beta}^M=\sigma^\mu_{\alpha\beta}A^M_\mu$,
\item
untwisted scalars $\phi^A_a$ and fermions $\psi^A_{\alpha \dot b}$,
\item
twisted scalars $\twist\phi^{\twist A}_{\dot a}$ and fermions $\twist\psi^{\twist A}_{\alpha b}$.
\end{itemize}
The following combinations of fields (field strength, covariant derivatives,
moments, currents) have proven useful
\[\begin{array}[b]{rclcrcl}
\multicolumn{7}{c}{
\mathcal{F}^M_{\alpha\beta}=
-\half\varepsilon^{\gamma\delta}\partial_{\alpha\gamma}\mathcal{A}_{\beta\delta}^M
-\half\varepsilon^{\gamma\delta}\partial_{\beta\gamma}\mathcal{A}_{\alpha\delta}^M
-\half\varepsilon^{\gamma\delta}F^M_{NP}\mathcal{A}^N_{\alpha\gamma}\mathcal{A}_{\beta\delta}^P,
}
\\[3pt]
\mathcal{D}_{\alpha\beta} \mathcal{X}^A\eq\partial_{\alpha\beta} \mathcal{X}^A+T^A_{MB}\mathcal{A}_{\alpha\beta}^M \mathcal{X}^B,
&&
\mathcal{D}_{\alpha\beta} \mathcal{X}^{\twist A}\eq\partial_{\alpha\beta} \mathcal{X}^{\twist A}+\twist T^{\twist A}_{M\twist B}\mathcal{A}_{\alpha\beta}^M \mathcal{X}^{\twist B},
\\[3pt]
\mathcal{M}^M_{ab}\eq M^M_{AB}\phi^A_a \phi^B_b,
&&
\mathcal{\twist M}^M_{\dot a\dot b}\eq \twist M^M_{\twist A\twist B}\twist\phi^{\twist A}_{\dot a}\twist\phi^{\twist B}_{\dot b},
\\[3pt]
\mathcal{J}^M_{\alpha b\dot c}\eq M^M_{AB}\phi^A_b \psi^B_{\alpha \dot c},
&&
\mathcal{\twist J}^M_{\alpha \dot bc}\eq \twist M^M_{\twist A\twist B}\twist\phi^{\twist A}_{\dot b} \twist\psi^{\twist B}_{\alpha c}.
\end{array}
\]
With respect to the conventions in \cite{Hosomichi:2008jd}
we have rescaled the fields for our convenience as follows:
\[\label{eq:fielddict}
\begin{array}[b]{rclcrcl}
\phi^A_a\earel{\to}+\sqrt{4\pi}\, q^A_a,&&
\twist\phi^{\twist A}_{\dot a}\earel{\to}-\sqrt{4\pi}\, \twist q^{\twist A}_{\dot a},
\\[3pt]
\psi^A_{\alpha\dot b}\earel{\to}+\sqrt{4\pi}\, \psi^A_{\alpha\dot b},&&
\twist\psi^{\twist A}_{\alpha b}\earel{\to}+\sqrt{4\pi}\, \twist\psi^{\twist A}_{\alpha b}.
\end{array}\]

\subsection{Action}
\label{app:susyinteract.action}


The action which appears in the path integral as $e^{iS}$
is defined as $S=\int d^3x\,\Lagr$ with the Lagrangian
$(4\pi/k)\Lagr$:
\begin{align}
\label{eqn:CSLag}
&
-\sfrac{1}{2}
 K_{MN}
 \varepsilon^{\beta\gamma}\varepsilon^{\delta\kappa}\varepsilon^{\lambda\alpha}
 \mathcal{A}^M_{\alpha\beta}\partial_{\gamma\delta}\mathcal{A}^N_{\kappa\lambda}
&&
\ifarxiv\smash{\raisebox{-6.1cm}{\makebox[0cm][c]{\hspace{+9.6cm}\includegraphics[height=2cm]{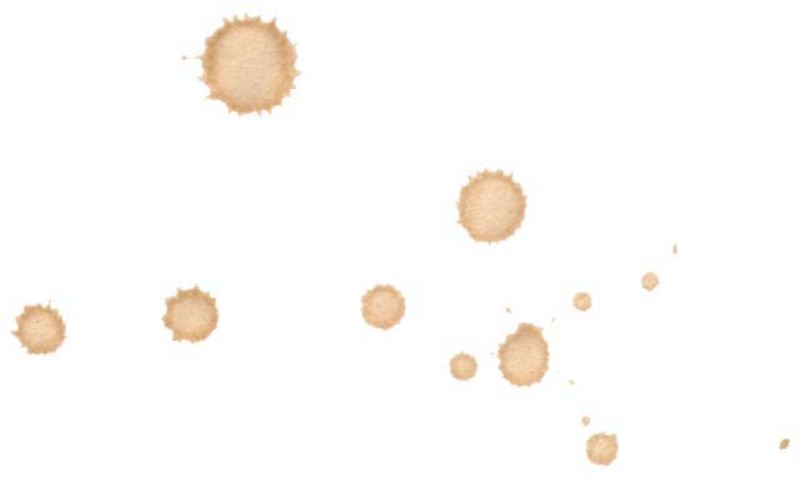}}}}\fi
\nln&
-\sfrac{1}{6}
 K_{MN}F^N_{PQ}
 \varepsilon^{\beta\gamma}\varepsilon^{\delta\kappa}\varepsilon^{\lambda\alpha}
 \mathcal{A}^M_{\alpha\beta}\mathcal{A}^P_{\gamma\delta}\mathcal{A}^Q_{\kappa\lambda}
\nln&
+\sfrac{1}{4}
 L_{AB}
 \varepsilon^{ab}\varepsilon^{\gamma\kappa}\varepsilon^{\delta\lambda}
 \mathcal{D}_{\gamma\delta}\phi^A_a\mathcal{D}_{\kappa\lambda}\phi^B_b
&&
+\sfrac{1}{4}
 \twist L_{\twist A\twist B}
 \varepsilon^{\dot a\dot b}\varepsilon^{\gamma\kappa}\varepsilon^{\delta\lambda}
 \mathcal{D}_{\gamma\delta}\twist \phi^{\twist A}_{\dot a}\mathcal{D}_{\kappa\lambda}\twist\phi^{\twist B}_{\dot b}
\nln&
-\sfrac{1}{2}m^2
 L_{AB}
 \varepsilon^{ab}
 \phi^A_a\phi^B_b
&&
-\sfrac{1}{2}m^2
 \twist L_{\twist A\twist B}
 \varepsilon^{\dot a\dot b}
 \twist\phi^{\twist A}_{\dot a}\twist\phi^{\twist B}_{\dot b}
\nln&
+\sfrac{i}{2}
 L_{AB}
 \varepsilon^{\dot a\dot b}\varepsilon^{\gamma\kappa}\varepsilon^{\lambda\delta}
 \psi^A_{\gamma \dot a}\mathcal{D}_{\kappa\lambda}\psi^B_{\delta\dot b}
&&
+\sfrac{i}{2}
 \twist L_{\twist A\twist B}\varepsilon^{ab}\varepsilon^{\gamma\kappa}\varepsilon^{\lambda\delta}
 \twist\psi^{\twist A}_{\gamma a}\mathcal{D}_{\kappa\lambda}\twist\psi^{\twist B}_{\delta b}
\nln&
+\sfrac{i}{2}m
 L_{AB}
 \varepsilon^{\dot a\dot b}\varepsilon^{\gamma\delta}
 \psi^A_{\gamma\dot a}\psi^B_{\delta\dot b}
&&
-\sfrac{i}{2}m
 \twist L_{\twist A\twist B}\varepsilon^{ab}\varepsilon^{\gamma\delta}
 \twist\psi^{\twist A}_{\gamma a}\twist\psi^{\twist B}_{\delta b}
\nln&
+\sfrac{i}{4}
 K_{MN}
 \varepsilon^{ab}\varepsilon^{\dot c\dot d}\varepsilon^{\kappa\lambda}
 \mathcal{J}^M_{\kappa a\dot c}\mathcal{J}^N_{\lambda b\dot d}
&&
+\sfrac{i}{4}
 K_{MN}
 \varepsilon^{\dot a\dot b}\varepsilon^{cd}\varepsilon^{\kappa\lambda}
 \mathcal{\twist J}^M_{\kappa \dot ac}\mathcal{\twist J}^N_{\lambda \dot bd}
\nln&
+i
 K_{MN}
 \varepsilon^{ad}\varepsilon^{\dot b\dot c}\varepsilon^{\kappa\lambda}
 \mathcal{J}^M_{\kappa a\dot b}\mathcal{\twist J}^N_{\lambda\dot cd}
&&
\nln&
-\sfrac{i}{4}
 L_{AC}T^C_{MB}
 \varepsilon^{\dot a\dot c}\varepsilon^{\dot b\dot d}\varepsilon^{\kappa\lambda}
 \mathcal{\twist M}^M_{\dot a\dot b}\psi^A_{\kappa\dot c}\psi^B_{\lambda \dot d}
&&
-\sfrac{i}{4}
 \twist L_{\twist A\twist C}\twist T^{\twist C}_{M\twist B}
 \varepsilon^{ac}\varepsilon^{bd}\varepsilon^{\kappa\lambda}
 \mathcal{M}^M_{ab}\twist\psi^{\twist A}_{\kappa c}\twist\psi^{\twist B}_{\lambda d}
\nln&
+\sfrac{1}{6}m
 K_{MN}
 \varepsilon^{bc}\varepsilon^{da}
 \mathcal{M}^M_{ab}\mathcal{M}^N_{cd}
&&
-\sfrac{1}{6}m
 K_{MN}
 \varepsilon^{\dot b\dot c}\varepsilon^{\dot d\dot a}
 \mathcal{\tilde M}^M_{\dot a\dot b}\mathcal{\tilde M}^N_{\dot c\dot d}
\nln&
+\sfrac{1}{96}
 K_{MN}F^N_{PQ}
 \varepsilon^{bc}\varepsilon^{de}\varepsilon^{fa}
 \mathcal{M}^M_{ab}\mathcal{M}^P_{cd}\mathcal{M}^Q_{ef}
&&
+\sfrac{1}{96}
 K_{MN}F^N_{PQ}
 \varepsilon^{\dot b\dot c}\varepsilon^{\dot d\dot e}\varepsilon^{\dot f\dot a}
 \mathcal{\twist M}^M_{\dot a\dot b}\mathcal{\twist M}^P_{\dot c\dot d}\mathcal{\twist M}^Q_{\dot e\dot f}
\nln&
+\sfrac{1}{16}
 \twist L_{\twist C\twist D} \twist T^{\twist C}_{M\twist A} \twist T^{\twist D}_{N\twist B}
 \varepsilon^{\dot a\dot b}\varepsilon^{ce}\varepsilon^{df}
 \mathcal{M}^M_{cd}\mathcal{M}^N_{ef}\twist\phi^{\twist A}_{\dot a}\twist\phi^{\twist B}_{\dot b}
&&
+\sfrac{1}{16}
 L_{CD} T^C_{MA} T^D_{NB}
 \varepsilon^{ab}\varepsilon^{\dot c\dot e}\varepsilon^{\dot d\dot f}
 \mathcal{\twist M}^M_{\dot c\dot d}\mathcal{\twist M}^N_{\dot e\dot f}\phi^{A}_{a}\phi^{B}_{b}.
\end{align}
This is equivalent to the action presented in \cite{Hosomichi:2008jd} after using the dictionary
in \eqref{eq:fielddict}.

\subsection{Symmetries}

Here we collect the global symmetries of the model.

\paragraph{Rotations.}

The $\alg{su}(2)\oplus\alg{su}(2)$ flavor and $\alg{sl}(2)$ Lorentz rotations act on the
corresponding indices of some field $\mathcal{X}$
as follows
\<
\gen{R}_{ab}\mathcal{X}_{c}\eq
\ihalf\varepsilon_{bc}\mathcal{X}_a
+\ihalf\varepsilon_{ac}\mathcal{X}_b,
\nln
\gen{\dot R}_{\dot a\dot b}\mathcal{X}_{\dot c}\eq
\ihalf\varepsilon_{\dot b\dot c}\mathcal{X}_{\dot a}
+\ihalf\varepsilon_{\dot a\dot c}\mathcal{X}_{\dot b},
\nln
\gen{L}_{\alpha\beta}\mathcal{X}_{\gamma}\eq
\half\varepsilon_{\beta\gamma}\mathcal{X}_{\alpha}
+\half\varepsilon_{\alpha\gamma}\mathcal{X}_{\beta}.
\>

\paragraph{Translations.}
The momentum generators act by covariant derivatives
\<
\gen{P}_{\alpha\beta}\mathcal{X}^A\eq
-i\mathcal{D}_{\alpha\beta}\mathcal{X}^A,
\nln
\gen{P}_{\alpha\beta}\mathcal{X}^{\twist A}\eq
-i\mathcal{D}_{\alpha\beta}\mathcal{X}^{\twist A},
\nln
\gen{P}_{\alpha\beta}\mathcal{A}^M_{\gamma\delta}\eq
\ihalf\varepsilon_{\beta\delta}\mathcal{F}^M_{\alpha\gamma}
+\ihalf\varepsilon_{\beta\gamma}\mathcal{F}^M_{\alpha\delta}
+\ihalf\varepsilon_{\alpha\delta}\mathcal{F}^M_{\beta\gamma}
+\ihalf\varepsilon_{\alpha\gamma}\mathcal{F}^M_{\beta\delta}.
\>

\paragraph{Supersymmetry.}
Supersymmetry generators act on the fields according to the rules
\<
\gen{Q}_{\alpha b\dot c} \phi^{A}_{d}\eq
\varepsilon_{bd}\psi^{A}_{\alpha\dot c},
\nln
\gen{Q}_{\alpha b\dot c}\twist\phi^{\twist A}_{\dot d}\eq
\varepsilon_{\dot c\dot d}\twist\psi^{\twist A}_{\alpha b},
\ifarxiv\smash{\raisebox{-0cm}{\makebox[0cm][c]{\hspace{18cm}\includegraphics[height=1cm]{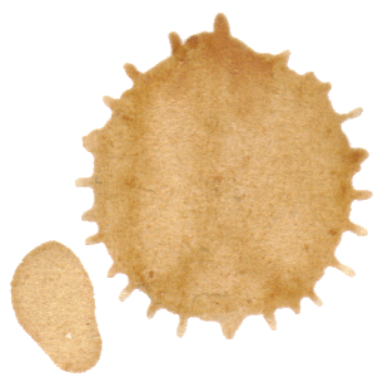}}}}\fi
\nln
\gen{Q}_{\alpha b\dot c} \psi^{A}_{\delta \dot e}\eq
i\varepsilon_{\dot c\dot e}\bigbrk{\mathcal{D}_{\alpha\delta}-m\varepsilon_{\alpha\delta}}\phi^{A}_{b}
+
i\varepsilon_{\alpha\delta}
\bigbrk{
\sfrac{1}{6}\mathcal{M}^M_{bf} \varepsilon^{fg}\varepsilon_{\dot c\dot e}
-\sfrac{1}{2}\delta_b^g\mathcal{\twist M}^M_{\dot c\dot e}
}T^A_{MB}\phi^{B}_{g},
\nln
\gen{Q}_{\alpha b\dot c} \twist\psi^{\twist A}_{\delta e}\eq
i\varepsilon_{be}\bigbrk{\mathcal{D}_{\alpha\delta}+m\varepsilon_{\alpha\delta}}\twist\phi^{\twist A}_{\dot c}
+
i\varepsilon_{\alpha\delta}\bigbrk{
\sfrac{1}{6}\varepsilon_{be}\mathcal{\twist M}^M_{\dot c\dot f}\varepsilon^{\dot f\dot g}
-\sfrac{1}{2}\mathcal{M}^M_{be}\delta^{\dot g}_{\dot c}
}
\twist T^{\twist A}_{M\twist B}\twist\phi^{\twist B}_{\dot g},
\nln
\gen{Q}_{\alpha b\dot c} \mathcal{A}^{A}_{\delta\epsilon}
\eq
\half\varepsilon_{\alpha\delta}
\mathcal{J}^M_{\epsilon b\dot c}
+\half\varepsilon_{\alpha\epsilon}\mathcal{J}^M_{\delta b\dot c}
+\half\varepsilon_{\alpha\delta}
\mathcal{\twist J}^M_{\epsilon \dot c b}
+\half\varepsilon_{\alpha\epsilon}
\mathcal{\twist J}^M_{\delta\dot c b}.
\>
The supersymmetry variation $\delta$ of \cite{Hosomichi:2008jd}
corresponds to the action of $\delta=i\eta^{\alpha b\dot c}\gen{Q}_{\alpha b\dot c}$
with a fermionic field $\eta$. Lowering some of the indices of this field
yields
\[
\eta^{\alpha}{}_{d}{}^{\dot c}=\eta^{\alpha b\dot c}\varepsilon_{bd},\quad
\eta^{\alpha b}{}_{\dot d}=\eta^{\alpha b\dot c}\varepsilon_{\dot c\dot d},\quad
\eta_{\delta}{}^{b}{}_{\dot e}=\eta^{\alpha b\dot c}\varepsilon_{\alpha\delta} \varepsilon_{\dot c\dot e},\quad
\eta_{\delta e}{}^{\dot c}=\eta^{\alpha b\dot c}\varepsilon_{\alpha\delta}\varepsilon_{be}.
\]

\subsection{Interacting Symmetry Algebra}

The symmetry algebra takes the form described in \secref{sec:superalg}.
However, it is well known that the symmetry algebra in an interacting
gauge theory closes only on shell and modulo (field-dependent)
gauge transformations. The additional terms form an ideal
of the algebra and thus can be factored out consistently
by acting only on on-shell, gauge-invariant states.

\paragraph{Commutators.}

The additional terms in the commutators can be written explicitly
as
\<
\acomm{\gen{Q}_{\alpha b\dot c}}{\gen{Q}_{\delta e\dot f}}\eq
\varepsilon_{be}\varepsilon_{\dot c\dot f}
(+\gen{P}_{\alpha\delta}+\gen{E}_{\alpha\delta})
\nl
+
\varepsilon_{\alpha\delta}\varepsilon_{\dot c\dot f}
(-2m\gen{R}_{be}-\half\gen{G}[\mathcal{M}_{be}]+\gen{E}_{be})
\nl
+\varepsilon_{\alpha\delta}\varepsilon_{be}
(+2m\gen{\dot R}_{\dot c\dot f}-\half\gen{\twist G}[\mathcal{\twist M}_{\dot c\dot f}]+\gen{E}_{\dot c\dot f}),
\nln
\comm{\gen{P}_{\alpha \beta}}{\gen{Q}_{\gamma d\dot e}}\eq
\half\varepsilon_{\beta\gamma}
\gen{G}[\mathcal{J}_{\alpha d\dot e}+\mathcal{\twist J}_{\alpha \dot e d}]
+\half\varepsilon_{\alpha\gamma}
\gen{G}[\mathcal{J}_{\beta d\dot e}+\mathcal{\twist J}_{\beta \dot e d}],
\nln
\comm{\gen{P}_{\alpha\beta}}{\gen{P}_{\gamma\delta}}\eq
\ihalf\varepsilon_{\beta\delta}\gen{G}[\mathcal{F}_{\alpha\gamma}]
+\ihalf\varepsilon_{\beta\gamma}\gen{G}[\mathcal{F}_{\alpha\delta}]
+\ihalf\varepsilon_{\alpha\delta}\gen{G}[\mathcal{F}_{\beta\gamma}]
+\ihalf\varepsilon_{\alpha\gamma}\gen{G}[\mathcal{F}_{\beta\delta}]
.
\>
The generator $\gen{E}$ annihilates on-shell fields and
the generators $\gen{G}[\mathcal{X}]$ are field-dependent gauge transformations.

\paragraph{Gauge Transformations.}

The generators $\gen{G}[\mathcal{X}]$
are gauge variations
with variation parameter $\mathcal{X}^M$
\<
\gen{G}[\mathcal{X}]\mathcal{Y}^A \eq -iT^A_{MB}\mathcal{X}^M \mathcal{Y}^B,
\nln
\gen{G}[\mathcal{X}]\mathcal{Y}^{\twist A} \eq -i\twist T^{\twist A}_{M\twist B}\mathcal{X}^M \mathcal{Y}^{\twist B},
\nln
\gen{G}[\mathcal{X}]\mathcal{A}^M_{\alpha\beta}\eq i \mathcal{D}_{\alpha\beta} \mathcal{X}^M.
\>
The gauge transformation generate an ideal of the full
symmetry algebra:
One can show that commutators of gauge transformations
close onto gauge transformations, explicitly
\[
\bigcomm{\gen{J}}{\gen{G}[\mathcal{X}]}=
\gen{G}[\gen{J}\mathcal{X}].
\]
%

\paragraph{Equation of Motion Generators.}

The action of the generators $\gen{E}$
is defined as
\<
\gen{E}_{\alpha\beta}\psi^A_{\gamma\dot d}\eq
\ihalf\varepsilon_{\beta\gamma}\eom{\psi^A_{\alpha\dot d}}
+\ihalf\varepsilon_{\alpha\gamma}\eom{\psi^A_{\beta\dot d}},
\nln
\gen{E}_{\alpha\beta}\twist\psi^{\twist A}_{\gamma d}\eq
\ihalf\varepsilon_{\beta\gamma}\eom{\twist\psi^A_{\alpha d}}
+\ihalf\varepsilon_{\alpha\gamma}\eom{\twist\psi^A_{\beta d}},
\nln
\gen{E}_{\alpha\beta}\mathcal{A}^{M}_{\gamma\delta}\eq
-\ihalf\varepsilon_{\beta\gamma}\eom{\mathcal{A}^{M}_{\alpha\delta}}
-\ihalf\varepsilon_{\beta\delta}\eom{\mathcal{A}^{M}_{\alpha\gamma}}
-\ihalf\varepsilon_{\alpha\gamma}\eom{\mathcal{A}^{M}_{\beta\delta}}
-\ihalf\varepsilon_{\alpha\delta}\eom{\mathcal{A}^{M}_{\beta\gamma}},
\nln
\gen{E}_{ab}\twist\psi^{\twist A}_{\gamma d}\eq
-\ihalf\varepsilon_{bd}\eom{\twist\psi^{\twist A}_{\gamma a}}
-\ihalf\varepsilon_{ad}\eom{\twist\psi^{\twist A}_{\gamma b}},
\nln
\gen{E}_{\dot a\dot b}\psi^{A}_{\gamma \dot d}\eq
-\ihalf\varepsilon_{\dot b\dot d}\eom{\psi^{A}_{\gamma \dot a}}
-\ihalf\varepsilon_{\dot a\dot d}\eom{\psi^{A}_{\gamma \dot b}}.
\>
They annihilate on-shell fields
because $\eom{\mathcal{X}}=0$ is defined as the equation of
motion for the field $\mathcal{X}$
\<
\eom{\phi^A_a}\eq \sfrac{1}{2}\varepsilon^{\gamma\kappa}\varepsilon^{\delta\lambda}
\mathcal{D}_{\gamma\delta}\mathcal{D}_{\kappa\lambda}\phi^A_a-m^2 \phi^A_a\nl
+\sfrac{i}{2}\varepsilon^{\dot c \dot d}\varepsilon ^{\kappa\lambda}T_{N}^{A}{}_B
\psi_{\kappa\dot c}^B\mathcal{J}^N_{\lambda a\dot d}
+i \varepsilon^{\dot b \dot c}\varepsilon^{\kappa \lambda}T_N^{A}{}_B
\psi^B_{\kappa \dot b}\mathcal{\twist J}^N_{\lambda \dot c a}\nl
-\sfrac{i}{2}{\tilde M}^M_{\bar C\bar D}\varepsilon^{bd}\varepsilon^{\kappa\lambda}T_M^A{}_{B}\phi^B_b
\tilde{\psi}^{\bar C}_{\kappa a}\tilde{\psi}^{\bar D}_{\lambda d}
+\sfrac{2}{3}m \varepsilon^{bd}T^A_N{}_{B}\phi^B_b\mathcal{M}^N_{ad}\nl
+\sfrac{1}{16}F^N_{PQ}T_N^A{}_B\varepsilon^{bc}\varepsilon^{de}\mathcal{M}_{ab}^P\mathcal{M}_{cd}\phi^B_e
+\sfrac{1}{4}{\tilde M}^M_{\bar A\bar D}{\tilde T}_N^{\bar D}{}_{\bar B}\varepsilon^{bf}T_M^A{}_{B}
\phi^B_b\mathcal{M}^M_{af}{\tilde \phi}^{\bar A}_{\dot c}{\tilde \phi}^{\bar B}_{\dot d}\nl
+\sfrac{1}{8}L_{CD}T_M^{CA}T^D_{N}{}_B\varepsilon^{\dot c\dot e}\varepsilon^{\dot d\dot f}
\tilde{\mathcal{M}}^M_{\dot c\dot d}\tilde{\mathcal{M}}^N_{\dot e\dot f}\phi^B_a,
\nln
\eom{\twist\phi^{\twist A}_{\dot a}}\eq \sfrac{1}{2}\varepsilon^{\gamma\kappa}\varepsilon^{\delta\lambda}
\mathcal{D}_{\gamma\delta}\mathcal{D}_{\kappa\lambda}\phi^A_{\dot a}-m^2 \phi^A_{\dot a}\nl
+\sfrac{i}{2}\varepsilon^{ c  d}\varepsilon ^{\kappa\lambda}{\tilde T}_N^{\bar A}{}_{\bar B}
{\tilde \psi}_{\kappa c}^{\bar B}\tilde{\mathcal{J}}^N_{\lambda \dot a d}
+i \varepsilon^{ b  c}\varepsilon^{\kappa \lambda}{\tilde T}_N^{\bar A}{}_{\bar B}
{\tilde \psi}^{\bar B}_{\kappa  b}\mathcal{ J}^N_{\lambda  c \dot a}\nl
-\sfrac{i}{2}{ M}^M_{ C D}\varepsilon^{\dot b\dot d}\varepsilon^{\kappa\lambda}{\tilde T}_M^{\bar A}{}_{\bar B}{\tilde \phi}^{\bar B}_{\dot b}
{\psi}^{ C}_{\kappa \dot a}{\psi}^{ D}_{\lambda \dot d}
-\sfrac{2}{3}m \varepsilon^{\dot b\dot d}{\tilde T}^{\bar A}_N{}_{\bar B}{\tilde \phi}^{\bar B}_{\dot b}\mathcal{\twist M}^N_{\dot a\dot d}\nl
+\sfrac{1}{16}F^N_{PQ}{\tilde T}_N^{\bar A}{}_{\bar B}\varepsilon^{\dot b \dot c}\varepsilon^{\dot d\dot e}\mathcal{\twist M}_{\dot a\dot b}^P\mathcal{\twist M}_{\dot c\dot d}{\tilde \phi}^{\bar B}_{\dot e}
+\sfrac{1}{4}{ M}^M_{ A D}{ T}_N^{ D}{}_{ B}\varepsilon^{\dot b\dot f} {\tilde T}_M^{\bar A}{}_{\bar B}
{\tilde \phi}^{\bar B}_{\dot b} \mathcal{\twist M}^M_{\dot a\dot f} {\phi}^{ A}_{ c} {\phi}^{B}_{ d}\nl
+\sfrac{1}{8}{\tilde L}_{\bar C\bar D} {\tilde T}_M^{\bar C\bar A} {\tilde T}^{\bar D}_{N}{}_{\bar B} \varepsilon^{ c e}\varepsilon^{ d f}
{\mathcal{M}}^M_{ c d}{\mathcal{M}}^N_{ ef}{\tilde \phi}^{\bar B}_{\dot a},
\nln
\eom{\psi^A_{\alpha\dot b}}
\eq
\varepsilon^{\gamma\delta}
\mathcal{D}_{\alpha\gamma}\psi^{A}_{\delta\dot b}
+m\psi^{A}_{\alpha\dot b}
+T^{A}_{MB}
(\mathcal{J}^M_{\alpha c \dot b}+\mathcal{\twist J}^M_{\alpha \dot b c})
\varepsilon^{cd}\phi^B_d ,
\nln
\eom{\twist\psi^{\twist A}_{\alpha b}}
\eq
\varepsilon^{\gamma\delta}
\mathcal{D}_{\alpha\gamma}\twist\psi^{\twist A}_{\delta b}
-m\twist\psi^{\twist A}_{\alpha b}
+\twist T^{\twist A}_{M\twist B}
(\mathcal{J}^M_{\alpha b \dot c}+\mathcal{\twist J}^M_{\alpha \dot c b})
\varepsilon^{\dot c\dot d}\twist \phi^{\twist B}_{\dot d},
\nln
\eom{\mathcal{A}^{M}_{\alpha\beta}}
\eq
\mathcal{F}^M_{\alpha\beta}
+\half M^M_{AB}\varepsilon^{cd}\phi^A_c\mathcal{D}_{\alpha\beta}\phi^B_d
+\half \twist M^M_{\twist A\twist B}\varepsilon^{\dot c\dot d}\twist\phi^{\twist A}_{\dot c}\mathcal{D}_{\alpha\beta}\twist\phi^{\twist B}_{\dot d}
\nl
+\ihalf M^M_{AB}\varepsilon^{\dot c\dot d}\psi^A_{\alpha \dot c}\psi^B_{\beta\dot d}
+\ihalf \twist M^M_{\twist A\twist B}\varepsilon^{cd}\twist\psi^{\twist A}_{\alpha c}\twist\psi^{\twist B}_{\beta d}.
\>
The commutators of the generators $\gen{E}$ close
onto further generators $\gen{E}$ which annihilate
all on-shell fields. Thus they also form an ideal
of the symmetry algebra.

\paragraph{Oscillator Expansion.}

We use the following oscillator expansions to the fields, using the basis of solutions found in \eqref{app:convpol}:
\<
\label{eqn:Mode_exp}
\phi^A_a(x) \eq \int \frac{d^2p}{\sqrt{2 E(p)}} \left(e^{-ip\cdot x}a^A_{a}(p) + e^{ip\cdot x}a^{A\dagger}_{a}( p)\right)\nln
\tilde{\phi}^A_{\dot a}(x) \eq \int \frac{d^2p}{\sqrt{2 E(p)}}\left(e^{-ip\cdot x}\tilde{a}^A_{\dot a}(p) + e^{ip\cdot x}\tilde{a}^{A\dagger}_{\dot a}(p)\right)\nln
\psi^A_{\dot{a}}(x) \eq \int \frac{d^2p}{\sqrt{2 E(p)}}\left(u(p)e^{ip\cdot x}b^{\dagger A}_{\dot{a}}(p) + v(p)e^{-ip\cdot x}b^{A}_{\dot{a}}(p)\right)\nln
\tilde{\psi}^A_{a}(x) \eq \int \frac{d^2p}{\sqrt{2 E(p)}}\left(v(p)e^{ip\cdot x}\tilde{b}^{\dagger A}_{a}(p) + u(p)e^{-ip\cdot x}\tilde{b}^{A}_{a}(p)\right).
\>
$\phi(x)$ and $\psi(x)$ are the bosonic and fermionic fields in the action found
in \Appref{app:susyinteract}
and the action \eqref{eqn:CSLag}.
Requiring that the linearized $\mathcal{N}=4$ algebra
be realized on bosonic/fermionic oscillator  states,
leads to \eqref{eq:susyrep}.
We note that the positive and negative energy frequency parts
for the twisted fermions are related to the conjugates of those for the untwisted ones.
This has to do with the fact that the twisted fermions have a negative mass compared to the untwisted ones.

\bibliographystyle{nb}
\bibliography{MassCS}

\end{document}